\documentclass[aip,jcp,reprint,noshowkeys,superscriptaddress]{revtex4-1}
\usepackage{graphicx,dcolumn,bm,xcolor,microtype,multirow,amscd,amsmath,amssymb,amsfonts,physics,longtable,wrapfig,txfonts}
\usepackage[version=4]{mhchem}

\usepackage[utf8]{inputenc}
\usepackage[T1]{fontenc}
\usepackage{txfonts}

\usepackage[
	colorlinks=true,
    citecolor=blue,
    breaklinks=true
	]{hyperref}
\newcommand{\ie}{\textit{i.e.}}

\newcommand{\alert}[1]{\textcolor{black}{#1}}
\usepackage[normalem]{ulem}

\newcommand{\mc}{\multicolumn}
\newcommand{\fnm}{\footnotemark}
\newcommand{\fnt}{\footnotetext}

\newcommand{\SI}{\textcolor{blue}{supporting information}}

\newcommand{\T}[1]{#1^{\intercal}}

\newcommand{\br}{\mathbf{r}}

\newcommand{\GOWO}{$G_0W_0$}	

\newcommand{\xc}{\text{xc}}

\newcommand{\co}{\text{c}}
\newcommand{\x}{\text{x}}

\newcommand{\hS}{\Hat{S}}

\newcommand{\KS}{\text{KS}}

\newcommand{\RPA}{\text{RPA}}
\newcommand{\BSE}{\text{BSE}}
\newcommand{\dBSE}{\text{dBSE}}

\newcommand{\stat}{\text{stat}}

\newcommand{\e}[1]{\eps_{#1}}

\newcommand{\eKS}[1]{\eps^\text{KS}_{#1}}

\newcommand{\eGW}[1]{\eps^{GW}_{#1}}

\newcommand{\Om}[2]{\Omega_{#1}^{#2}}

\newcommand{\A}[2]{A_{#1}^{#2}}

\newcommand{\B}[2]{B_{#1}^{#2}}

\newcommand{\Sig}[2]{\Sigma_{#1}^{#2}}

\newcommand{\MO}[1]{\phi_{#1}}
\newcommand{\ERI}[2]{(#1|#2)}

\newcommand{\bO}{\mathbf{0}}

\newcommand{\bOm}[1]{\mathbf{\Omega}^{#1}}
\newcommand{\bA}[2]{\mathbf{A}_{#1}^{#2}}
\newcommand{\bB}[2]{\mathbf{B}_{#1}^{#2}}
\newcommand{\bX}[2]{\mathbf{X}_{#1}^{#2}}
\newcommand{\bY}[2]{\mathbf{Y}_{#1}^{#2}}
\newcommand{\bZ}[2]{\mathbf{Z}_{#1}^{#2}}

\newcommand{\eps}{\varepsilon}

\newcommand{\sig}{\sigma}
\newcommand{\bsig}{{\Bar{\sigma}}}
\newcommand{\sigp}{{\sigma'}}

\newcommand{\taup}{{\tau'}}

\newcommand{\up}{\uparrow}
\newcommand{\dw}{\downarrow}
\newcommand{\upup}{\uparrow\uparrow}
\newcommand{\updw}{\uparrow\downarrow}
\newcommand{\dwup}{\downarrow\uparrow}
\newcommand{\dwdw}{\downarrow\downarrow}
\newcommand{\spc}{\text{sc}}
\newcommand{\spf}{\text{sf}}

\newcommand{\LCPQ}{Laboratoire de Chimie et Physique Quantiques (UMR 5626), Universit\'e de Toulouse, CNRS, UPS, France}

\begin{document}	

\title{Spin-Conserved and Spin-Flip Optical Excitations From the Bethe-Salpeter Equation Formalism}

\author{Enzo \surname{Monino}}
	\affiliation{\LCPQ}	 
\author{Pierre-Fran\c{c}ois \surname{Loos}}
	\email{loos@irsamc.ups-tlse.fr}
	\affiliation{\LCPQ}

\begin{abstract}
Like adiabatic time-dependent density-functional theory (TD-DFT), the Bethe-Salpeter equation (BSE) formalism of many-body perturbation theory, in its static approximation, is ``blind'' to double (and higher) excitations, which are ubiquitous, for example, in conjugated molecules like polyenes. 
Here, we apply the spin-flip \textit{ansatz} (which considers the lowest triplet state as the reference configuration instead of the singlet ground state) to the BSE formalism in order to access, in particular, double excitations.
The present scheme is based on a spin-unrestricted version of the $GW$ approximation employed to compute the charged excitations and screened Coulomb potential required for the BSE calculations. 
Dynamical corrections to the static BSE optical excitations are taken into account via an unrestricted generalization of our recently developed (renormalized) perturbative treatment.
The performance of the present spin-flip BSE formalism is illustrated by computing excited-state energies of the beryllium atom, the hydrogen molecule at various bond lengths, and cyclobutadiene in its rectangular and square-planar geometries.
\bigskip
\begin{center}
	\boxed{\includegraphics[width=0.4\linewidth]{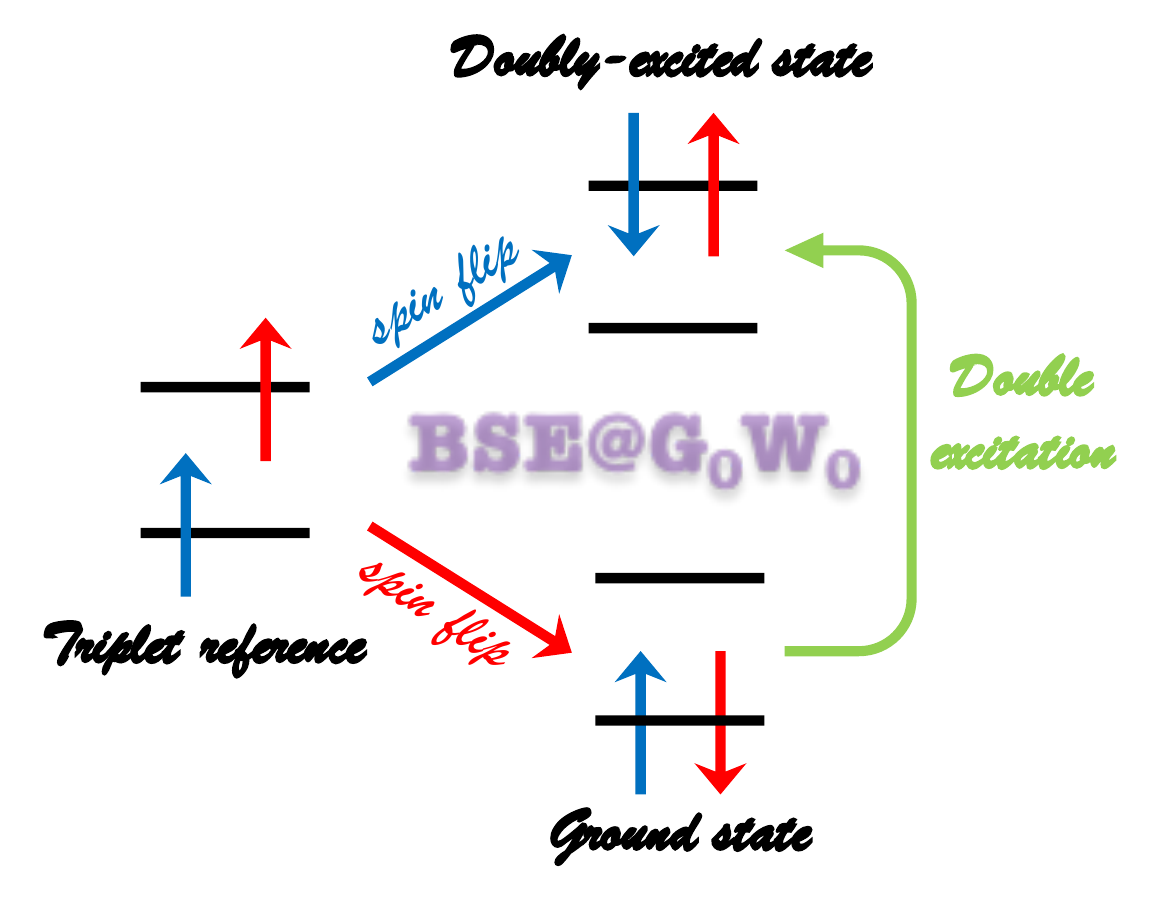}}
\end{center}
\bigskip
\end{abstract}

\maketitle

\section{Introduction}
\label{sec:intro}

Due to the ubiquitous influence of processes involving electronic excited states in physics, chemistry, and biology, their faithful description from first principles has been one of the grand challenges faced by theoretical chemists since the dawn of computational chemistry. 
Accurately predicting ground- and excited-state energies (hence excitation energies) is particularly valuable in this context, and it has concentrated most of the efforts within the community.
An armada of theoretical and computational methods have been developed to this end, each of them being plagued by its own flaws. \cite{Roos_1996,Piecuch_2002,Dreuw_2005,Krylov_2006,Sneskov_2012,Gonzales_2012,Laurent_2013,Adamo_2013,Ghosh_2018,Blase_2020,Loos_2020d,Casanova_2020}
The fact that none of these methods is successful in every chemical scenario has encouraged chemists to carry on the development of new excited-state methodologies, their main goal being to get the most accurate excitation energies (and properties) at the lowest possible computational cost in the most general context. \cite{Loos_2020d}

Originally developed in the framework of nuclear physics, \cite{Salpeter_1951} and popularized in condensed-matter physics, \cite{Sham_1966,Strinati_1984,Delerue_2000} one of the new emerging method in the computational chemistry landscape is the Bethe-Salpeter equation (BSE) formalism \cite{Salpeter_1951,Strinati_1988,Albrecht_1998,Rohlfing_1998,Benedict_1998,vanderHorst_1999,Blase_2018,Blase_2020} from many-body perturbation theory \cite{Onida_2002,Martin_2016} which, based on an underlying $GW$ calculation to compute accurate charged excitations (\ie, ionization potentials and electron affinities) and the dynamically-screened Coulomb potential, \cite{Hedin_1965,Golze_2019} is able to provide accurate optical (\ie, neutral) excitations for molecular systems at a rather modest computational cost.\cite{Rohlfing_1999a,Horst_1999,Puschnig_2002,Tiago_2003,Boulanger_2014,Jacquemin_2015a,Bruneval_2015,Jacquemin_2015b,Hirose_2015,Jacquemin_2017a,Jacquemin_2017b,Rangel_2017,Krause_2017,Gui_2018,Blase_2018,Liu_2020,Blase_2020,Holzer_2018a,Holzer_2018b,Loos_2020e}
Most of BSE implementations rely on the so-called static approximation, \cite{Blase_2018,Bruneval_2016,Krause_2017,Liu_2020} which approximates the dynamical (\ie, frequency-dependent) BSE kernel by its static limit.
Like adiabatic time-dependent density-functional theory (TD-DFT), \cite{Runge_1984,Casida_1995,Petersilka_1996,UlrichBook} the static BSE formalism is plagued by the lack of double (and higher) excitations, which are, for example, ubiquitous in conjugated molecules like polyenes \cite{Maitra_2004,Cave_2004,Saha_2006,Watson_2012,Shu_2017,Barca_2018a,Barca_2018b,Loos_2019} or the ground state of open-shell molecules. \cite{Casida_2005,Huix-Rotllant_2011,Loos_2020f} 
Indeed, both adiabatic TD-DFT \cite{Levine_2006,Tozer_2000,Elliott_2011,Maitra_2012,Maitra_2016} and static BSE \cite{ReiningBook,Romaniello_2009b,Sangalli_2011,Loos_2020h,Authier_2020} can only access (singlet and triplet) single excitations with respect to the reference determinant usually taken as the closed-shell singlet ground state.
Double excitations are even challenging for state-of-the-art methods, \cite{Loos_2018a,Loos_2019,Loos_2020c,Loos_2020d,Veril_2020} like the approximate third-order coupled-cluster (CC3) method \cite{Christiansen_1995b,Koch_1997} or equation-of-motion coupled-cluster with singles, doubles and triples (EOM-CCSDT). \cite{Kucharski_1991,Kallay_2004,Hirata_2000,Hirata_2004}

One way to access double excitations is via the spin-flip formalism established by Krylov in 2001, \cite{Krylov_2001a,Krylov_2001b,Krylov_2002} with earlier attempts by Bethe, \cite{Bethe_1931} as well as Shibuya and McKoy. \cite{Shibuya_1970}
The idea behind the spin-flip \textit{ansatz} is rather simple: instead of considering the singlet ground state as reference, the reference configuration is taken as the lowest triplet state.
In such a way, one can access the singlet ground state and the singlet doubly-excited state via a spin-flip deexcitation and excitation (respectively), the difference of these two excitation energies providing an estimate of the double excitation. 
We refer the interested reader to Refs.~\onlinecite{Krylov_2006,Krylov_2008,Casanova_2020} for detailed reviews on spin-flip methods.
Note that a similar idea has been exploited by the group of Yang to access double excitations in the context of the particle-particle random-phase approximation. \cite{Peng_2013,Yang_2013b,Yang_2014a,Peng_2014,Zhang_2016,Sutton_2018}

One obvious issue of spin-flip methods is that not all double excitations are accessible in such a way.
Moreover, spin-flip methods are usually hampered by spin contamination \cite{Casanova_2020} (\ie, artificial mixing with configurations of different spin multiplicities) due to spin incompleteness of the configuration interaction expansion as well as the possible spin contamination of the reference configuration. \cite{Krylov_2000b}
This issue can be alleviated by increasing the excitation order at a significant cost or by selectively complementing the spin-incomplete configuration set with the missing configurations. \cite{Sears_2003,Casanova_2008,Huix-Rotllant_2010,Li_2010,Li_2011a,Li_2011b,Zhang_2015,Lee_2018}

Nowadays, spin-flip techniques are widely available for many types of methods such as equation-of-motion coupled cluster (EOM-CC), \cite{Krylov_2001a,Levchenko_2004,Manohar_2008,Casanova_2009a,Dutta_2013} configuration interaction (CI), \cite{Krylov_2001b,Krylov_2002,Mato_2018,Casanova_2008,Casanova_2009b} TD-DFT, \cite{Shao_2003,Wang_2004,Li_2011a,Bernard_2012,Zhang_2015} the algebraic-diagrammatic construction (ADC) scheme,\cite{Lefrancois_2015,Lefrancois_2016} and others \cite{Mayhall_2014a,Mayhall_2014b,Bell_2013,Mayhall_2014c} with successful applications in bond breaking processes, \cite{Golubeva_2007} radical chemistry, \cite{Slipchenko_2002,Wang_2005,Slipchenko_2003,Rinkevicius_2010,Ibeji_2015,Hossain_2017,Orms_2018,Luxon_2018} and photochemistry in general \cite{Casanova_2012,Gozem_2013,Nikiforov_2014,Lefrancois_2016} to mention a few. 

Here we apply the spin-flip technique to the BSE formalism in order to access, in particular, double excitations, \cite{Authier_2020} but not only.
The present BSE calculations are based on the spin-unrestricted version of both $GW$ (Sec.~\ref{sec:UGW}) and BSE (Sec.~\ref{sec:UBSE}).
To the best of our knowledge, the present study is the first to apply the spin-flip formalism to the BSE method.
Moreover, we also go beyond the static approximation by taking into account dynamical effects (Sec.~\ref{sec:dBSE}) via an unrestricted generalization of our recently developed (renormalized) perturbative correction which builds on the seminal work of Strinati, \cite{Strinati_1982,Strinati_1984,Strinati_1988} Romaniello and collaborators, \cite{Romaniello_2009b,Sangalli_2011} and Rohlfing and coworkers. \cite{Rohlfing_2000,Ma_2009a,Ma_2009b,Baumeier_2012b,Lettmann_2019}
We also discuss the computation of oscillator strengths (Sec.~\ref{sec:os}) and the expectation value of the spin operator $\expval{\hS^2}$ as a diagnostic of the spin contamination for both ground and excited states (Sec.~\ref{sec:spin}).
Computational details are reported in Sec.~\ref{sec:compdet} and our results for the beryllium atom \ce{Be} (Subsec.~\ref{sec:Be}), the hydrogen molecule \ce{H2} (Subsec.~\ref{sec:H2}), and cyclobutadiene \ce{C4H4} (Subsec.~\ref{sec:CBD}) are discussed in Sec.~\ref{sec:res}.
Finally,  we draw our conclusions in Sec.~\ref{sec:ccl}.
Unless otherwise stated, atomic units are used.

\section{Unrestricted $GW$ formalism}
\label{sec:UGW}
Let us consider an electronic system consisting of $n = n_\up + n_\dw$ electrons (where $n_\up$ and $n_\dw$ are the number of spin-up and spin-down electrons, respectively) and $N$ one-electron basis functions.
The number of spin-up and spin-down occupied orbitals are $O_\up = n_\up$ and $O_\dw = n_\dw$, respectively, and, assuming the absence of linear dependencies in the one-electron basis set, there is $V_\up = N - O_\up$ and $V_\dw = N - O_\dw$ spin-up and spin-down virtual (\ie, unoccupied) orbitals.
The number of spin-conserved (sc) single excitations is then $S^\spc = S_{\up\up}^\spc + S_{\dw\dw}^\spc = O_\up V_\up + O_\dw V_\dw$, while the number of spin-flip (sf) excitations is $S^\spf = S_{\up\dw}^\spf + S_{\dw\up}^\spf = O_\up V_\dw + O_\dw V_\up$.
Let us denote as $\MO{p_\sig}(\br)$ the $p$th spatial orbital associated with the spin-$\sig$ electrons (where $\sig =$ $\up$ or $\dw$) and $\e{p_\sig}{}$ its one-electron energy.
It is important to understand that, in a spin-conserved excitation the hole orbital $\MO{i_\sig}$ and particle orbital $\MO{a_\sig}$ have the same spin $\sig$.
In a spin-flip excitation, the hole and particle states, $\MO{i_\sig}$ and $\MO{a_\bsig}$, have opposite spins, $\sig$ and $\bsig$.
We assume real quantities throughout this manuscript, $i$ and $j$ are occupied orbitals, $a$ and $b$ are unoccupied orbitals, $p$, $q$, $r$, and $s$ indicate arbitrary orbitals, and $m$ labels single excitations.
Moreover, we consider systems with collinear spins and a spin-independent Hamiltonian without contributions such as spin-orbit interaction.

\subsection{The dynamical screening}
The pillar of Green's function many-body perturbation theory is the (time-ordered) one-body Green's function, which has poles at the charged excitations (i.e., ionization potentials and electron affinities) of the system. \cite{ReiningBook}
The spin-$\sig$ component of the one-body Green's function reads \cite{ReiningBook,Bruneval_2016}
\begin{equation}
\label{eq:G}
	G^{\sig}(\br_1,\br_2;\omega) 
	= \sum_i \frac{\MO{i_\sig}(\br_1) \MO{i_\sig}(\br_2)}{\omega - \e{i_\sig}{} - i\eta}	
	+ \sum_a \frac{\MO{a_\sig}(\br_1) \MO{a_\sig}(\br_2)}{\omega - \e{a_\sig}{} + i\eta}	
\end{equation}
where $\eta$ is a positive infinitesimal.
As readily seen in Eq.~\eqref{eq:G}, the Green's function can be evaluated at different levels of theory depending on the choice of orbitals and energies, $\MO{p_\sig}$ and $\e{p_\sig}{}$. 
For example, $G_{\KS}^{\sig}$ is the independent-particle Green's function built with Kohn-Sham (KS) orbitals $\MO{p_\sig}^{\KS}(\br)$ and one-electron energies $\e{p_\sig}^{\KS}$. \cite{Hohenberg_1964,Kohn_1965,ParrBook} 
Within self-consistent schemes, these quantities can be replaced by quasiparticle energies and orbitals evaluated within the $GW$ approximation (see below). \cite{Hedin_1965,Golze_2019}

Based on the spin-up and spin-down components of $G$ defined in Eq.~\eqref{eq:G}, one can easily compute the non-interacting polarizability (which is a sum over spins)
\begin{equation}
\label{eq:chi0}
	\chi_0(\br_1,\br_2;\omega) = - \frac{i}{2\pi} \sum_\sig \int G^{\sig}(\br_1,\br_2;\omega+\omega') G^{\sig}(\br_1,\br_2;\omega') d\omega'
\end{equation}
and subsequently the dielectric function
\begin{equation}
\label{eq:eps}
	\epsilon(\br_1,\br_2;\omega) = \delta(\br_1 - \br_2) - \int \frac{\chi_0(\br_1,\br_3;\omega) }{\abs{\br_2 - \br_3}} d\br_3
\end{equation}
where $\delta(\br)$ is the Dirac delta function.
Based on this latter ingredient, one can access the dynamically-screened Coulomb potential
\begin{equation}
\label{eq:W}
	W(\br_1,\br_2;\omega) =  \int \frac{\epsilon^{-1}(\br_1,\br_3;\omega) }{\abs{\br_2 - \br_3}} d\br_3
\end{equation}
which is naturally spin independent as the bare Coulomb interaction $\abs{\br_1 - \br_2}^{-1}$ does not depend on spin coordinates.

Within the $GW$ formalism, \cite{Hedin_1965,Onida_2002,Golze_2019} the dynamical screening is computed at the random-phase approximation (RPA) level by considering only the manifold of the spin-conserved neutral excitations.
In the orbital basis, the spectral representation of $W$ is
\begin{multline}
\label{eq:W_spectral}
	W_{p_\sig q_\sig,r_\sigp s_\sigp}(\omega) = \ERI{p_\sig q_\sig}{r_\sigp s_\sigp} 
	+ \sum_m \ERI{p_\sig q_\sig}{m}\ERI{r_\sigp s_\sigp}{m} 
	\\
	\times \qty[ \frac{1}{\omega - \Om{m}{\spc,\RPA} + i \eta} - \frac{1}{\omega + \Om{m}{\spc,\RPA} - i \eta} ]
\end{multline}
where the bare two-electron integrals are \cite{Gill_1994}
\begin{equation}
		\ERI{p_\sig q_\tau}{r_\sigp s_\taup} = \int  \frac{\MO{p_\sig}(\br_1) \MO{q_\tau}(\br_1) \MO{r_\sigp}(\br_2) \MO{s_\taup}(\br_2)}{\abs{\br_1 - \br_2}} d\br_1 d\br_2
\end{equation}
and the screened two-electron integrals (or spectral weights) are explicitly given by
\begin{equation}
\label{eq:sERI}
		\ERI{p_\sig q_\sig}{m} = \sum_{ia\sigp} \ERI{p_\sig q_\sig}{i_\sigp a_\sigp} (\bX{m}{\spc,\RPA}+\bY{m}{\spc,\RPA})_{i_\sigp a_\sigp}
\end{equation}
In Eqs.~\eqref{eq:W_spectral} and \eqref{eq:sERI}, the spin-conserved RPA neutral excitations $\Om{m}{\spc,\RPA}$ and their corresponding eigenvectors, $\bX{m}{\spc,\RPA}$ and $\bY{m}{\spc,\RPA}$, are obtained by solving a linear response system of the form
\begin{equation}
\label{eq:LR-RPA}
	\begin{pmatrix}
		\bA{}{}	&	\bB{}{}	\\
		-\bB{}{}	&	-\bA{}{}	\\
	\end{pmatrix}
	\cdot
	\begin{pmatrix}
		\bX{m}{}	\\
		\bY{m}{}	\\
	\end{pmatrix}
	=
	\Om{m}{}
	\begin{pmatrix}
		\bX{m}{}	\\
		\bY{m}{}	\\
	\end{pmatrix}
\end{equation}
where the expressions of the matrix elements of $\bA{}{}$ and $\bB{}{}$ are specific of the method and of the spin manifold. 
The spin structure of these matrices, though, is general 
\begin{subequations}
\begin{align}
\label{eq:LR-RPA-AB-sc}
	\bA{}{\spc} & = \begin{pmatrix}
		\bA{}{\upup,\upup}	&	\bA{}{\upup,\dwdw}	\\
		\bA{}{\dwdw,\upup}	&	\bA{}{\dwdw,\dwdw}	\\
	\end{pmatrix}
	&
	\bB{}{\spc} & = \begin{pmatrix}
		\bB{}{\upup,\upup}	&	\bB{}{\upup,\dwdw}	\\
		\bB{}{\dwdw,\upup}	&	\bB{}{\dwdw,\dwdw}	\\
	\end{pmatrix}
\\
\label{eq:LR-RPA-AB-sf}
	\bA{}{\spf} & = \begin{pmatrix}
		\bA{}{\updw,\updw}	&	\bO	\\
		\bO					&	\bA{}{\dwup,\dwup}	\\
	\end{pmatrix}
	&
	\bB{}{\spf} & = \begin{pmatrix}
		\bO					&	\bB{}{\updw,\dwup}	\\
		\bB{}{\dwup,\updw}	&	\bO	\\
	\end{pmatrix}
\end{align}
\end{subequations}
In the absence of instabilities, the linear eigenvalue problem \eqref{eq:LR-RPA} has particle-hole symmetry which means that the eigenvalues are obtained by pairs $\pm \Om{m}{}$.
In such a case, $(\bA{}{}-\bB{}{})^{1/2}$ is positive definite, and Eq.~\eqref{eq:LR-RPA} can be recast as a Hermitian problem of half its original dimension 
\begin{equation}
\label{eq:small-LR}
	(\bA{}{} - \bB{}{})^{1/2} \cdot (\bA{}{} + \bB{}{}) \cdot (\bA{}{} - \bB{}{})^{1/2} \cdot \bZ{}{} = \bOm{2} \cdot \bZ{}{}
\end{equation}
where the excitation amplitudes are
\begin{equation}
	\bX{}{} + \bY{}{} = \bOm{-1/2} \cdot (\bA{}{} - \bB{}{})^{1/2} \cdot \bZ{}{}
\end{equation}
Within the Tamm-Dancoff approximation (TDA), the coupling terms between the resonant and anti-resonant parts, $\bA{}{}$ and $-\bA{}{}$, are neglected, which consists in setting $\bB{}{} = \bO$.
In such a case, Eq.~\eqref{eq:LR-RPA} reduces to a straightforward Hermitian problem of the form:
\begin{equation}
	\bA{}{} \cdot \bX{m}{} = \Om{m}{} \bX{m}{}
\end{equation}
Note that, for spin-flip excitations, it is quite common to enforce the TDA especially when one considers a triplet reference as the first ``excited-state'' is usually the ground state of the closed-shell system (hence, corresponding to a negative excitation energy).

At the RPA level, the matrix elements of $\bA{}{}$ and $\bB{}{}$ are
\begin{subequations}
\begin{align}
	\label{eq:LR_RPA-A}
	\A{i_\sig a_\tau,j_\sigp b_\taup}{\RPA} & = \delta_{ij} \delta_{ab} \delta_{\sig \sigp}  \delta_{\tau \taup} (\e{a_\tau} - \e{i_\sig}) + \ERI{i_\sig a_\tau}{b_\sigp j_\taup}
	\\ 
	\label{eq:LR_RPA-B}
	\B{i_\sig a_\tau,j_\sigp b_\taup}{\RPA} & = \ERI{i_\sig a_\tau}{j_\sigp b_\taup}
\end{align}
\end{subequations}
from which we obtain the following expressions
\begin{subequations}
\begin{align}
	\label{eq:LR_RPA-Asc}
	\A{i_\sig a_\sig,j_\sigp b_\sigp}{\spc,\RPA} & = \delta_{ij} \delta_{ab} \delta_{\sig \sigp} (\e{a_\sig} - \e{i_\sig}) + \ERI{i_\sig a_\sig}{b_\sigp j_\sigp}
	\\ 
	\label{eq:LR_RPA-Bsc}
	\B{i_\sig a_\sig,j_\sigp b_\sigp}{\spc,\RPA} & = \ERI{i_\sig a_\sig}{j_\sigp b_\sigp}
\end{align}
\end{subequations}
for the spin-conserved excitations and 
\begin{subequations}
\begin{align}
	\label{eq:LR_RPA-Asf}
	\A{i_\sig a_\bsig,j_\sig b_\bsig}{\spf,\RPA} & = \delta_{ij} \delta_{ab} (\e{a_\bsig} - \e{i_\sig})
	\\ 
	\label{eq:LR_RPA-Bsf}
	\B{i_\sig a_\bsig,j_\bsig b_\sig}{\spf,\RPA} & = 0
\end{align}
\end{subequations}
for the spin-flip excitations.

\subsection{The $GW$ self-energy}
Within the acclaimed $GW$ approximation, \cite{Hedin_1965,Golze_2019} the exchange-correlation (xc) part of the self-energy
\begin{equation}
\label{eq:Sig}
\begin{split}
	\Sig{}{\xc,\sig}(\br_1,\br_2;\omega) 
	& = \Sig{}{\x,\sig}(\br_1,\br_2) + \Sig{}{\co,\sig}(\br_1,\br_2;\omega) 
	\\
	& = \frac{i}{2\pi} \int G^{\sig}(\br_1,\br_2;\omega+\omega') W(\br_1,\br_2;\omega') e^{i \eta \omega'} d\omega'
\end{split}
\end{equation}
is, like the one-body Green's function, spin-diagonal, and its spectral representation reads
\begin{subequations}
\begin{gather}
	\Sig{p_\sig q_\sig}{\x}
	= - \sum_{i} \ERI{p_\sig i_\sig}{i_\sig q_\sig}	 
	\\
\begin{split}
	\Sig{p_\sig q_\sig}{\co}(\omega) 
	& = \sum_{im} \frac{\ERI{p_\sig i_\sig}{m} \ERI{q_\sig i_\sig}{m}}{\omega - \e{i_\sig} + \Om{m}{\spc,\RPA} - i \eta}
	\\
	& + \sum_{am} \frac{\ERI{p_\sig a_\sig}{m} \ERI{q_\sig a_\sig}{m}}{\omega - \e{a_\sig} - \Om{m}{\spc,\RPA} + i \eta}			 
\end{split}
\end{gather}
\end{subequations}
where the self-energy has been split in its exchange (x) and correlation (c) contributions.
The Dyson equation linking the Green's function and the self-energy holds separately for each spin component 
\begin{equation}
\label{eq:Dyson_G}
\begin{split}
	\qty[ G^{\sig}(\br_1,\br_2;\omega) ]^{-1} 
	& = \qty[ G_{\KS}^{\sig}(\br_1,\br_2;\omega) ]^{-1} 
	\\
	& + \Sig{}{\xc,\sig}(\br_1,\br_2;\omega) - v^{\xc}(\br_1) \delta(\br_1 - \br_2)
\end{split}
\end{equation}
where $v^{\xc}(\br)$ is the KS (local) exchange-correlation potential.
The target quantities here are the quasiparticle energies $\eGW{p_\sig}$, \ie, the poles of $G$ [see Eq.~\eqref{eq:G}], which correspond to well-defined addition/removal energies (unlike the KS orbital energies).
Because the exchange-correlation part of the self-energy is, itself, constructed with the Green's function [see Eq.~\eqref{eq:Sig}], the present process is, by nature, self-consistent.
The same comment applies to the dynamically-screened Coulomb potential $W$ entering the definition of $\Sig{}{\xc}$ [see Eq.~\eqref{eq:Sig}] which is also constructed from $G$ [see Eqs.~\eqref{eq:chi0}, \eqref{eq:eps}, and \eqref{eq:W}].

\subsection{Level of self-consistency}
This is where $GW$ schemes differ.
In its simplest perturbative (\ie, one-shot) version, known as {\GOWO}, \cite{Strinati_1980,Hybertsen_1985a,Hybertsen_1986,Godby_1988,Linden_1988,Northrup_1991,Blase_1994,Rohlfing_1995,Shishkin_2007} a single iteration is performed, and the quasiparticle energies $\eGW{p_\sig}$ are obtained by solving the frequency-dependent quasiparticle equation
\begin{equation}
\label{eq:QP-eq}
	\omega = \eKS{p_\sig} + \Sig{p_\sig}{\xc}(\omega) - V_{p_\sig}^{\xc} 
\end{equation}
where $\Sig{p_\sig}{\xc}(\omega) \equiv \Sig{p_\sig p_\sig}{\xc}(\omega)$ and its offspring quantities have been constructed at the KS level, and
\begin{equation}
	V_{p_\sigma}^{\xc} = \int \MO{p_\sig}(\br) v^{\xc}(\br) \MO{p_\sig}(\br) d\br 
\end{equation}
Because, from a practical point of view, one is usually interested by the so-called quasiparticle solution (or peak), the quasiparticle equation \eqref{eq:QP-eq} is often linearized around $\omega = \e{p_\sig}^{\KS}$, yielding
\begin{equation}
\label{eq:G0W0_lin}
    \eGW{p_\sig}
    = \e{p_\sig}^{\KS} + Z_{p_\sig} [\Sig{p_\sig}{\xc}(\e{p_\sig}^{\KS}) - V_{p_\sig}^{\xc} ]
\end{equation}
where
\begin{equation}
\label{eq:Z_GW}
	Z_{p_\sig} = \qty[ 1 - \left. \pdv{\Sig{p_\sig}{\xc}(\omega)}{\omega} \right|_{\omega = \e{p_\sig}^{\KS}} ]^{-1}
\end{equation}
is a renormalization factor (with $0 \le Z_{p_\sig} \le 1$) which also represents the spectral weight of the quasiparticle solution.
In addition to the principal quasiparticle peak which, in a well-behaved case, contains most of the spectral weight, the frequency-dependent quasiparticle equation \eqref{eq:QP-eq} generates a finite number of satellite resonances with smaller weights. \cite{Loos_2018b}

Within the ``eigenvalue'' self-consistent $GW$ scheme (known as ev$GW$), \cite{Hybertsen_1986,Shishkin_2007,Blase_2011,Faber_2011,Rangel_2016,Gui_2018}  several iterations are performed during which only the one-electron energies entering the definition of the Green's function [see Eq.~\eqref{eq:G}] are updated by the quasiparticle energies obtained at the previous iteration (the corresponding orbitals remain evaluated at the KS level).

Finally, within the quasiparticle self-consistent $GW$ (qs$GW$) scheme, \cite{Faleev_2004,vanSchilfgaarde_2006,Kotani_2007,Ke_2011,Kaplan_2016} both the one-electron energies and the orbitals are updated until convergence is reached.
These are obtained via the diagonalization of an effective Fock matrix which includes explicitly a frequency-independent and Hermitian self-energy defined as
\begin{equation}
	\Tilde{\Sigma}_{p_\sig q_\sig}^{\xc} = \frac{1}{2} \qty[ \Sig{p_\sig q_\sig}{\xc}(\e{p_\sig}{}) + \Sig{q_\sig p_\sig}{\xc}(\e{p_\sig}{}) ]
\end{equation}

\section{Unrestricted Bethe-Salpeter equation formalism}
\label{sec:UBSE}
Like its TD-DFT cousin, \cite{Runge_1984,Casida_1995,Petersilka_1996,Dreuw_2005} the BSE formalism \cite{Salpeter_1951,Strinati_1988,Albrecht_1998,Rohlfing_1998,Benedict_1998,vanderHorst_1999} deals with the calculation of (neutral) optical excitations as measured by absorption spectroscopy. \cite{Rohlfing_1999a,Horst_1999,Puschnig_2002,Tiago_2003,Boulanger_2014,Jacquemin_2015a,Bruneval_2015,Jacquemin_2015b,Hirose_2015,Jacquemin_2017a,Jacquemin_2017b,Rangel_2017,Krause_2017,Gui_2018} 
Using the BSE formalism, one can access the spin-conserved and spin-flip excitations.
In a nutshell, BSE builds on top of a $GW$ calculation by adding up excitonic effects (\ie, the electron-hole binding energy) to the $GW$ fundamental gap which is itself a corrected version of the KS gap. 
The purpose of the underlying $GW$ calculation is to provide quasiparticle energies and a dynamically-screened Coulomb potential that are used to build the BSE Hamiltonian from which the vertical excitations of the system are extracted.

\subsection{Static approximation}
\label{sec:BSE}

Within the so-called static approximation of BSE, the Dyson equation that links the generalized four-point susceptibility $L^{\sig\sigp}(\br_1,\br_2;\br_1',\br_2';\omega)$ and the BSE kernel $\Xi^{\sig\sigp}(\br_3,\br_5;\br_4,\br_6)$ is \cite{ReiningBook,Bruneval_2016}
\begin{multline}
	L^{\sig\sigp}(\br_1,\br_2;\br_1',\br_2';\omega) 
	= L_{0}^{\sig\sigp}(\br_1,\br_2;\br_1',\br_2';\omega)
	\\
	+ \int L_{0}^{\sig\sigp}(\br_1,\br_4;\br_1',\br_3;\omega) 
	\Xi^{\sig\sigp}(\br_3,\br_5;\br_4,\br_6)
	\\
	\times L^{\sig\sigp}(\br_6,\br_2;\br_5,\br_2';\omega)
	d\br_3 d\br_4 d\br_5 d\br_6
\end{multline}
where 
\begin{multline}
	L_{0}^{\sig\sigp}(\br_1,\br_2;\br_1',\br_2';\omega) 
	\\
	= \frac{1}{2\pi} \int G^{\sig}(\br_1,\br_2';\omega+\omega') G^{\sig}(\br_1',\br_2;\omega') d\omega'
\end{multline}
is the non-interacting analog of the two-particle correlation function $L$.

Within the $GW$ approximation, the static BSE kernel is
\begin{multline}
	\label{eq:kernel}
	i \Xi^{\sig\sigp}(\br_3,\br_5;\br_4,\br_6) 
	= \frac{\delta(\br_3 - \br_4) \delta(\br_5 - \br_6) }{\abs{\br_3-\br_6}} 
	\\
	- \delta_{\sig\sigp} W(\br_3,\br_4;\omega = 0) \delta(\br_3 - \br_6) \delta(\br_4 - \br_6) 
\end{multline}
where, as usual, we have not considered the higher-order terms in $W$ by neglecting the derivative $\partial W/\partial G$. \cite{Hanke_1980,Strinati_1982,Strinati_1984,Strinati_1988}

As readily seen in Eq.~\eqref{eq:kernel}, the static approximation consists in neglecting the frequency dependence of the dynamically-screened Coulomb potential.
In this case, the spin-conserved and spin-flip BSE optical excitations are obtained by solving the usual Casida-like linear response (eigen)problem:
\begin{equation}
\label{eq:LR-BSE}
	\begin{pmatrix}
		\bA{}{\BSE}		&	\bB{}{\BSE}	\\
		-\bB{}{\BSE}	&	-\bA{}{\BSE}	\\
	\end{pmatrix}
	\cdot
	\begin{pmatrix}
		\bX{m}{\BSE}	\\
		\bY{m}{\BSE}	\\
	\end{pmatrix}
	=
	\Om{m}{\BSE}
	\begin{pmatrix}
		\bX{m}{\BSE}	\\
		\bY{m}{\BSE}	\\
	\end{pmatrix}
\end{equation}
Defining the elements of the static screening as $W^{\stat}_{p_\sig q_\sig,r_\sigp s_\sigp} = W_{p_\sig q_\sig,r_\sigp s_\sigp}(\omega = 0)$, the general expressions of the BSE matrix elements are
\begin{subequations}
\begin{align}
	\label{eq:LR_BSE-A}
	\A{i_\sig a_\tau,j_\sigp b_\taup}{\BSE} & = \A{i_\sig a_\tau,j_\sigp b_\taup}{\RPA} - \delta_{\sig \sigp} W^{\stat}_{i_\sig j_\sigp,b_\taup a_\tau}
	\\ 
	\label{eq:LR_BSE-B}
	\B{i_\sig a_\tau,j_\sigp b_\taup}{\BSE} & = \B{i_\sig a_\tau,j_\sigp b_\taup}{\RPA} - \delta_{\sig \sigp} W^{\stat}_{i_\sig b_\taup,j_\sigp a_\tau}
\end{align}
\end{subequations}
from which we obtain the following expressions for the spin-conserved and spin-flip BSE excitations: 
\begin{subequations}
\begin{align}
	\label{eq:LR_BSE-Asc}
	\A{i_\sig a_\sig,j_\sigp b_\sigp}{\spc,\BSE} & = \A{i_\sig a_\sig,j_\sigp b_\sigp}{\spc,\RPA} - \delta_{\sig \sigp} W^{\stat}_{i_\sig j_\sigp,b_\sigp a_\sig}
	\\ 
	\label{eq:LR_BSE-Bsc}
	\B{i_\sig a_\sig,j_\sigp b_\sigp}{\spc,\BSE} & = \B{i_\sig a_\sig,j_\sigp b_\sigp}{\spc,\RPA} - \delta_{\sig \sigp} W^{\stat}_{i_\sig b_\sigp,j_\sigp a_\sig}
	\\
	\label{eq:LR_BSE-Asf}
	\A{i_\sig a_\bsig,j_\sig b_\bsig}{\spf,\BSE} & = \A{i_\sig a_\bsig,j_\sig b_\bsig}{\spf,\RPA} - W^{\stat}_{i_\sig j_\sig,b_\bsig a_\bsig}
	\\ 
	\label{eq:LR_BSE-Bsf}
	\B{i_\sig a_\bsig,j_\bsig b_\sig}{\spf,\BSE} & = - W^{\stat}_{i_\sig b_\sig,j_\bsig a_\bsig}
\end{align}
\end{subequations}

At this stage, it is of particular interest to discuss the form of the spin-flip matrix elements defined in Eqs.~\eqref{eq:LR_BSE-Asf} and \eqref{eq:LR_BSE-Bsf}.
As readily seen from Eq.~\eqref{eq:LR_RPA-Asf}, at the RPA level, the spin-flip excitations are given by the difference of one-electron energies, hence missing out on key exchange and correlation effects.
This is also the case at the TD-DFT level when one relies on (semi-)local functionals. 
This explains why most of spin-flip TD-DFT calculations are performed with global hybrid functionals containing a substantial amount of Hartree-Fock exchange as only the exact exchange integral of the form $\ERI{i_\sig j_\sig}{b_\bsig a_\bsig}$ survive spin-symmetry requirements.
At the BSE level, these matrix elements are, of course, also present thanks to the contribution of $W^{\stat}_{i_\sig j_\sig,b_\bsig a_\bsig}$ as evidenced in Eq.~\eqref{eq:W_spectral} but it also includes correlation effects.

\subsection{Dynamical correction}
\label{sec:dBSE}
In order to go beyond the ubiquitous static approximation of BSE \cite{Strinati_1988,Rohlfing_2000,Sottile_2003,Myohanen_2008,Ma_2009a,Ma_2009b,Romaniello_2009b,Sangalli_2011,Huix-Rotllant_2011,Sakkinen_2012,Zhang_2013,Rebolini_2016,Olevano_2019,Lettmann_2019} (which is somehow similar to the adiabatic approximation of TD-DFT  \cite{Casida_2005,Huix-Rotllant_2011,Casida_2016,Maitra_2004,Cave_2004,Elliott_2011,Maitra_2012}), we have recently implemented, following Strinati's seminal work \cite{Strinati_1982,Strinati_1984,Strinati_1988} (see also the work of Romaniello \textit{et al.} \cite{Romaniello_2009b} and Sangalli \textit{et al.} \cite{Sangalli_2011}), a renormalized first-order perturbative correction in order to take into consideration the dynamical nature of the screened Coulomb potential $W$. \cite{Loos_2020h,Authier_2020}
This dynamical correction to the static BSE kernel (dubbed as dBSE in the following) does permit to recover additional relaxation effects coming from higher excitations.

Our implementation follows closely the work of Rohlfing and co-workers \cite{Rohlfing_2000,Ma_2009a,Ma_2009b,Baumeier_2012b} in which they computed the dynamical correction in the TDA and plasmon-pole approximation. 
However, our scheme goes beyond the plasmon-pole approximation as the spectral representation of the dynamically-screened Coulomb potential is computed exactly at the RPA level consistently with the underlying $GW$ calculation:
\begin{multline}
	\widetilde{W}_{p_\sig q_\sig,r_\sigp s_\sigp}(\omega) = \ERI{p_\sig q_\sig}{r_\sigp s_\sigp} 
	+ \sum_m \ERI{p_\sig q_\sig}{m}\ERI{r_\sigp s_\sigp}{m} 
	\\
	\times \Bigg[ \frac{1}{\omega - (\alert{\eGW{s_\sigp}{} - \eGW{q_\sig}{}}) - \Om{m}{\spc,\RPA} + i \eta} 
	\\
	+ \frac{1}{\omega - (\alert{\eGW{r_\sigp}{} - \eGW{p_\sig}{}}) - \Om{m}{\spc,\RPA} + i \eta} \Bigg]
\end{multline}
The dBSE non-linear response problem is
\begin{multline}
\label{eq:LR-dyn}
	\begin{pmatrix}
		\bA{}{\dBSE}(\Om{m}{\dBSE})		&	\bB{}{\dBSE}(\Om{m}{\dBSE})	
		\\
		-\bB{}{\dBSE}(-\Om{m}{\dBSE})	&	-\bA{}{\dBSE}(-\Om{m}{\dBSE})	
		\\
	\end{pmatrix}
	\cdot
	\begin{pmatrix}
		\bX{m}{\dBSE}	\\
		\bY{m}{\dBSE}	\\
	\end{pmatrix}
	\\
	=
	\Om{m}{\dBSE}
	\begin{pmatrix}
		\bX{m}{\dBSE}	\\
		\bY{m}{\dBSE}	\\
	\end{pmatrix}
\end{multline}
where the dynamical matrices are generally defined as
\begin{subequations}
\begin{align}
	\label{eq:LR_dBSE-A}
	\A{i_\sig a_\tau,j_\sigp b_\taup}{\dBSE}(\omega) & = \A{i_\sig a_\tau,j_\sigp b_\taup}{\RPA} - \delta_{\sig \sigp} \widetilde{W}_{i_\sig j_\sigp,b_\taup a_\tau}(\omega) 
	\\ 
	\label{eq:LR_dBSE-B}
	\B{i_\sig a_\tau,j_\sigp b_\taup}{\dBSE}(\omega) & = \B{i_\sig a_\tau,j_\sigp b_\taup}{\RPA} - \delta_{\sig \sigp} \widetilde{W}_{i_\sig b_\taup,j_\sigp a_\tau}(\omega) 
\end{align}
\end{subequations}
from which one can easily obtained the matrix elements for the spin-conserved and spin-flip manifolds similarly to Eqs.~\eqref{eq:LR_BSE-Asc}, \eqref{eq:LR_BSE-Bsc}, \eqref{eq:LR_BSE-Asf}, and \eqref{eq:LR_BSE-Bsf}. 
Following Rayleigh-Schr\"odinger perturbation theory, we then decompose the non-linear eigenproblem \eqref{eq:LR-dyn} as a zeroth-order static (\ie, linear) reference and a first-order dynamic (\ie, non-linear) perturbation such that
\begin{multline}
\label{eq:LR-PT}
	\begin{pmatrix}
		\bA{}{\dBSE}(\omega)		&	\bB{}{\dBSE}(\omega)	\\
		-\bB{}{\dBSE}(-\omega)	&	-\bA{}{\dBSE}(-\omega)	\\
	\end{pmatrix}
	\\
	=
	\begin{pmatrix}
		\bA{}{(0)}	&	\bB{}{(0)}	
		\\
		-\bB{}{(0)}	&	-\bA{}{(0)}	
		\\
	\end{pmatrix}
	+
	\begin{pmatrix}
		\bA{}{(1)}(\omega)		&	\bB{}{(1)}(\omega)	\\
		-\bB{}{(1)}(-\omega)	&	-\bA{}{(1)}(-\omega)	\\
	\end{pmatrix}
\end{multline}
with
\begin{subequations}
\begin{align}
	\label{eq:BSE-A0}
	\A{i_\sig a_\tau,j_\sigp b_\taup}{(0)} & = \A{i_\sig a_\tau,j_\sigp b_\taup}{\BSE}
	\\ 
	\label{eq:BSE-B0}
	\B{i_\sig a_\tau,j_\sigp b_\taup}{(0)} & = \B{i_\sig a_\tau,j_\sigp b_\taup}{\BSE}
\end{align}
\end{subequations}
and 
\begin{subequations}
\begin{align}
	\label{eq:BSE-A1}
	\A{i_\sig a_\tau,j_\sigp b_\taup}{(1)}(\omega) & = - \delta_{\sig \sigp} \widetilde{W}_{i_\sig j_\sigp,b_\taup a_\tau}(\omega) + \delta_{\sig \sigp} W^{\stat}_{i_\sig j_\sigp,b_\taup a_\tau}
	\\ 
	\label{eq:BSE-B1}
	\B{i_\sig a_\tau,j_\sigp b_\taup}{(1)}(\omega) & = - \delta_{\sig \sigp} \widetilde{W}_{i_\sig b_\taup,j_\sigp a_\tau}(\omega) + \delta_{\sig \sigp} W^{\stat}_{i_\sig b_\taup,j_\sigp a_\tau}
\end{align}
\end{subequations}
The dBSE excitation energies are then obtained via
\begin{equation}
	\Om{m}{\dBSE} = \Om{m}{\BSE} + \zeta_{m} \Om{m}{(1)}
\end{equation}
where $\Om{m}{\BSE} \equiv \Om{m}{(0)}$ are the static (zeroth-order) BSE excitation energies obtained by solving Eq.~\eqref{eq:LR-BSE}, and  
\begin{equation}
\label{eq:Om1-TDA}
	\Om{m}{(1)} = \T{(\bX{m}{\BSE})} \cdot \bA{}{(1)}(\Om{m}{\BSE}) \cdot \bX{m}{\BSE}
\end{equation}
are first-order corrections (with $\bX{m}{\BSE} \equiv \bX{m}{(0)}$) obtained within the dynamical TDA (dTDA) with the renormalization factor
\begin{equation}
\label{eq:Z}
	\zeta_{m} = \qty[ 1 - \T{(\bX{m}{\BSE})} \cdot \left. \pdv{\bA{}{(1)}(\omega)}{\omega} \right|_{\omega = \Om{m}{\BSE}} \cdot \bX{m}{\BSE} ]^{-1}
\end{equation}
which, unlike the $GW$ case [see Eq.~\eqref{eq:Z_GW}], is not restricted to be between $0$ and $1$.
In most cases, the value of $\zeta_{m}$ is close to unity which indicates that the perturbative expansion behaves nicely.

\subsection{Oscillator strengths}
\label{sec:os}
Oscillator strengths, \ie, transition dipole moments from the ground to the corresponding excited state, are key quantities that are linked to experimental intensities and are usually used to probe the quality of excited-state calculations. \cite{Harbach_2014,Kannar_2014,Chrayteh_2021,Sarkar_2021}

For the spin-conserved transitions, the $x$ component of the transition dipole moment is
\begin{equation}
	\mu_{x,m}^{\spc} = \sum_{ia\sig} (i_\sig|x|a_\sig)(\bX{m}{\spc}+\bY{m}{\spc})_{i_\sig a_\sig}
\end{equation}
where 
\begin{equation}
	(p_\sig|x|q_\sigp) = \int \MO{p_\sig}(\br) \, x \, \MO{q_\sigp}(\br) d\br
\end{equation}
are one-electron integrals in the orbital basis.
The total oscillator strength in the so-called length gauge \cite{Sarkar_2021} is given by
\begin{equation}
	f_{m}^{\spc} = \frac{2}{3} \Om{m}{\spc} \qty[ \qty(\mu_{x,m}^{\spc})^2 + \qty(\mu_{x,m}^{\spc})^2 + \qty(\mu_{x,m}^{\spc})^2 ]
\end{equation}
For spin-flip transitions, we have $f_{m}^{\spf} = 0$ as the transition matrix elements $(i_\sig|x|a_\bsig)$ vanish via integration over the spin coordinate.

\subsection{Spin contamination}
\label{sec:spin}
One of the key issues of linear response formalism based on unrestricted references is spin contamination or the artificial mixing with configurations of different spin multiplicities.
As nicely explained in Ref.~\onlinecite{Casanova_2020}, there are two sources of spin contamination: i) spin contamination of the reference configuration for which, for example, $\expval{\hS^2} > 2$ for high-spin triplets, and ii) spin contamination of the excited states due to spin incompleteness of the CI expansion.
The latter issue is an important source of spin contamination in the present context as BSE is limited to single excitations with respect to the reference configuration.
Specific schemes have been developed to palliate these shortcomings and we refer the interested reader to Ref.~\onlinecite{Casanova_2020} for a detailed discussion on this matter.

In order to monitor closely how contaminated are these states, we compute
\begin{equation}
	\expval{\hS^2}_m = \expval{\hS^2}_0 + \Delta \expval{\hS^2}_m
\end{equation}
where  
\begin{equation}
	\expval{\hS^2}_{0} 
	= \frac{n_{\up} - n_{\dw}}{2} \qty( \frac{n_{\up} - n_{\dw}}{2} + 1 )
	+ n_{\dw} - \sum_p (p_{\up}|p_{\dw})^2
\end{equation}
is the expectation value of $\hS^2$ for the reference configuration, the first term corresponding to the exact value of $\expval{\hS^2}$, and
\begin{equation}
\label{eq:OV}
	(p_\sig|q_\sigp) = \int \MO{p_\sig}(\br) \MO{q_\sigp}(\br) d\br
\end{equation}
are overlap integrals between spin-up and spin-down orbitals.

For a given single excitation $m$, the explicit expressions of $\Delta \expval{\hS^2}_m^{\spc}$ and $\Delta \expval{\hS^2}_m^{\spf}$ can be found in the Appendix of Ref.~\onlinecite{Li_2011a} for spin-conserved and spin-flip excitations, and are functions of the vectors $\bX{m}{}$ and $\bY{m}{}$ as well as the orbital overlaps defined in Eq.~\eqref{eq:OV}.

\section{Computational details}
\label{sec:compdet}
All the systems under investigation here have a closed-shell singlet ground state and we consider the lowest triplet state as reference for the spin-flip calculations adopting the unrestricted formalism throughout this work.
The {\GOWO} calculations performed to obtain the screened Coulomb potential and the quasiparticle energies required to compute the BSE neutral excitations are performed using an unrestricted Hartree-Fock (UHF) starting point, and the {\GOWO} quasiparticle energies are obtained by linearizing the frequency-dependent quasiparticle equation [see Eq.~\eqref{eq:G0W0_lin}].
Note that the entire set of orbitals and energies is corrected.
Further details about our implementation of {\GOWO} can be found in Refs.~\onlinecite{Loos_2018b,Veril_2018,Loos_2020e,Loos_2020h,Berger_2021}.

Here, we do not investigate how the starting orbitals affect the BSE@{\GOWO} excitation energies.
This is left for future work.
However, it is worth mentioning that, for the present (small) molecular systems, Hartree-Fock is usually a good starting point, \cite{Loos_2020a,Loos_2020e,Loos_2020h} although improvements could certainly be obtained with starting orbitals and energies computed with, for example, optimally-tuned range-separated hybrid (RSH) functionals. \cite{Stein_2009,Stein_2010,Refaely-Abramson_2012,Kronik_2012}
Besides, {\GOWO}@UHF and ev$GW$@UHF yield similar quasiparticle energies, while {\GOWO} allows us to avoid rather laborious iterations as well as the significant additional computational effort of ev$GW$. \cite{Loos_2020e,Loos_2020h,Berger_2021}
In the following, all linear response calculations are performed within the TDA to ensure consistency between the spin-conserved and spin-flip results.
Finally, the infinitesimal $\eta$ is set to $100$ meV for all calculations.
 
All the static and dynamic BSE calculations (labeled in the following as SF-BSE and SF-dBSE respectively) are performed with the software \texttt{QuAcK}, \cite{QuAcK} developed in our group and freely available on \texttt{github}.
The standard and extended spin-flip ADC(2) calculations [SF-ADC(2)-s and SF-ADC(2)-x, respectively] as well as the SF-ADC(3) \cite{Lefrancois_2015} are performed with Q-CHEM 5.2.1. \cite{qchem4}
Spin-flip TD-DFT calculations \cite{Shao_2003} (also performed with Q-CHEM 5.2.1) considering the BLYP, \cite{Becke_1988,Lee_1988} B3LYP, \cite{Becke_1988,Lee_1988,Becke_1993a} and BH\&HLYP \cite{Lee_1988,Becke_1993b} functionals with contains $0\%$, $20\%$, and $50\%$ of exact exchange are labeled as SF-TD-BLYP, SF-TD-B3LYP, and SF-TD-BH\&HLYP, respectively.
\alert{Additionally, we have performed spin-flip TD-DFT calculations considering the following the RSH functionals: CAM-B3LYP, \cite{Yanai_2004} LC-$\omega$PBE08, \cite{Weintraub_2009} and $\omega$B97X-D. \cite{Chai_2008a,Chai_2008b}
In the present context, the main difference between these RSHs is their amount of exact exchange at long range: 75\% for CAM-B3LYP and 100\% for both LC-$\omega$PBE08 and $\omega$B97X-D.}
EOM-CCSD excitation energies \cite{Koch_1990,Stanton_1993,Koch_1994} are computed with Gaussian 09. \cite{g09}
As a consistency check, we systematically perform SF-CIS calculations \cite{Krylov_2001a} with both \texttt{QuAcK} and Q-CHEM, and make sure that they yield identical excitation energies.
Throughout this work, all spin-flip and spin-conserved calculations are performed with a UHF reference.

\section{Results}
\label{sec:res}

\subsection{Beryllium atom}
\label{sec:Be}

\begin{figure*}
	\includegraphics[width=0.8\linewidth]{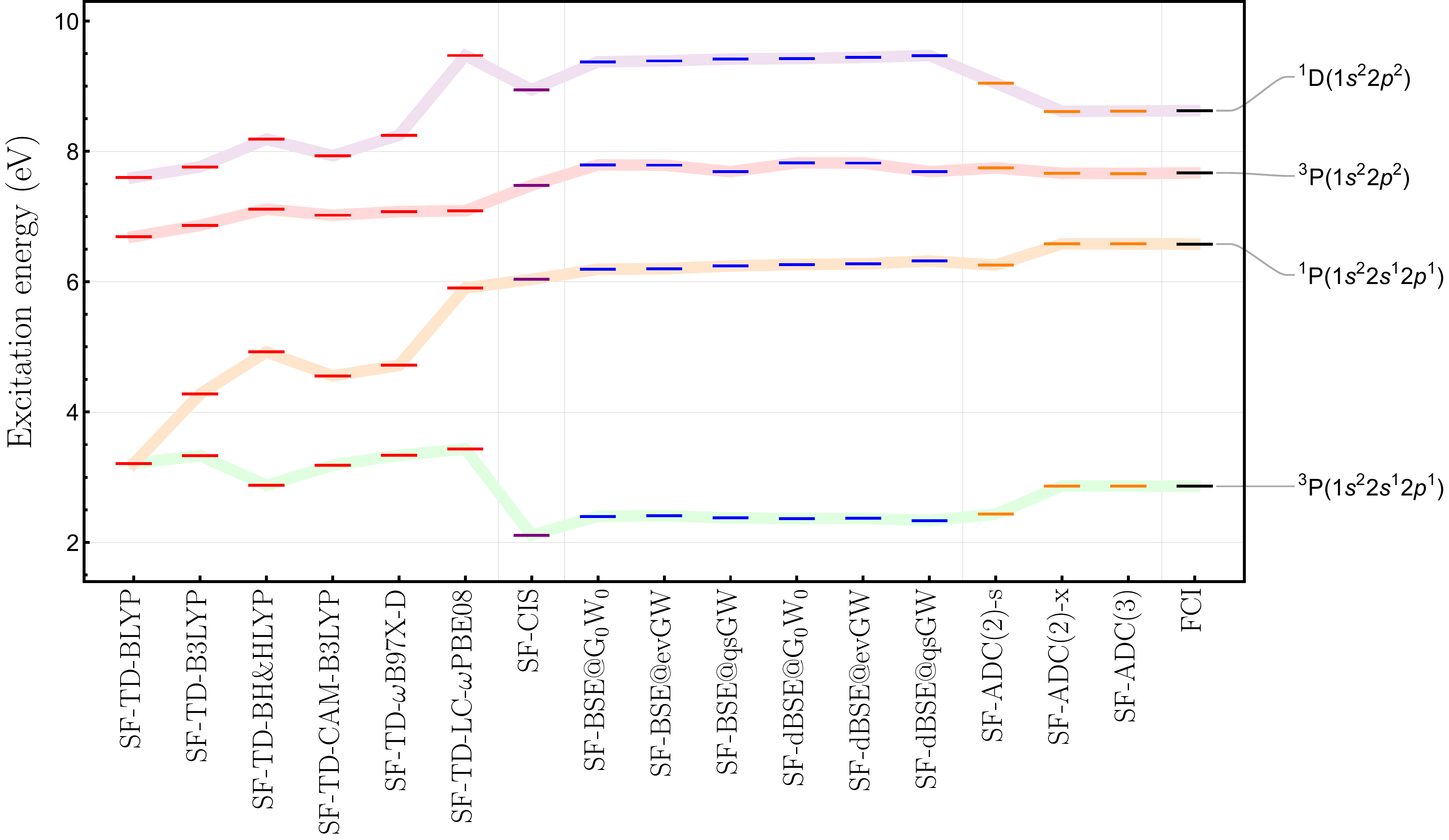}
	\caption{
	Excitation energies (in eV) with respect to the $^1S(1s^2 2s^2)$ singlet ground state of \ce{Be} obtained with the 6-31G basis at various levels of theory:
	SF-TD-DFT (red), SF-CIS (purple), SF-BSE (blue), SF-ADC (orange), and FCI (black). 
	All the spin-flip calculations have been performed with an unrestricted reference.
	\label{fig:Be}}
\end{figure*}

As a first example, we consider the simple case of the beryllium atom in a small basis (6-31G) which was considered by Krylov in two of her very first papers on spin-flip methods. \cite{Krylov_2001a,Krylov_2001b} 
It was also considered in later studies thanks to its pedagogical value. \cite{Sears_2003,Casanova_2020} 
Beryllium has a $^1S$ ground state with $1s^2 2s^2$ configuration. 
The excitation energies corresponding to the first singlet and triplet single excitations $2s \to 2p$ with $P$ spatial symmetries as well as the first singlet and triplet double excitations $2s^2 \to 2p^2$ with $D$ and $P$ spatial symmetries (respectively) are reported in Table \ref{tab:Be} and depicted in Fig.~\ref{fig:Be}.

On the left side of Fig.~\ref{fig:Be}, we report SF-TD-DFT excitation energies (red lines) obtained with the BLYP, B3LYP, and BH\&HLYP functionals, which correspond to an increase of exact exchange from 0\% to 50\%. 
As mentioned in Ref.~\onlinecite{Casanova_2020}, the $^3P(1s^2 2s^1 2p^1)$ and the $^1P(1s^2 2s^1 2p^1)$ states are degenerate at the SF-TD-BLYP level.
Indeed, due to the lack of coupling terms in the spin-flip block of the SD-TD-DFT equations (see Subsec.~\ref{sec:BSE}), their excitation energies are given by the energy difference between the $2s$ and $2p$ orbitals and both states are strongly spin contaminated.
Including exact exchange, like in SF-TD-B3LYP and SF-TD-BH\&HLYP, lifts this degeneracy and improves the description of both states.
However, the SF-TD-BH\&HLYP excitation energy of the $^1P(1s^2 2s^1 2p^1)$ state is still off by $1.6$ eV as compared to the FCI reference. 
For the other states, the agreement between SF-TD-BH\&HLYP and FCI is significantly improved. 
\alert{Spin-flip TD-DFT calculations performed with CAM-B3LYP and $\omega$B97X-D are only slightly more accurate than their global hybrid counterparts, while SF-TD-LC-$\omega$PBE08 yields more significant improvements although it does not reach the accuracy of SF-(d)BSE.}

The center part of Fig.~\ref{fig:Be} shows the SF-(d)BSE results (blue lines) alongside the SF-CIS excitation energies (purple lines). 
All of these are computed with 100\% of exact exchange with the additional inclusion of correlation in the case of SF-BSE and SF-dBSE thanks to the introduction of static and dynamical screening, respectively.
Overall, the SF-CIS and SF-BSE excitation energies are closer to FCI than the SF-TD-DFT ones, except for the lowest triplet state where the SF-TD-BH\&HLYP excitation energy is more accurate probably due to error compensation.
At the exception of the $^1D$ state, SF-BSE improves over SF-CIS with a rather small contribution from the additional dynamical effects included in the SF-dBSE scheme.
Note that the exact exchange seems to spin purified the $^3P(1s^2 2s^1 2p^1)$ state while the singlet states at the SF-BSE level are slightly more spin contaminated than their SF-CIS counterparts.

\alert{Table \ref{tab:Be} and Fig.~\ref{fig:Be} also gathers results obtained at the partially self-consistent SF-(d)BSE@ev$GW$ and fully self-consistent SF-(d)BSE@qs$GW$ levels.
The SF-(d)BSE excitation energies are quite stable with respect to the underlying $GW$ scheme which nicely illustrates that UHF eigenstates are actually an excellent starting point in this particular case.}

The right side of Fig.~\ref{fig:Be} illustrates the performance of the SF-ADC methods.
Interestingly, SF-BSE and SF-ADC(2)-s have rather similar accuracies, except again for the $^1D$ state where SF-ADC(2)-s has clearly the edge over SF-BSE.
Finally, both SF-ADC(2)-x and SF-ADC(3) yield excitation energies very close to FCI for this simple system with significant improvements for the lowest $^3P$ state and the $^1D$ doubly-excited state.
\alert{Although the (d)BSE and ADC(2)-s have obvious theoretical similarities, we would like to mention that they are not strictly identical as ADC(2) includes key second-order exchange contributions that are not included at the $GW$ level even in the case of more elaborate schemes like ev$GW$ and qs$GW$.}

\begin{table*}
	\caption{
	Excitation energies (in eV) with respect to the $^1S(1s^2 2s^2)$ singlet ground state of \ce{Be} obtained at various methods with the 6-31G basis set.
	All the spin-flip calculations have been performed with an unrestricted reference.
	The $\expval{\hS^2}$ value associated with each state is reported in parenthesis (when available).
	\label{tab:Be}}
	\begin{ruledtabular}
		\begin{tabular}{llllll}
									&	\mc{5}{c}{Excitation energies (eV)}		\\
										\cline{2-6}					
			Method					&	$^1S(1s^2 2s^2)$		&	$^3P(1s^2 2s^1 2p^1)$	&	$^1P(1s^2 2s^1 2p^1)$	
									&	$^3P(1s^22 p^2)$		&	$^1D(1s^22p^2)$			\\
			\hline
			SF-TD-BLYP\fnm[1]		&	(0.002)	&	3.210(1.000)	&	3.210(1.000)	&	6.691(1.000)	&	7.598(0.013)	\\
			SF-TD-B3LYP\fnm[1]		&	(0.001)	&	3.332(1.839)	&	4.275(0.164)	&	6.864(1.000)	&	7.762(0.006)	\\	
			SF-TD-BH\&HLYP\fnm[1]	&	(0.000)	&	2.874(1.981)	&	4.922(0.023)	&	7.112(1.000)	&	8.188(0.002)	\\
			\alert{SF-TD-CAM-B3LYP} &	(0.001) &	3.186(1.960)	&	4.554(0.043)	&	7.020(1.000)	&	7.933(0.008)	\\
			\alert{SF-TD-$\omega$B97X-D} & (0.006) & 3.337(1.867) & 4.717(0.147)	& 7.076(1.000) & 8.247(0.040)\\
			\alert{SF-TD-LC-$\omega$PBE08} & (0.014) & 3.434(1.720) &5.904(0.287)& 7.088(1.000) & 9.471(0.073) \\
			SF-CIS\fnm[2]			&	(0.002)	&	2.111(2.000)	&	6.036(0.014)	&	7.480(1.000)	&	8.945(0.006)	\\
			SF-BSE@{\GOWO}			&	(0.004)	&	2.399(1.999)	&	6.191(0.023)	&	7.792(1.000)	&	9.373(0.013)	\\
			\alert{SF-BSE@ev$GW$}	&	(0.004)	&	2.407(1.999)	&	6.199(0.023)	&	7.788(1.000)	&	9.388(0.013)	\\
			\alert{SF-BSE@qs$GW$}	&	(0.057)	&	2.376(1.963)	&	6.241(0.048)	&	7.668(1.000)	&	9.417(0.004)	\\
			SF-dBSE@{\GOWO}			&			&	2.363			&	6.263			&	7.824			&	9.424			\\
			\alert{SF-dBSE@ev$GW$}	&			&	2.369			&	6.273			&	7.820			&	9.441	\\
			\alert{SF-dBSE@qs$GW$}	&			&	2.335			&	6.317			&	7.689			&	9.470	\\
			SF-ADC(2)-s				&			&	2.433			&	6.255			&	7.745			&	9.047 		\\
			SF-ADC(2)-x				&			&	2.866			&	6.581			&	7.664			&	8.612		\\
			SF-ADC(3)				&			&	2.863			&	6.579			&	7.658			&	8.618 		\\
			FCI\fnm[2]				&	(0.000)	&	2.862(2.000)	&	6.577(0.000)	&	7.669(2.000)	&	8.624(0.000)	\\	
		\end{tabular}
	\end{ruledtabular}
	\fnt[1]{Excitation energies taken from Ref.~\onlinecite{Casanova_2020}.}
	\fnt[2]{Excitation energies taken from Ref.~\onlinecite{Krylov_2001a}.}
\end{table*}

\subsection{Hydrogen molecule}
\label{sec:H2}

Our second example deals with the dissociation of the \ce{H2} molecule, which is a prototypical system for testing new electronic structure methods and, specifically, their accuracy in the presence of strong correlation (see, for example, Refs.~\onlinecite{Caruso_2013,Barca_2014,Vuckovic_2017,Li_2021}, and references therein).
The $\text{X}\,{}^1 \Sigma_g^+$ ground state of \ce{H2} has an electronic configuration $(1\sigma_g)^2$ configuration. 
The variation of the excitation energies associated with the three lowest singlet excited states with respect to the elongation of the \ce{H-H} bond are of particular interest here. 
The lowest singly excited state $\text{B}\,{}^1 \Sigma_u^+$ has a $(1\sigma_g )(1\sigma_u)$ configuration, while the singly excited state $\text{E}\,{}^1 \Sigma_g^+$ and the doubly excited state $\text{F}\,{}^1 \Sigma_g^+$ have $(1\sigma_g ) (2\sigma_g)$ and $(1\sigma_u )^2$ configurations, respectively. 
Because these latter two excited states interact strongly and form an avoided crossing around $R(\ce{H-H}) = 1.4$ \AA, they are usually labeled as the $\text{EF}\,{}^1 \Sigma_g^+$ state.
Note that this avoided crossing is not visible with non-spin-flip methods restricted to single excitations (such as CIS, TD-DFT, and BSE) as these are ``blind'' to double excitations.
Three methods, in their standard and spin-flip versions, are studied here (CIS, TD-BH\&HLYP and BSE) and are compared to the reference EOM-CCSD excitation energies (that is equivalent to FCI in the case of \ce{H2}).
All these calculations are performed with the cc-pVQZ basis.

\begin{figure*}
	\includegraphics[width=0.45\linewidth]{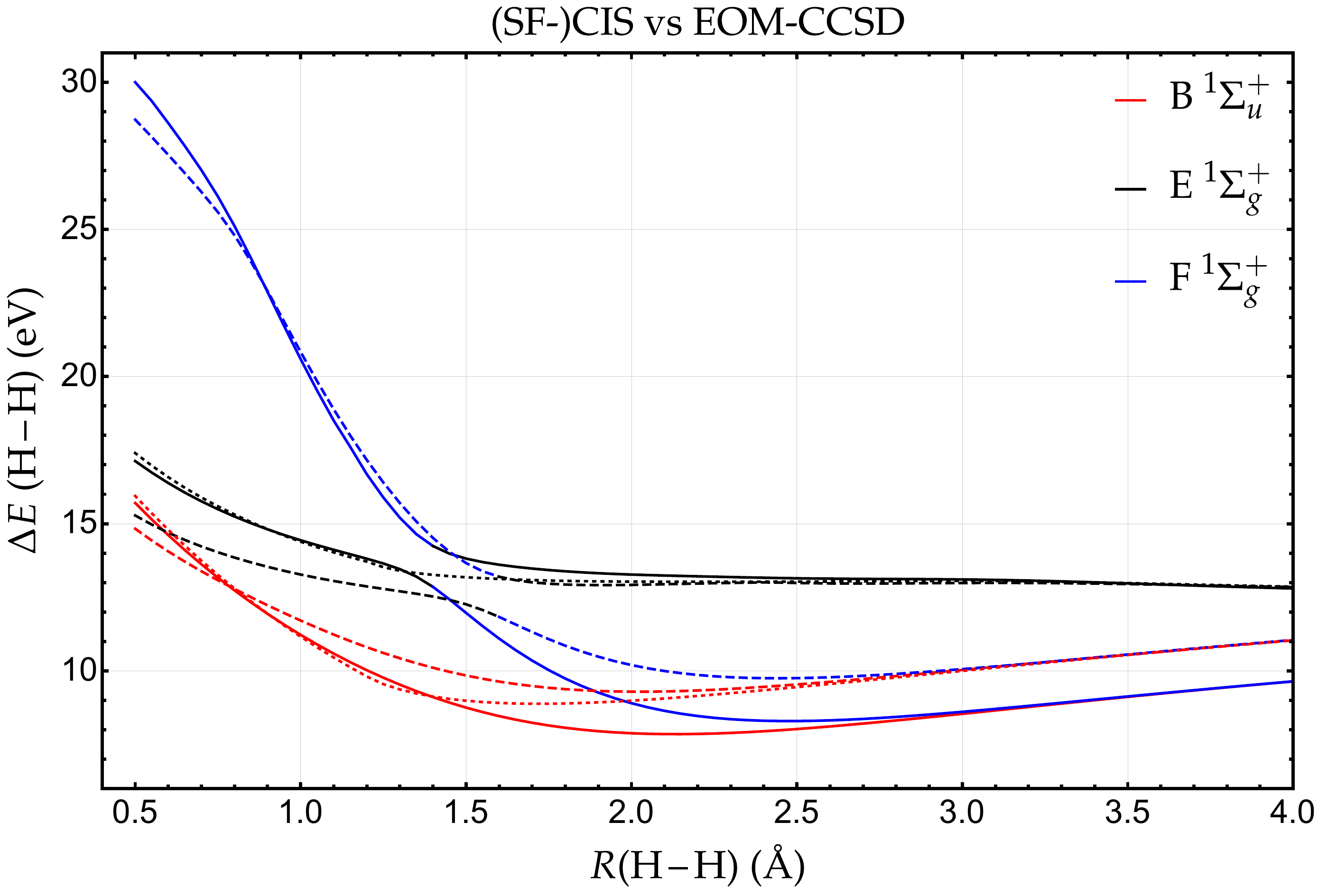}
	\hspace{0.05\linewidth}
	\includegraphics[width=0.45\linewidth]{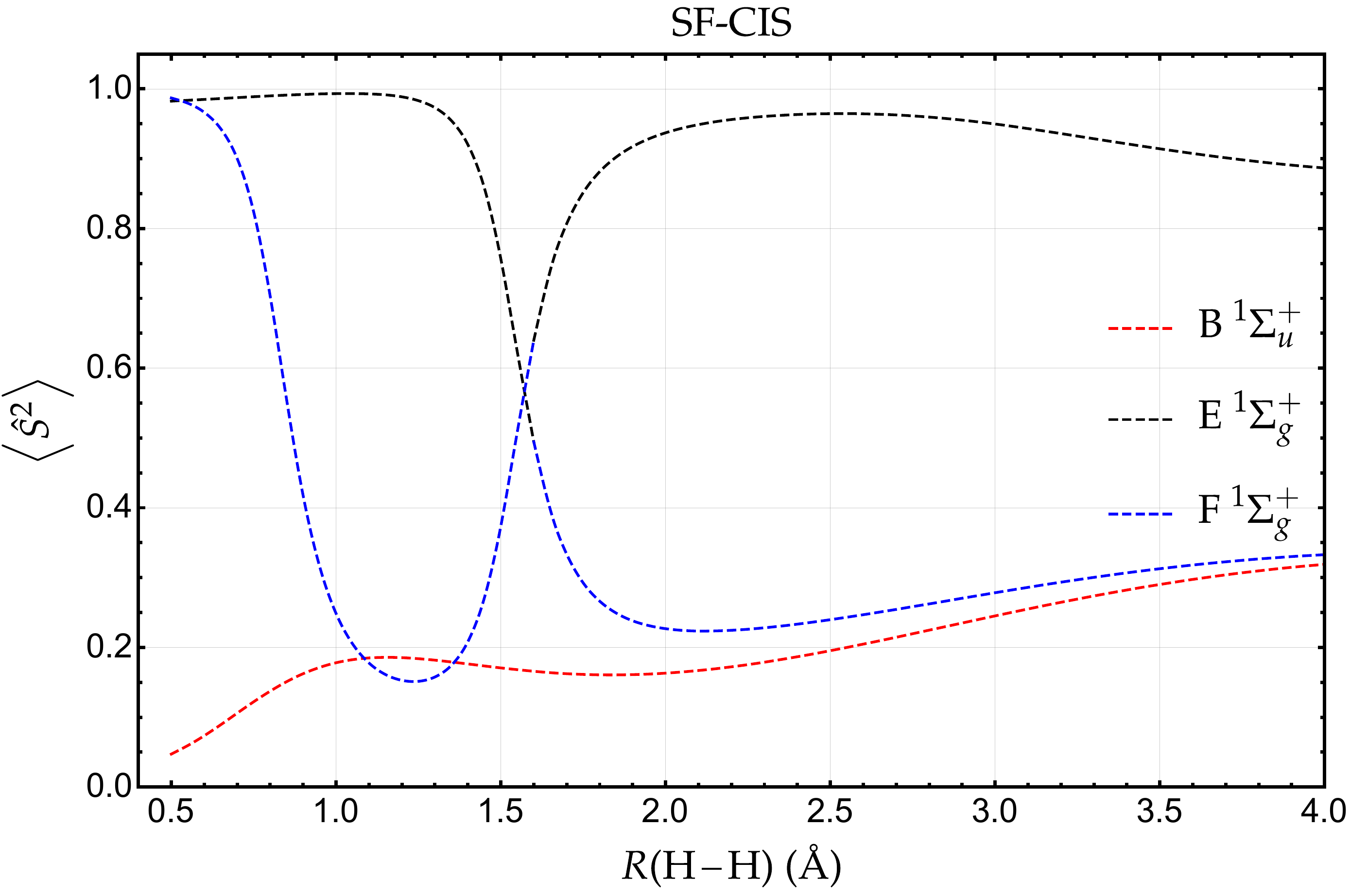}
	\vspace{0.025\linewidth}
	\\
	\includegraphics[width=0.45\linewidth]{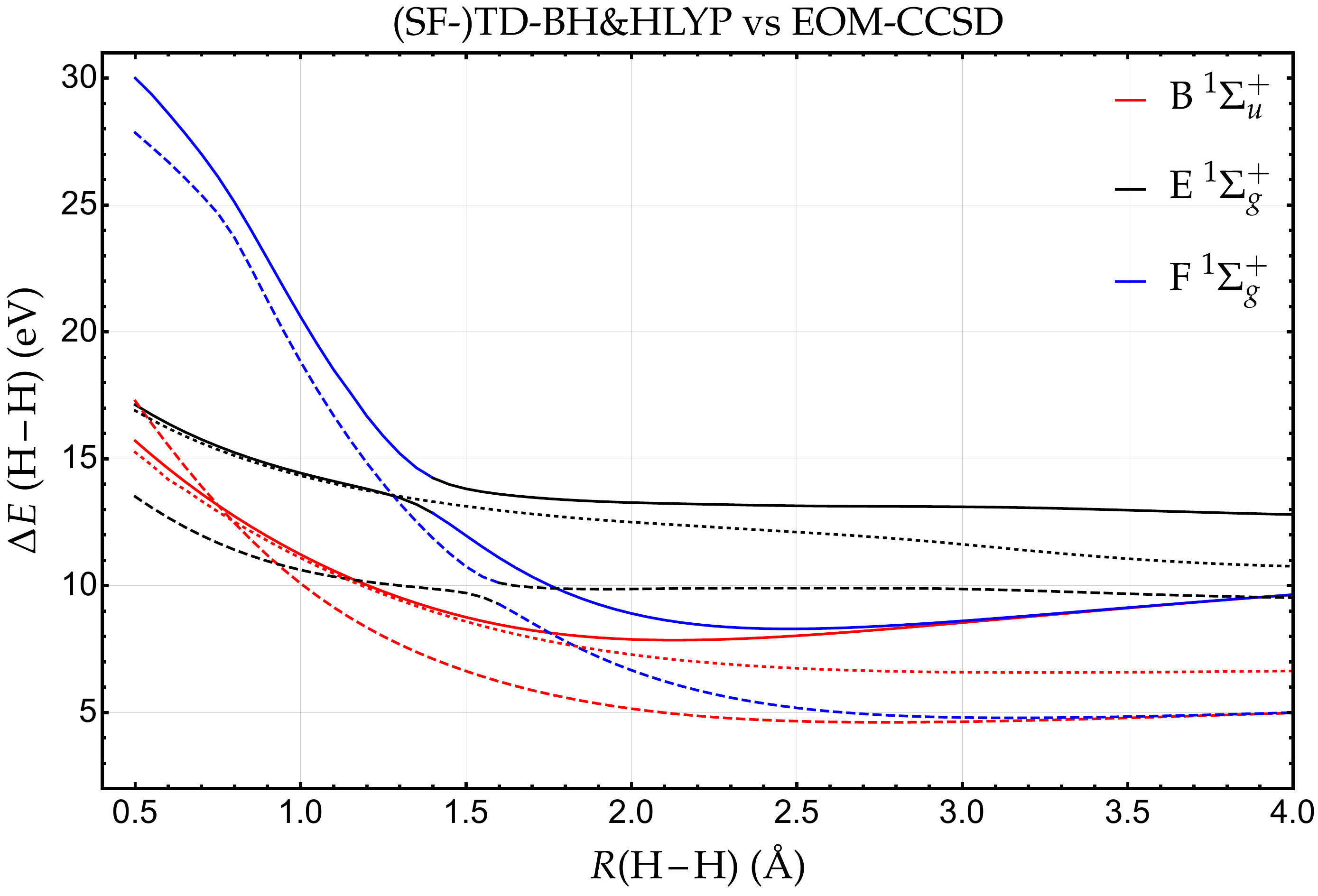}
	\hspace{0.05\linewidth}
	\includegraphics[width=0.45\linewidth]{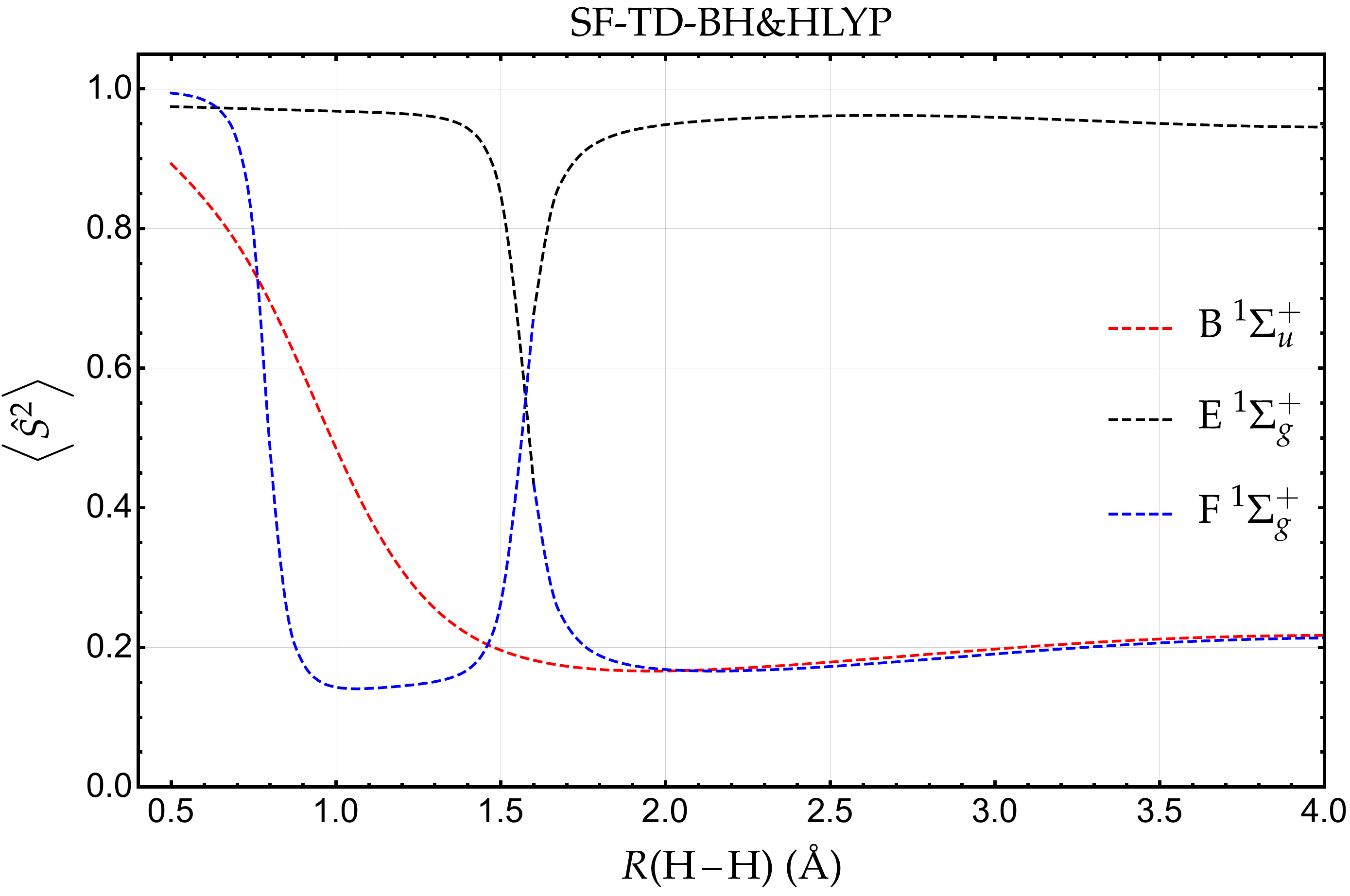}
	\vspace{0.025\linewidth}
	\\
	\includegraphics[width=0.45\linewidth]{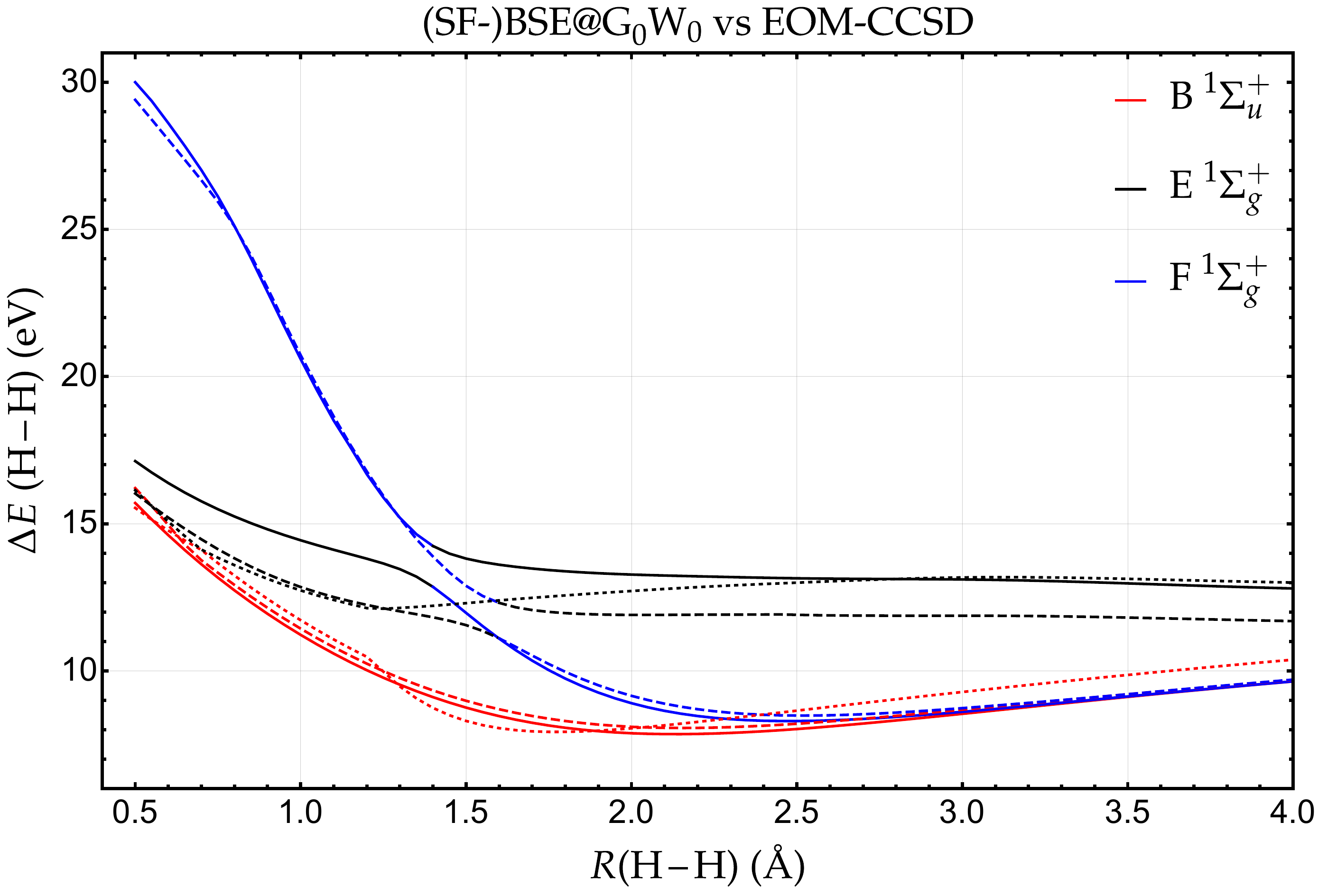}
	\hspace{0.05\linewidth}
	\includegraphics[width=0.45\linewidth]{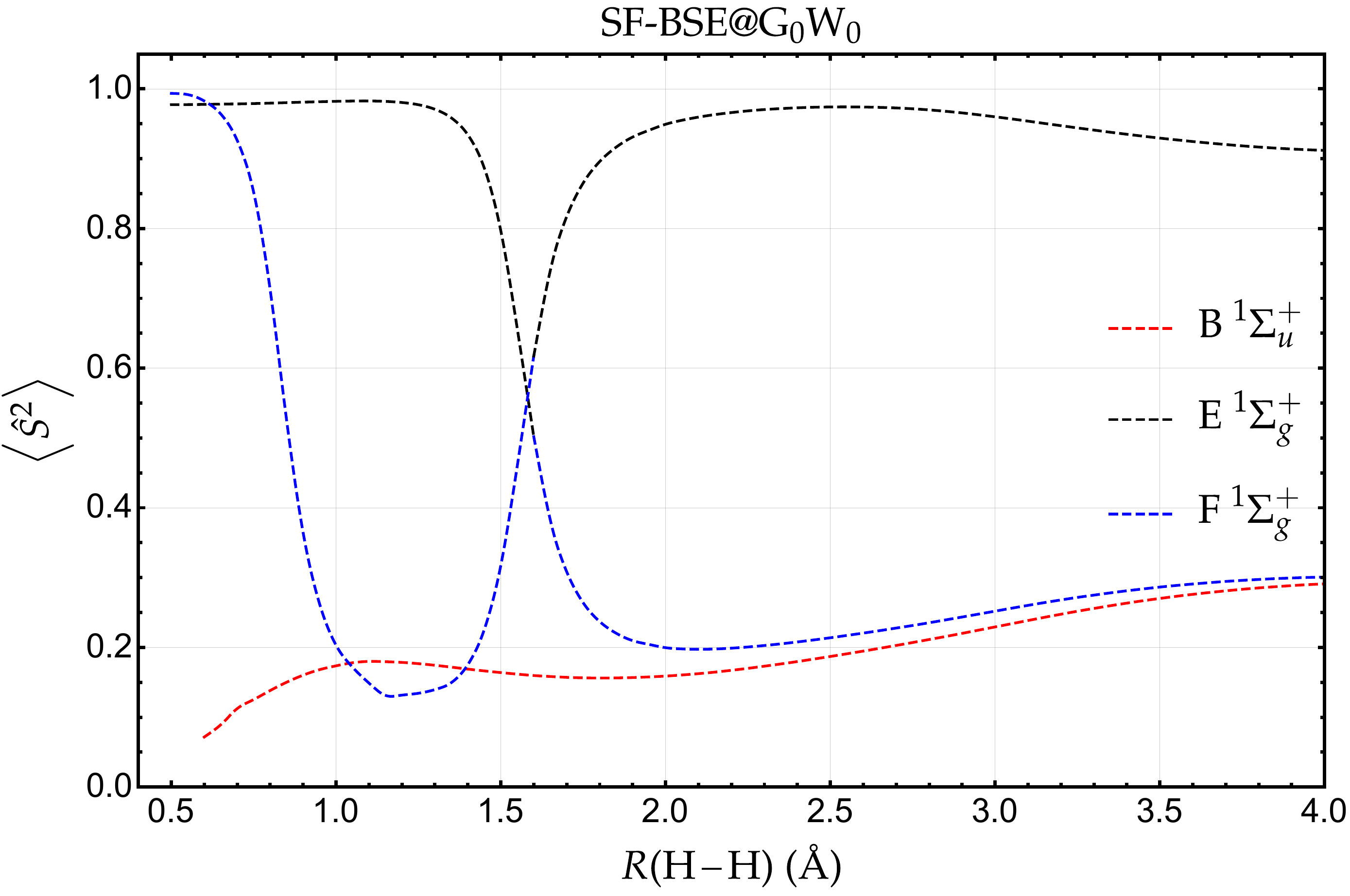}
\caption{
	Excitation energies with respect to the $\text{X}\,{}^1 \Sigma_g^+$ ground state (left) and expectation value of the spin operator $\expval{\hS^2}$ (right) of the $\text{B}\,{}^1\Sigma_u^+$ (red), $\text{E}\,{}^1\Sigma_g^+$ (black), and $\text{F}\,{}^1\Sigma_g^+$ (blue) states of \ce{H2} obtained with the cc-pVQZ basis at the (SF-)CIS (top), (SF-)TD-BH\&HLYP (middle), and (SF-)BSE (bottom) levels of theory.  
	The reference EOM-CCSD excitation energies are represented as solid lines, while the results obtained with and without spin-flip are represented as dashed and dotted lines, respectively.
	All the spin-conserved and spin-flip calculations have been performed with an unrestricted reference.
	The raw data are reported in the {\SI}.
	\label{fig:H2}}
\end{figure*}

The top panel of Fig.~\ref{fig:H2} shows the CIS (dotted lines) and SF-CIS (dashed lines) excitation energies as functions of $R(\ce{H-H})$.
The EOM-CCSD reference energies are represented by solid lines.
We observe that both CIS and SF-CIS poorly describe the $\text{B}\,{}^1\Sigma_u^+$ state in the dissociation limit with an error greater than $1$ eV, while CIS, unlike SF-CIS, is much more accurate around the equilibrium geometry.
Similar observations can be made for the $\text{E}\,{}^1\Sigma_g^+$ state with a good description at the CIS level for all bond lengths.
SF-CIS does not model accurately the $\text{E}\,{}^1\Sigma_g^+$ state before the avoided crossing, but the agreement between SF-CIS and EOM-CCSD is much satisfactory for bond length greater than $1.6$ \AA. 
Oppositely, SF-CIS describes better the $\text{F}\,{}^1\Sigma_g^+$ state before the avoided crossing than after, while this state is completely absent at the CIS level.
Indeed, as mentioned earlier, CIS is unable to locate any avoided crossing as it cannot access double excitations.
At the SF-CIS level, the avoided crossing between the $\text{E}$ and $\text{F}$ states is qualitatively reproduced and placed at a slightly larger bond length [$R(\ce{H-H}) \approx 1.5$ \AA] than at the EOM-CCSD level.

In the central panel of Fig.~\ref{fig:H2}, we report the (SF-)TD-BH\&HLYP results.
SF-TD-BH\&HLYP shows, at best, qualitative agreement with EOM-CCSD, while the TD-BH\&HLYP excitation energies of the $\text{B}$ and $\text{E}$ states are only trustworthy around equilibrium but inaccurate at dissociation.
Note that \ce{H2} is a rather challenging system for (SF)-TD-DFT from a general point of view. \cite{Vuckovic_2017,Cohen_2008a,Cohen_2008c,Cohen_2012}
Similar graphs for (SF-)TD-BLYP and (SF-)TD-B3LYP are reported in the {\SI} from which one can draw similar conclusions.
Notably, one can see that the $\text{E}\,{}^1\Sigma_g^+$ and $\text{F}\,{}^1 \Sigma_g^+$ states crossed without interacting at the SF-TD-BLYP level due to the lack of Hartree-Fock exchange.
\alert{In the {\SI}, we also report the potential energy curves of \ce{H2} obtained with three RSHs (CAM-B3LYP, $\omega$B97X-D, and LC-$\omega$PBE08), which only brought a
modest improvement.}

In the bottom panel of Fig.~\ref{fig:H2}, (SF-)BSE excitation energies for the same three singlet states are represented. 
SF-BSE provides surprisingly accurate excitation energies for the $\text{B}\,{}^1\Sigma_u^+$ state with errors between $0.05$ and $0.3$ eV, outperforming in the process the standard BSE formalism.
However SF-BSE does not describe well the $\text{E}\,{}^1\Sigma_g^+$ state with error ranging from half an eV to $1.6$ eV. 
Similar performances are observed at the BSE level around equilibrium with a clear improvement in the dissociation limit.
Remarkably, SF-BSE shows a good agreement with EOM-CCSD for the $\text{F}\,{}^1\Sigma_g^+$ doubly-excited state, resulting in an avoided crossing around $R(\ce{H-H}) = 1.6$ \AA.
A similar graph comparing (SF-)dBSE and EOM-CCSD excitation energies can be found in the {\SI} where it is shown that dynamical effects do not affect the present conclusions.
\alert{One would also notice a little ``kink'' in the potential energy curves of the $\text{B}\,{}^1\Sigma_u^+$ and $\text{E}\,{}^1\Sigma_g^+$ states around $R(\ce{H-H}) = 1.2~\AA$ computed at the (d)BSE@{\GOWO} level. 
This unfortunate feature is due to the appearance of the symmetry-broken UHF solution and the lack of self-consistent in {\GOWO}.	
Indeed, $R = 1.2~\AA$ corresponds to the location of the well-known Coulson-Fischer point. \cite{Coulson_1949}
Note that, as mentioned earlier, all the calculations are performed with a UHF reference even the ones based on a closed-shell singlet reference.
If one relies solely on the restricted HF solution, this kink disappears and one obtains smooth potential energy curves (see {\SI}).} 
	
The right side of Fig.~\ref{fig:H2} shows the amount of spin contamination as a function of the bond length for SF-CIS (top), SF-TD-BH\&HLYP (center), and SF-BSE (bottom).
Overall, one can see that $\expval{\hS^2}$ behaves similarly for SF-CIS and SF-BSE with a small spin contamination of the $\text{B}\,{}^1\Sigma_u^+$ at short bond length. In contrast, the $\text{B}$ state is much more spin contaminated at the SF-TD-BH\&HLYP level.
For all spin-flip methods, the $\text{E}$ state is strongly spin contaminated as expected, while the $\expval{\hS^2}$ values associated with the $\text{F}$ state 
only deviate significantly from zero for short bond length and around the avoided crossing where it strongly couples with the spin contaminated $\text{E}$ state.

\subsection{Cyclobutadiene}
\label{sec:CBD}

Cyclobutadiene (CBD) is an interesting example as the electronic character of its ground state can be tuned via geometrical deformation. \cite{Balkova_1994,Levchenko_2004,Manohar_2008,Karadakov_2008,Li_2009,Shen_2012,Lefrancois_2015,Casanova_2020,Vitale_2020}
In the $D_{2h}$ rectangular geometry of the $A_g$ singlet ground state, the highest occupied molecular orbital (HOMO) and lowest unoccupied molecular orbital (LUMO) are non-degenerate, and the singlet ground state can be safely labeled as single-reference with well-defined doubly-occupied orbitals.
However, in the $D_{4h}$ square-planar geometry of the $A_{2g}$ triplet state, the HOMO and LUMO are strictly degenerate, and the electronic ground state, which is still of singlet nature with $B_{1g}$ spatial symmetry (hence violating Hund's rule), is strongly multi-reference with singly occupied orbitals (\ie, singlet open-shell state).
In this case, single-reference methods notoriously fail.
Nonetheless, the lowest triplet state of symmetry $^3 A_{2g}$ remains of single-reference character and is then a perfect starting point for spin-flip calculations.
The $D_{2h}$ and $D_{4h}$ optimized geometries of the $^1 A_g$ and $^3 A_{2g}$ states of CBD have been extracted from Ref.~\onlinecite{Manohar_2008} and have been obtained at the CCSD(T)/cc-pVTZ level.
For comparison purposes, EOM-SF-CCSD and SF-ADC excitation energies have been extracted from Ref.~\onlinecite{Manohar_2008} and Ref.~\onlinecite{Lefrancois_2015}, respectively.
All of them have been obtained with a UHF reference like the SF-BSE calculations performed here.

\begin{figure*}
	\includegraphics[width=0.45\linewidth]{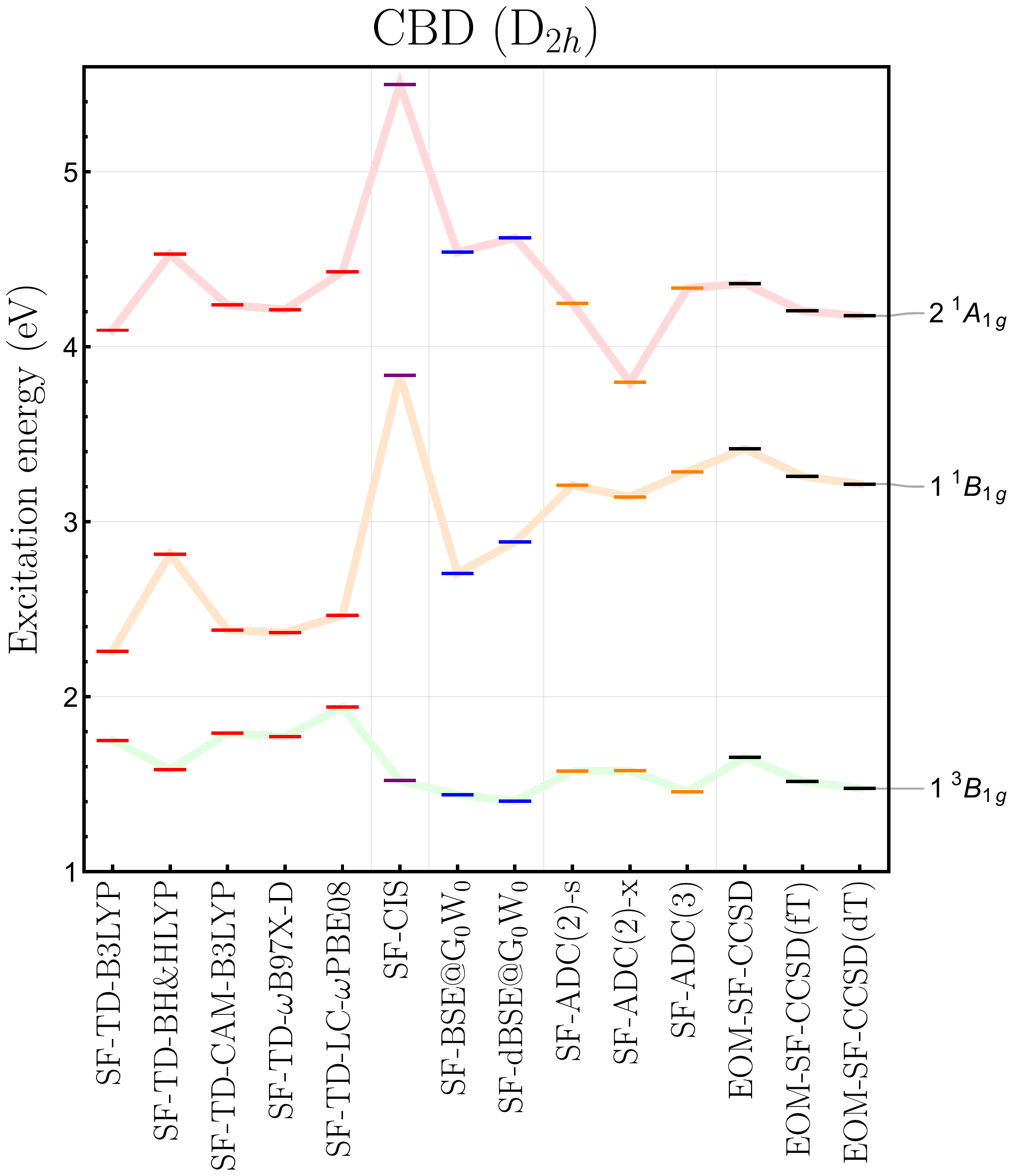}
	\hspace{0.05\linewidth}
	\includegraphics[width=0.45\linewidth]{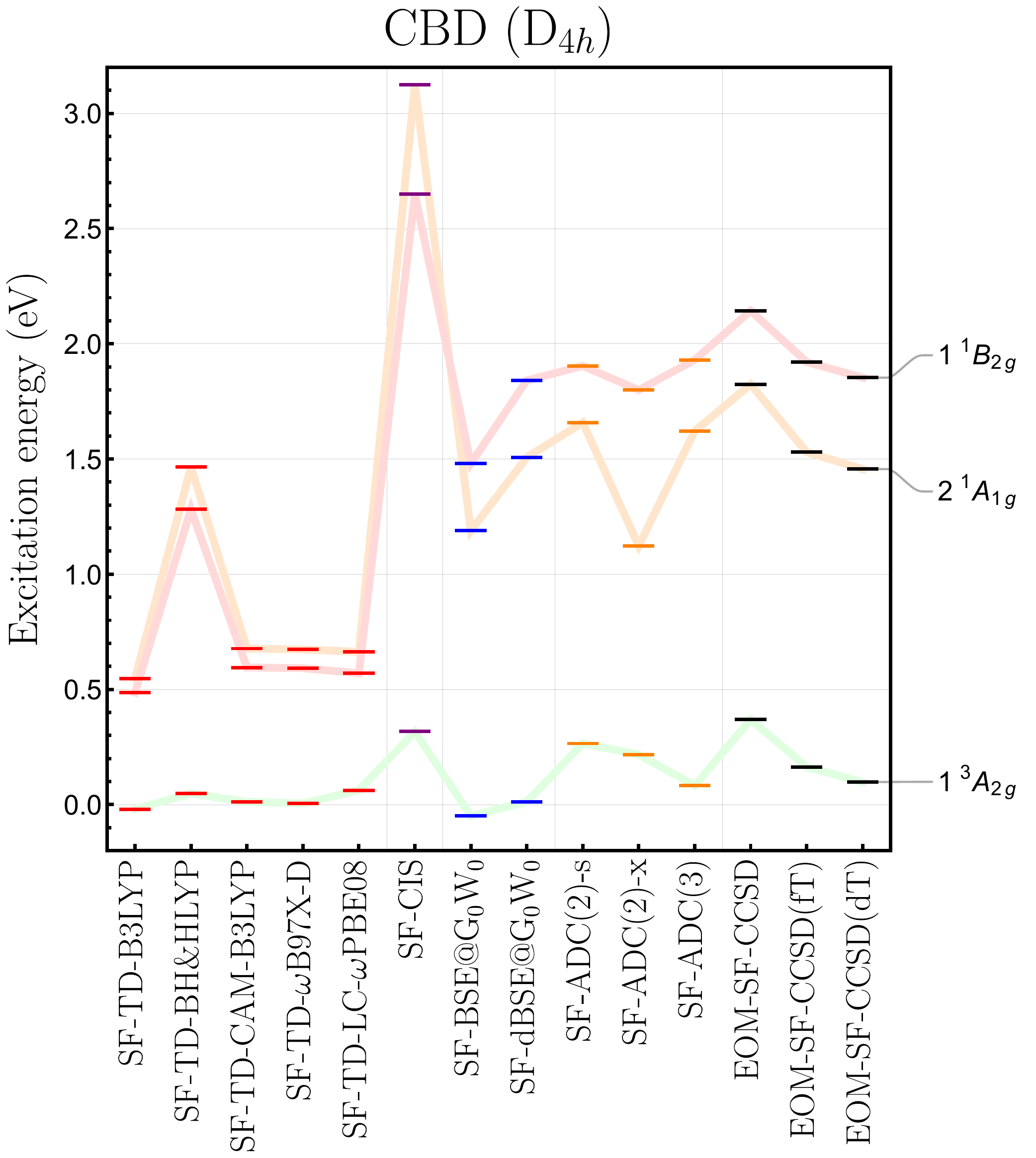}
\caption{
	Vertical excitation energies of CBD at various levels of theory:
	SF-TD-DFT (red), SF-CIS (purple), SF-BSE (blue), SF-ADC (orange), and EOM-SF-CCSD (black).
	Left: $1\,{}^3B_{1g}$, $1\,{}^1B_{1g}$, and $2\,{}^1A_{1g}$ states at the $D_{2h}$ rectangular equilibrium geometry of the $\text{X}\,{}^1 A_{g}$ ground state (see Table \ref{tab:CBD_D2h} for the raw data).
	Right: $1\,{}^3A_{2g}$, $2\,{}^1A_{1g}$, and $1\,{}^1B_{2g}$ states at the $D_{4h}$ square-planar equilibrium geometry of the $1\,{}^3 A_{2g}$ state (see Table \ref{tab:CBD_D4h} for the raw data).
	All the spin-flip calculations have been performed with an unrestricted reference and the cc-pVTZ basis set.
	\label{fig:CBD}}
\end{figure*}

Tables~\ref{tab:CBD_D2h} and \ref{tab:CBD_D4h} report excitation energies (with respect to the singlet ground state) obtained at the $D_{2h}$ and $D_{4h}$ geometries, respectively, for several methods using the spin-flip \textit{ansatz}.
All these results are represented in Fig.~\ref{fig:CBD}.
For each geometry, three excited states are under investigation:
i) the $1\,{}^3B_{1g}$, $1\,{}^1B_{1g}$, and $2\,{}^1A_{g}$ states of the $D_{2h}$ geometry;
ii) the $1\,{}^3 A_{2g}$, $2\,{}^1 A_{1g}$, and $1\,{}^1 B_{2g}$ states of the $D_{4h}$ geometry.
It is important to mention that the $2\,{}^1A_{1g}$ state of the rectangular geometry has a significant double excitation character, \cite{Loos_2019} and is then hardly described by second-order methods [such as CIS(D), \cite{Head-Gordon_1994,Head-Gordon_1995} ADC(2), \cite{Trofimov_1997,Dreuw_2015} CC2, \cite{Christiansen_1995a} or EOM-CCSD \cite{Koch_1990,Stanton_1993,Koch_1994}] and remains a real challenge for third-order methods [as, for example, ADC(3), \cite{Trofimov_2002,Harbach_2014,Dreuw_2015} CC3, \cite{Christiansen_1995b} or EOM-CCSDT \cite{Kucharski_1991,Kallay_2004,Hirata_2000,Hirata_2004}].

Comparing the present SF-BSE@{\GOWO} results for the rectangular geometry (see Table \ref{tab:CBD_D2h}) to the most accurate ADC level, \ie, SF-ADC(3), we have a difference in excitation energy of $0.017$ eV for the $1\,^3B_{1g}$ state. 
This difference grows to $0.572$ eV for the $1\,^1B_{1g}$ state and then shrinks to $0.212$ eV for the $2\,^1A_{g}$ state. 
Overall, adding dynamical corrections via the SF-dBSE@{\GOWO} scheme does not improve the accuracy of the excitation energies [as compared to SF-ADC(3)] with errors of $0.052$, $0.393$, and $0.293$ eV for the $1\,^3B_{1g}$, $1\,^1B_{1g}$, and $2\,^1 A_{g}$ states, respectively.

Now, looking at Table \ref{tab:CBD_D4h} which gathers the results for the square-planar geometry, we see that, at the SF-BSE@{\GOWO} level, the first two states are wrongly ordered with the triplet $1\,^3B_{1g}$ state lower than the singlet $1\,^1A_g$ state.
(The same observation can be made at the SF-TD-B3LYP level.)
This is certainly due to the poor Hartree-Fock reference which lacks opposite-spin correlation and this issue could be potentially alleviated by using a better starting point for the $GW$ calculation, as discussed in Sec.~\ref{sec:compdet}. 
Nonetheless, it is pleasing to see that adding the dynamical correction in SF-dBSE@{\GOWO} not only improves the agreement with SF-ADC(3) but also retrieves the right state ordering. 
Then, CBD stands as an excellent example for which dynamical corrections are necessary to get the right chemistry at the SF-BSE level.
Another interesting feature is the wrong ordering of the $2\,{}^1A_{1g}$ and $1\,{}^1B_{2g}$ states at the SF-B3LYP, SF-BH\&HLYP, and SF-CIS levels which give the former higher in energy than the latter.
This issue does not appear at the SF-BSE, SF-ADC, and SF-EOM-SF-CCSD levels.
\alert{Here again, one does not observe a clear improvement by considering RSHs instead of global hybrids (BH\&HLYP seems to perform particularly well in the case of CBD), although it is worth mentioning that RSH-based SF-TD-DFT calculations yield accurate excitation for the double excitation $1\,{}^1A_{g} \to 2\,{}^1A_{g}$ in the $D_{2h}$ geometry.}

\begin{table}
	\caption{
	Vertical excitation energies (with respect to the singlet $\text{X}\,{}^1A_{g}$ ground state) of the $1\,{}^3B_{1g}$, $1\,{}^1B_{1g}$, and $2\,{}^1A_{g}$ states of CBD at the $D_{2h}$ rectangular equilibrium geometry of the $\text{X}\,{}^1 A_{g}$ ground state.
	All the spin-flip calculations have been performed with an unrestricted reference and the cc-pVTZ basis set.
	\label{tab:CBD_D2h}}
	\begin{ruledtabular}
		\begin{tabular}{lrrr}
									&	\mc{3}{c}{Excitation energies (eV)}	\\
									\cline{2-4}
			Method					&	$1\,{}^3B_{1g}$	&	$1\,{}^1B_{1g}$	&	$2\,{}^1A_{g}$	\\
			\hline
			SF-TD-B3LYP\fnm[1] 		&	$1.750$	&	$2.260$	&	$4.094$	\\
			SF-TD-BH\&HLYP\fnm[1] 	&	$1.583$	&	$2.813$	&	$4.528$	\\
			\alert{SF-TD-CAM-B3LYP} &	$1.790$ &	$2.379$ &	$4.238$ \\
			\alert{SF-TD-$\omega$B97X-D} & $1.771$ & $2.366$ & $4.212$ \\
			\alert{SF-TD-LC-$\omega$PBE08} & $1.941$ & $2.464$ & $4.428$  \\
			SF-CIS\fnm[2]			&	$1.521$	&	$3.836$	&	$5.499$	\\
			EOM-SF-CCSD\fnm[3]		&	$1.654$	&	$3.416$	&	$4.360$	\\
			EOM-SF-CCSD(fT)\fnm[3]	&	$1.516$	&	$3.260$	&	$4.205$	\\
			EOM-SF-CCSD(dT)\fnm[3]	&	$1.475$	&	$3.215$	&	$4.176$	\\
			SF-ADC(2)-s\fnm[4]		&	$1.573$	&	$3.208$	&	$4.247$	\\
			SF-ADC(2)-x\fnm[4]		&	$1.576$	&	$3.141$	&	$3.796$	\\
			SF-ADC(3)\fnm[2]		&	$1.456$	&	$3.285$	&	$4.334$	\\
			SF-BSE@{\GOWO}\fnm[1] 	&	$1.438$	&	$2.704$	&	$4.540$ \\
			SF-dBSE@{\GOWO}\fnm[1]	&	$1.403$	&	$2.883$	&	$4.621$	\\
		\end{tabular}
	\end{ruledtabular}
	\fnt[1]{This work.}
	\fnt[2]{Values from Ref.~\onlinecite{Casanova_2020}.}
	\fnt[3]{Values from Ref.~\onlinecite{Manohar_2008}.}
	\fnt[4]{Values from Ref.~\onlinecite{Lefrancois_2015}.}
\end{table}

\begin{table}
	\caption{
	Vertical excitation energies (with respect to the singlet $\text{X}\,{}^1B_{1g}$ ground state) of the $1\,{}^3A_{2g}$, $2\,{}^1A_{1g}$, and $1\,{}^1B_{2g}$ states of CBD at the $D_{4h}$ square-planar equilibrium geometry of the $1\,{}^3A_{2g}$ state.
	All the spin-flip calculations have been performed with an unrestricted reference and the cc-pVTZ basis set.
	\label{tab:CBD_D4h}}
	\begin{ruledtabular}
		\begin{tabular}{lrrr}
									&	\mc{3}{c}{Excitation energies (eV)}	\\
									\cline{2-4}
			Method					&	$1\,{}^3A_{2g}$	&	$2\,{}^1A_{1g}$	&	$1\,{}^1B_{2g}$	\\
			\hline
			SF-TD-B3LYP\fnm[1]		&	$-0.020$	&	$0.547$	&	$0.486$	\\
			SF-TD-BH\&HLYP\fnm[1]	&	$0.048$		&	$1.465$	&	$1.282$	\\
			\alert{SF-TD-CAM-B3LYP} &	$0.012$		&	$0.677$ &	$0.595$ \\
			\alert{SF-TD-$\omega$B97X-D} & $0.005$ & $0.673$ & $0.592$  \\
			\alert{SF-TD-LC-$\omega$PBE08} & $0.062$ & $0.663$ & $0.570$   \\
			SF-CIS\fnm[2]			&	$0.317$		&	$3.125$	&	$2.650$	\\
			EOM-SF-CCSD\fnm[3]		&	$0.369$		&	$1.824$	&	$2.143$	\\
			EOM-SF-CCSD(fT)\fnm[3]	&	$0.163$		&	$1.530$	&	$1.921$	\\
			EOM-SF-CCSD(dT)\fnm[3]	&	$0.098$		&	$1.456$	&	$1.853$	\\
			SF-ADC(2)-s\fnm[4]		&	$0.266$		&	$1.664$	&	$1.910$	\\
			SF-ADC(2)-x\fnm[4]		&	$0.217$		&	$1.123$	&	$1.799$	\\
			SF-ADC(3)\fnm[4]		&	$0.083$		&	$1.621$	&	$1.930$	\\
			SF-BSE@{\GOWO}\fnm[1] 	&	$-0.092$	&	$1.189$	&	$1.480$	\\
			SF-dBSE@{\GOWO}\fnm[1]	&	$0.012$		&	$1.507$	&	$1.841$	\\
		\end{tabular}
	\end{ruledtabular}
	\fnt[1]{This work.}
	\fnt[2]{Values from Ref.~\onlinecite{Casanova_2020}.}
	\fnt[3]{Values from Ref.~\onlinecite{Manohar_2008}.}
	\fnt[4]{Values from Ref.~\onlinecite{Lefrancois_2015}.}
\end{table}

\section{Conclusion}
\label{sec:ccl}
In this article, we have presented the extension of the BSE approach of many-body perturbation theory to the spin-flip formalism in order to access double excitations in realistic molecular systems.
The present spin-flip calculations rely on a spin-unrestricted version of the $GW$ approximation and the BSE formalism with, on top of this, a dynamical correction to the static BSE optical excitations via an unrestricted generalization of our recently developed renormalized perturbative treatment.
Taking the beryllium atom, the dissociation of the hydrogen molecule, and cyclobutadiene in two different geometries as examples, we have shown that the spin-flip BSE formalism can accurately model double excitations and seems to surpass systematically its spin-flip TD-DFT parent.
Further improvements could be obtained thanks to a better choice of the starting orbitals and their energies and we hope to investigate this in a forthcoming paper.
Techniques to alleviate the spin contamination in spin-flip BSE will also be explored in the near future.
We hope to these new encouraging results will stimulate new developments around the BSE formalism to further establish it as a valuable \textit{ab inito} alternative to TD-DFT for the study of molecular excited states.

\acknowledgements{
We would like to thank Pina Romaniello, Xavier Blase, and Denis Jacquemin for insightful discussions. 
This project has received funding from the European Research Council (ERC) under the European Union's Horizon 2020 research and innovation programme (Grant agreement No.~863481).}

\section*{Supporting information available}
Additional graphs comparing (SF-)TD-DFT and (SF-)dBSE with EOM-CCSD for the \ce{H2} molecule and raw data associated with Fig.~\ref{fig:H2}.

\bibliography{sfBSE}

\begin{thebibliography}{207}%
\makeatletter
\providecommand \@ifxundefined [1]{%
 \@ifx{#1\undefined}
}%
\providecommand \@ifnum [1]{%
 \ifnum #1\expandafter \@firstoftwo
 \else \expandafter \@secondoftwo
 \fi
}%
\providecommand \@ifx [1]{%
 \ifx #1\expandafter \@firstoftwo
 \else \expandafter \@secondoftwo
 \fi
}%
\providecommand \natexlab [1]{#1}%
\providecommand \enquote  [1]{``#1''}%
\providecommand \bibnamefont  [1]{#1}%
\providecommand \bibfnamefont [1]{#1}%
\providecommand \citenamefont [1]{#1}%
\providecommand \href@noop [0]{\@secondoftwo}%
\providecommand \href [0]{\begingroup \@sanitize@url \@href}%
\providecommand \@href[1]{\@@startlink{#1}\@@href}%
\providecommand \@@href[1]{\endgroup#1\@@endlink}%
\providecommand \@sanitize@url [0]{\catcode `\\12\catcode `\$12\catcode
  `\&12\catcode `\#12\catcode `\^12\catcode `\_12\catcode `\%12\relax}%
\providecommand \@@startlink[1]{}%
\providecommand \@@endlink[0]{}%
\providecommand \url  [0]{\begingroup\@sanitize@url \@url }%
\providecommand \@url [1]{\endgroup\@href {#1}{\urlprefix }}%
\providecommand \urlprefix  [0]{URL }%
\providecommand \Eprint [0]{\href }%
\providecommand \doibase [0]{http://dx.doi.org/}%
\providecommand \selectlanguage [0]{\@gobble}%
\providecommand \bibinfo  [0]{\@secondoftwo}%
\providecommand \bibfield  [0]{\@secondoftwo}%
\providecommand \translation [1]{[#1]}%
\providecommand \BibitemOpen [0]{}%
\providecommand \bibitemStop [0]{}%
\providecommand \bibitemNoStop [0]{.\EOS\space}%
\providecommand \EOS [0]{\spacefactor3000\relax}%
\providecommand \BibitemShut  [1]{\csname bibitem#1\endcsname}%
\let\auto@bib@innerbib\@empty
\bibitem [{\citenamefont {Roos}\ \emph {et~al.}(1996)\citenamefont {Roos},
  \citenamefont {Andersson}, \citenamefont {Fulscher}, \citenamefont
  {Malmqvist},\ and\ \citenamefont {{Serrano-Andr\'es}}}]{Roos_1996}%
  \BibitemOpen
  \bibfield  {author} {\bibinfo {author} {\bibfnamefont {B.~O.}\ \bibnamefont
  {Roos}}, \bibinfo {author} {\bibfnamefont {K.}~\bibnamefont {Andersson}},
  \bibinfo {author} {\bibfnamefont {M.~P.}\ \bibnamefont {Fulscher}}, \bibinfo
  {author} {\bibfnamefont {P.-A.}\ \bibnamefont {Malmqvist}}, \ and\ \bibinfo
  {author} {\bibfnamefont {L.}~\bibnamefont {{Serrano-Andr\'es}}},\ }\enquote
  {\bibinfo {title} {Multiconfigurational perturbation theory: Applications in
  electronic spectroscopy},}\ \ (\bibinfo  {publisher} {Wiley, New York},\
  \bibinfo {year} {1996})\ pp.\ \bibinfo {pages} {219--331}\BibitemShut
  {NoStop}%
\bibitem [{\citenamefont {Piecuch}\ \emph {et~al.}(2002)\citenamefont
  {Piecuch}, \citenamefont {Kowalski}, \citenamefont {Pimienta},\ and\
  \citenamefont {Mcguire}}]{Piecuch_2002}%
  \BibitemOpen
  \bibfield  {author} {\bibinfo {author} {\bibfnamefont {P.}~\bibnamefont
  {Piecuch}}, \bibinfo {author} {\bibfnamefont {K.}~\bibnamefont {Kowalski}},
  \bibinfo {author} {\bibfnamefont {I.~S.~O.}\ \bibnamefont {Pimienta}}, \ and\
  \bibinfo {author} {\bibfnamefont {M.~J.}\ \bibnamefont {Mcguire}},\ }\href
  {\doibase 10.1080/0144235021000053811} {\bibfield  {journal} {\bibinfo
  {journal} {Int. Rev. Phys. Chem.}\ }\textbf {\bibinfo {volume} {21}},\
  \bibinfo {pages} {527} (\bibinfo {year} {2002})}\BibitemShut {NoStop}%
\bibitem [{\citenamefont {Dreuw}\ and\ \citenamefont
  {Head-Gordon}(2005)}]{Dreuw_2005}%
  \BibitemOpen
  \bibfield  {author} {\bibinfo {author} {\bibfnamefont {A.}~\bibnamefont
  {Dreuw}}\ and\ \bibinfo {author} {\bibfnamefont {M.}~\bibnamefont
  {Head-Gordon}},\ }\href {\doibase 10.1021/cr0505627} {\bibfield  {journal}
  {\bibinfo  {journal} {Chem. Rev.}\ }\textbf {\bibinfo {volume} {105}},\
  \bibinfo {pages} {4009} (\bibinfo {year} {2005})}\BibitemShut {NoStop}%
\bibitem [{\citenamefont {Krylov}(2006)}]{Krylov_2006}%
  \BibitemOpen
  \bibfield  {author} {\bibinfo {author} {\bibfnamefont {A.~I.}\ \bibnamefont
  {Krylov}},\ }\href {\doibase 10.1021/ar0402006} {\bibfield  {journal}
  {\bibinfo  {journal} {Acc. Chem. Res.}\ }\textbf {\bibinfo {volume} {39}},\
  \bibinfo {pages} {83} (\bibinfo {year} {2006})}\BibitemShut {NoStop}%
\bibitem [{\citenamefont {Sneskov}\ and\ \citenamefont
  {Christiansen}(2012)}]{Sneskov_2012}%
  \BibitemOpen
  \bibfield  {author} {\bibinfo {author} {\bibfnamefont {K.}~\bibnamefont
  {Sneskov}}\ and\ \bibinfo {author} {\bibfnamefont {O.}~\bibnamefont
  {Christiansen}},\ }\href {\doibase https://doi.org/10.1002/wcms.99}
  {\bibfield  {journal} {\bibinfo  {journal} {WIREs Comput. Mol. Sci.}\
  }\textbf {\bibinfo {volume} {2}},\ \bibinfo {pages} {566} (\bibinfo {year}
  {2012})}\BibitemShut {NoStop}%
\bibitem [{\citenamefont {Gonz{\'a}lez}, \citenamefont {Escudero},\ and\
  \citenamefont {Serrano-Andr\`es}(2012)}]{Gonzales_2012}%
  \BibitemOpen
  \bibfield  {author} {\bibinfo {author} {\bibfnamefont {L.}~\bibnamefont
  {Gonz{\'a}lez}}, \bibinfo {author} {\bibfnamefont {D.}~\bibnamefont
  {Escudero}}, \ and\ \bibinfo {author} {\bibfnamefont {L.}~\bibnamefont
  {Serrano-Andr\`es}},\ }\href {\doibase 10.1002/cphc.201100200} {\bibfield
  {journal} {\bibinfo  {journal} {ChemPhysChem}\ }\textbf {\bibinfo {volume}
  {13}},\ \bibinfo {pages} {28} (\bibinfo {year} {2012})}\BibitemShut {NoStop}%
\bibitem [{\citenamefont {Laurent}\ and\ \citenamefont
  {Jacquemin}(2013)}]{Laurent_2013}%
  \BibitemOpen
  \bibfield  {author} {\bibinfo {author} {\bibfnamefont {A.~D.}\ \bibnamefont
  {Laurent}}\ and\ \bibinfo {author} {\bibfnamefont {D.}~\bibnamefont
  {Jacquemin}},\ }\href {\doibase 10.1002/qua.24438} {\bibfield  {journal}
  {\bibinfo  {journal} {Int. J. Quantum Chem.}\ }\textbf {\bibinfo {volume}
  {113}},\ \bibinfo {pages} {2019} (\bibinfo {year} {2013})}\BibitemShut
  {NoStop}%
\bibitem [{\citenamefont {Adamo}\ and\ \citenamefont
  {Jacquemin}(2013)}]{Adamo_2013}%
  \BibitemOpen
  \bibfield  {author} {\bibinfo {author} {\bibfnamefont {C.}~\bibnamefont
  {Adamo}}\ and\ \bibinfo {author} {\bibfnamefont {D.}~\bibnamefont
  {Jacquemin}},\ }\href {\doibase 10.1039/C2CS35394F} {\bibfield  {journal}
  {\bibinfo  {journal} {Chem. Soc. Rev.}\ }\textbf {\bibinfo {volume} {42}},\
  \bibinfo {pages} {845} (\bibinfo {year} {2013})}\BibitemShut {NoStop}%
\bibitem [{\citenamefont {Ghosh}\ \emph {et~al.}(2018)\citenamefont {Ghosh},
  \citenamefont {Verma}, \citenamefont {Cramer}, \citenamefont {Gagliardi},\
  and\ \citenamefont {Truhlar}}]{Ghosh_2018}%
  \BibitemOpen
  \bibfield  {author} {\bibinfo {author} {\bibfnamefont {S.}~\bibnamefont
  {Ghosh}}, \bibinfo {author} {\bibfnamefont {P.}~\bibnamefont {Verma}},
  \bibinfo {author} {\bibfnamefont {C.~J.}\ \bibnamefont {Cramer}}, \bibinfo
  {author} {\bibfnamefont {L.}~\bibnamefont {Gagliardi}}, \ and\ \bibinfo
  {author} {\bibfnamefont {D.~G.}\ \bibnamefont {Truhlar}},\ }\href {\doibase
  10.1021/acs.chemrev.8b00193} {\bibfield  {journal} {\bibinfo  {journal}
  {Chem. Rev.}\ }\textbf {\bibinfo {volume} {118}},\ \bibinfo {pages} {7249}
  (\bibinfo {year} {2018})}\BibitemShut {NoStop}%
\bibitem [{\citenamefont {Blase}\ \emph {et~al.}(2020)\citenamefont {Blase},
  \citenamefont {Duchemin}, \citenamefont {Jacquemin},\ and\ \citenamefont
  {Loos}}]{Blase_2020}%
  \BibitemOpen
  \bibfield  {author} {\bibinfo {author} {\bibfnamefont {X.}~\bibnamefont
  {Blase}}, \bibinfo {author} {\bibfnamefont {I.}~\bibnamefont {Duchemin}},
  \bibinfo {author} {\bibfnamefont {D.}~\bibnamefont {Jacquemin}}, \ and\
  \bibinfo {author} {\bibfnamefont {P.-F.}\ \bibnamefont {Loos}},\ }\href
  {\doibase 10.1021/acs.jpclett.0c01875} {\bibfield  {journal} {\bibinfo
  {journal} {J. Phys. Chem. Lett.}\ }\textbf {\bibinfo {volume} {11}},\
  \bibinfo {pages} {7371} (\bibinfo {year} {2020})}\BibitemShut {NoStop}%
\bibitem [{\citenamefont {Loos}, \citenamefont {Scemama},\ and\ \citenamefont
  {Jacquemin}(2020)}]{Loos_2020d}%
  \BibitemOpen
  \bibfield  {author} {\bibinfo {author} {\bibfnamefont {P.-F.}\ \bibnamefont
  {Loos}}, \bibinfo {author} {\bibfnamefont {A.}~\bibnamefont {Scemama}}, \
  and\ \bibinfo {author} {\bibfnamefont {D.}~\bibnamefont {Jacquemin}},\ }\href
  {\doibase 10.1021/acs.jpclett.0c00014} {\bibfield  {journal} {\bibinfo
  {journal} {J. Phys. Chem. Lett.}\ }\textbf {\bibinfo {volume} {11}},\
  \bibinfo {pages} {2374} (\bibinfo {year} {2020})}\BibitemShut {NoStop}%
\bibitem [{\citenamefont {Casanova}\ and\ \citenamefont
  {Krylov}(2020)}]{Casanova_2020}%
  \BibitemOpen
  \bibfield  {author} {\bibinfo {author} {\bibfnamefont {D.}~\bibnamefont
  {Casanova}}\ and\ \bibinfo {author} {\bibfnamefont {A.~I.}\ \bibnamefont
  {Krylov}},\ }\href {\doibase 10.1039/c9cp06507e} {\bibfield  {journal}
  {\bibinfo  {journal} {Phys. Chem. Chem. Phys.}\ }\textbf {\bibinfo {volume}
  {22}},\ \bibinfo {pages} {4326} (\bibinfo {year} {2020})}\BibitemShut
  {NoStop}%
\bibitem [{\citenamefont {Salpeter}\ and\ \citenamefont
  {Bethe}(1951)}]{Salpeter_1951}%
  \BibitemOpen
  \bibfield  {author} {\bibinfo {author} {\bibfnamefont {E.~E.}\ \bibnamefont
  {Salpeter}}\ and\ \bibinfo {author} {\bibfnamefont {H.~A.}\ \bibnamefont
  {Bethe}},\ }\href {\doibase 10.1103/PhysRev.84.1232} {\bibfield  {journal}
  {\bibinfo  {journal} {Phys. Rev.}\ }\textbf {\bibinfo {volume} {84}},\
  \bibinfo {pages} {1232} (\bibinfo {year} {1951})}\BibitemShut {NoStop}%
\bibitem [{\citenamefont {Sham}\ and\ \citenamefont {Rice}(1966)}]{Sham_1966}%
  \BibitemOpen
  \bibfield  {author} {\bibinfo {author} {\bibfnamefont {L.~J.}\ \bibnamefont
  {Sham}}\ and\ \bibinfo {author} {\bibfnamefont {T.~M.}\ \bibnamefont
  {Rice}},\ }\href {\doibase 10.1103/PhysRev.144.708} {\bibfield  {journal}
  {\bibinfo  {journal} {Phys. Rev.}\ }\textbf {\bibinfo {volume} {144}},\
  \bibinfo {pages} {708} (\bibinfo {year} {1966})}\BibitemShut {NoStop}%
\bibitem [{\citenamefont {Strinati}(1984)}]{Strinati_1984}%
  \BibitemOpen
  \bibfield  {author} {\bibinfo {author} {\bibfnamefont {G.}~\bibnamefont
  {Strinati}},\ }\href {\doibase 10.1103/PhysRevB.29.5718} {\bibfield
  {journal} {\bibinfo  {journal} {Phys. Rev. B}\ }\textbf {\bibinfo {volume}
  {29}},\ \bibinfo {pages} {5718} (\bibinfo {year} {1984})}\BibitemShut
  {NoStop}%
\bibitem [{\citenamefont {Delerue}, \citenamefont {Lannoo},\ and\ \citenamefont
  {Allan}(2000)}]{Delerue_2000}%
  \BibitemOpen
  \bibfield  {author} {\bibinfo {author} {\bibfnamefont {C.}~\bibnamefont
  {Delerue}}, \bibinfo {author} {\bibfnamefont {M.}~\bibnamefont {Lannoo}}, \
  and\ \bibinfo {author} {\bibfnamefont {G.}~\bibnamefont {Allan}},\ }\href
  {\doibase 10.1103/PhysRevLett.84.2457} {\bibfield  {journal} {\bibinfo
  {journal} {Phys. Rev. Lett.}\ }\textbf {\bibinfo {volume} {84}},\ \bibinfo
  {pages} {2457} (\bibinfo {year} {2000})}\BibitemShut {NoStop}%
\bibitem [{\citenamefont {Strinati}(1988)}]{Strinati_1988}%
  \BibitemOpen
  \bibfield  {author} {\bibinfo {author} {\bibfnamefont {G.}~\bibnamefont
  {Strinati}},\ }\href {\doibase 10.1007/BF02725962} {\bibfield  {journal}
  {\bibinfo  {journal} {Riv. Nuovo Cimento}\ }\textbf {\bibinfo {volume}
  {11}},\ \bibinfo {pages} {1} (\bibinfo {year} {1988})}\BibitemShut {NoStop}%
\bibitem [{\citenamefont {Albrecht}\ \emph {et~al.}(1998)\citenamefont
  {Albrecht}, \citenamefont {Reining}, \citenamefont {Del~Sole},\ and\
  \citenamefont {Onida}}]{Albrecht_1998}%
  \BibitemOpen
  \bibfield  {author} {\bibinfo {author} {\bibfnamefont {S.}~\bibnamefont
  {Albrecht}}, \bibinfo {author} {\bibfnamefont {L.}~\bibnamefont {Reining}},
  \bibinfo {author} {\bibfnamefont {R.}~\bibnamefont {Del~Sole}}, \ and\
  \bibinfo {author} {\bibfnamefont {G.}~\bibnamefont {Onida}},\ }\href
  {\doibase 10.1103/PhysRevLett.80.4510} {\bibfield  {journal} {\bibinfo
  {journal} {Phys. Rev. Lett.}\ }\textbf {\bibinfo {volume} {80}},\ \bibinfo
  {pages} {4510} (\bibinfo {year} {1998})}\BibitemShut {NoStop}%
\bibitem [{\citenamefont {Rohlfing}\ and\ \citenamefont
  {Louie}(1998)}]{Rohlfing_1998}%
  \BibitemOpen
  \bibfield  {author} {\bibinfo {author} {\bibfnamefont {M.}~\bibnamefont
  {Rohlfing}}\ and\ \bibinfo {author} {\bibfnamefont {S.~G.}\ \bibnamefont
  {Louie}},\ }\href {\doibase 10.1103/PhysRevLett.81.2312} {\bibfield
  {journal} {\bibinfo  {journal} {Phys. Rev. Lett.}\ }\textbf {\bibinfo
  {volume} {81}},\ \bibinfo {pages} {2312} (\bibinfo {year}
  {1998})}\BibitemShut {NoStop}%
\bibitem [{\citenamefont {Benedict}, \citenamefont {Shirley},\ and\
  \citenamefont {Bohn}(1998)}]{Benedict_1998}%
  \BibitemOpen
  \bibfield  {author} {\bibinfo {author} {\bibfnamefont {L.~X.}\ \bibnamefont
  {Benedict}}, \bibinfo {author} {\bibfnamefont {E.~L.}\ \bibnamefont
  {Shirley}}, \ and\ \bibinfo {author} {\bibfnamefont {R.~B.}\ \bibnamefont
  {Bohn}},\ }\href {\doibase 10.1103/PhysRevLett.80.4514} {\bibfield  {journal}
  {\bibinfo  {journal} {Phys. Rev. Lett.}\ }\textbf {\bibinfo {volume} {80}},\
  \bibinfo {pages} {4514} (\bibinfo {year} {1998})}\BibitemShut {NoStop}%
\bibitem [{\citenamefont {van~der Horst}\ \emph
  {et~al.}(1999{\natexlab{a}})\citenamefont {van~der Horst}, \citenamefont
  {Bobbert}, \citenamefont {Michels}, \citenamefont {Brocks},\ and\
  \citenamefont {Kelly}}]{vanderHorst_1999}%
  \BibitemOpen
  \bibfield  {author} {\bibinfo {author} {\bibfnamefont {J.-W.}\ \bibnamefont
  {van~der Horst}}, \bibinfo {author} {\bibfnamefont {P.~A.}\ \bibnamefont
  {Bobbert}}, \bibinfo {author} {\bibfnamefont {M.~A.~J.}\ \bibnamefont
  {Michels}}, \bibinfo {author} {\bibfnamefont {G.}~\bibnamefont {Brocks}}, \
  and\ \bibinfo {author} {\bibfnamefont {P.~J.}\ \bibnamefont {Kelly}},\ }\href
  {\doibase 10.1103/PhysRevLett.83.4413} {\bibfield  {journal} {\bibinfo
  {journal} {Phys. Rev. Lett.}\ }\textbf {\bibinfo {volume} {83}},\ \bibinfo
  {pages} {4413} (\bibinfo {year} {1999}{\natexlab{a}})}\BibitemShut {NoStop}%
\bibitem [{\citenamefont {Blase}, \citenamefont {Duchemin},\ and\ \citenamefont
  {Jacquemin}(2018)}]{Blase_2018}%
  \BibitemOpen
  \bibfield  {author} {\bibinfo {author} {\bibfnamefont {X.}~\bibnamefont
  {Blase}}, \bibinfo {author} {\bibfnamefont {I.}~\bibnamefont {Duchemin}}, \
  and\ \bibinfo {author} {\bibfnamefont {D.}~\bibnamefont {Jacquemin}},\ }\href
  {\doibase 10.1039/C7CS00049A} {\bibfield  {journal} {\bibinfo  {journal}
  {Chem. Soc. Rev.}\ }\textbf {\bibinfo {volume} {47}},\ \bibinfo {pages}
  {1022} (\bibinfo {year} {2018})}\BibitemShut {NoStop}%
\bibitem [{\citenamefont {Onida}, \citenamefont {Reining},\ and\ \citenamefont
  {and}(2002)}]{Onida_2002}%
  \BibitemOpen
  \bibfield  {author} {\bibinfo {author} {\bibfnamefont {G.}~\bibnamefont
  {Onida}}, \bibinfo {author} {\bibfnamefont {L.}~\bibnamefont {Reining}}, \
  and\ \bibinfo {author} {\bibfnamefont {A.~R.}\ \bibnamefont {and}},\ }\href
  {\doibase 10.1103/RevModPhys.74.601} {\bibfield  {journal} {\bibinfo
  {journal} {Rev. Mod. Phys.}\ }\textbf {\bibinfo {volume} {74}},\ \bibinfo
  {pages} {601} (\bibinfo {year} {2002})}\BibitemShut {NoStop}%
\bibitem [{\citenamefont {Martin}, \citenamefont {Reining},\ and\ \citenamefont
  {Ceperley}(2016{\natexlab{a}})}]{Martin_2016}%
  \BibitemOpen
  \bibfield  {author} {\bibinfo {author} {\bibfnamefont {R.~M.}\ \bibnamefont
  {Martin}}, \bibinfo {author} {\bibfnamefont {L.}~\bibnamefont {Reining}}, \
  and\ \bibinfo {author} {\bibfnamefont {D.~M.}\ \bibnamefont {Ceperley}},\
  }\href@noop {} {\emph {\bibinfo {title} {Interacting Electrons: Theory and
  Computational Approaches}}}\ (\bibinfo  {publisher} {Cambridge University
  Press},\ \bibinfo {year} {2016})\BibitemShut {NoStop}%
\bibitem [{\citenamefont {Hedin}(1965)}]{Hedin_1965}%
  \BibitemOpen
  \bibfield  {author} {\bibinfo {author} {\bibfnamefont {L.}~\bibnamefont
  {Hedin}},\ }\href {\doibase 10.1103/PhysRev.139.A796} {\bibfield  {journal}
  {\bibinfo  {journal} {Phys. Rev.}\ }\textbf {\bibinfo {volume} {139}},\
  \bibinfo {pages} {A796} (\bibinfo {year} {1965})}\BibitemShut {NoStop}%
\bibitem [{\citenamefont {Golze}, \citenamefont {Dvorak},\ and\ \citenamefont
  {Rinke}(2019)}]{Golze_2019}%
  \BibitemOpen
  \bibfield  {author} {\bibinfo {author} {\bibfnamefont {D.}~\bibnamefont
  {Golze}}, \bibinfo {author} {\bibfnamefont {M.}~\bibnamefont {Dvorak}}, \
  and\ \bibinfo {author} {\bibfnamefont {P.}~\bibnamefont {Rinke}},\ }\href
  {\doibase 10.3389/fchem.2019.00377} {\bibfield  {journal} {\bibinfo
  {journal} {Front. Chem.}\ }\textbf {\bibinfo {volume} {7}},\ \bibinfo {pages}
  {377} (\bibinfo {year} {2019})}\BibitemShut {NoStop}%
\bibitem [{\citenamefont {Rohlfing}\ and\ \citenamefont
  {Louie}(1999)}]{Rohlfing_1999a}%
  \BibitemOpen
  \bibfield  {author} {\bibinfo {author} {\bibfnamefont {M.}~\bibnamefont
  {Rohlfing}}\ and\ \bibinfo {author} {\bibfnamefont {S.~G.}\ \bibnamefont
  {Louie}},\ }\href {\doibase 10.1103/PhysRevLett.82.1959} {\bibfield
  {journal} {\bibinfo  {journal} {Phys. Rev. Lett.}\ }\textbf {\bibinfo
  {volume} {82}},\ \bibinfo {pages} {1959} (\bibinfo {year}
  {1999})}\BibitemShut {NoStop}%
\bibitem [{\citenamefont {van~der Horst}\ \emph
  {et~al.}(1999{\natexlab{b}})\citenamefont {van~der Horst}, \citenamefont
  {Bobbert}, \citenamefont {Michels}, \citenamefont {Brocks},\ and\
  \citenamefont {Kelly}}]{Horst_1999}%
  \BibitemOpen
  \bibfield  {author} {\bibinfo {author} {\bibfnamefont {J.-W.}\ \bibnamefont
  {van~der Horst}}, \bibinfo {author} {\bibfnamefont {P.~A.}\ \bibnamefont
  {Bobbert}}, \bibinfo {author} {\bibfnamefont {M.~A.~J.}\ \bibnamefont
  {Michels}}, \bibinfo {author} {\bibfnamefont {G.}~\bibnamefont {Brocks}}, \
  and\ \bibinfo {author} {\bibfnamefont {P.~J.}\ \bibnamefont {Kelly}},\ }\href
  {\doibase 10.1103/PhysRevLett.83.4413} {\bibfield  {journal} {\bibinfo
  {journal} {Phys. Rev. Lett.}\ }\textbf {\bibinfo {volume} {83}},\ \bibinfo
  {pages} {4413} (\bibinfo {year} {1999}{\natexlab{b}})}\BibitemShut {NoStop}%
\bibitem [{\citenamefont {Puschnig}\ and\ \citenamefont
  {Ambrosch-Draxl}(2002)}]{Puschnig_2002}%
  \BibitemOpen
  \bibfield  {author} {\bibinfo {author} {\bibfnamefont {P.}~\bibnamefont
  {Puschnig}}\ and\ \bibinfo {author} {\bibfnamefont {C.}~\bibnamefont
  {Ambrosch-Draxl}},\ }\href {\doibase 10.1103/PhysRevLett.89.056405}
  {\bibfield  {journal} {\bibinfo  {journal} {Phys. Rev. Lett.}\ }\textbf
  {\bibinfo {volume} {89}},\ \bibinfo {pages} {056405} (\bibinfo {year}
  {2002})}\BibitemShut {NoStop}%
\bibitem [{\citenamefont {Tiago}, \citenamefont {Northrup},\ and\ \citenamefont
  {Louie}(2003)}]{Tiago_2003}%
  \BibitemOpen
  \bibfield  {author} {\bibinfo {author} {\bibfnamefont {M.~L.}\ \bibnamefont
  {Tiago}}, \bibinfo {author} {\bibfnamefont {J.~E.}\ \bibnamefont {Northrup}},
  \ and\ \bibinfo {author} {\bibfnamefont {S.~G.}\ \bibnamefont {Louie}},\
  }\href {\doibase 10.1103/PhysRevB.67.115212} {\bibfield  {journal} {\bibinfo
  {journal} {Phys. Rev. B}\ }\textbf {\bibinfo {volume} {67}},\ \bibinfo
  {pages} {115212} (\bibinfo {year} {2003})}\BibitemShut {NoStop}%
\bibitem [{\citenamefont {Boulanger}\ \emph {et~al.}(2014)\citenamefont
  {Boulanger}, \citenamefont {Jacquemin}, \citenamefont {Duchemin},\ and\
  \citenamefont {Blase}}]{Boulanger_2014}%
  \BibitemOpen
  \bibfield  {author} {\bibinfo {author} {\bibfnamefont {P.}~\bibnamefont
  {Boulanger}}, \bibinfo {author} {\bibfnamefont {D.}~\bibnamefont
  {Jacquemin}}, \bibinfo {author} {\bibfnamefont {I.}~\bibnamefont {Duchemin}},
  \ and\ \bibinfo {author} {\bibfnamefont {X.}~\bibnamefont {Blase}},\ }\href
  {\doibase 10.1021/ct401101u} {\bibfield  {journal} {\bibinfo  {journal} {J.
  Chem. Theory Comput.}\ }\textbf {\bibinfo {volume} {10}},\ \bibinfo {pages}
  {1212} (\bibinfo {year} {2014})}\BibitemShut {NoStop}%
\bibitem [{\citenamefont {Jacquemin}, \citenamefont {Duchemin},\ and\
  \citenamefont {Blase}(2015{\natexlab{a}})}]{Jacquemin_2015a}%
  \BibitemOpen
  \bibfield  {author} {\bibinfo {author} {\bibfnamefont {D.}~\bibnamefont
  {Jacquemin}}, \bibinfo {author} {\bibfnamefont {I.}~\bibnamefont {Duchemin}},
  \ and\ \bibinfo {author} {\bibfnamefont {X.}~\bibnamefont {Blase}},\ }\href
  {\doibase 10.1021/acs.jctc.5b00304} {\bibfield  {journal} {\bibinfo
  {journal} {J. Chem. Theory Comput.}\ }\textbf {\bibinfo {volume} {11}},\
  \bibinfo {pages} {3290} (\bibinfo {year} {2015}{\natexlab{a}})}\BibitemShut
  {NoStop}%
\bibitem [{\citenamefont {Bruneval}, \citenamefont {Hamed},\ and\ \citenamefont
  {Neaton}(2015)}]{Bruneval_2015}%
  \BibitemOpen
  \bibfield  {author} {\bibinfo {author} {\bibfnamefont {F.}~\bibnamefont
  {Bruneval}}, \bibinfo {author} {\bibfnamefont {S.~M.}\ \bibnamefont {Hamed}},
  \ and\ \bibinfo {author} {\bibfnamefont {J.~B.}\ \bibnamefont {Neaton}},\
  }\href {\doibase 10.1063/1.4922489} {\bibfield  {journal} {\bibinfo
  {journal} {J. Chem. Phys.}\ }\textbf {\bibinfo {volume} {142}},\ \bibinfo
  {pages} {244101} (\bibinfo {year} {2015})}\BibitemShut {NoStop}%
\bibitem [{\citenamefont {Jacquemin}, \citenamefont {Duchemin},\ and\
  \citenamefont {Blase}(2015{\natexlab{b}})}]{Jacquemin_2015b}%
  \BibitemOpen
  \bibfield  {author} {\bibinfo {author} {\bibfnamefont {D.}~\bibnamefont
  {Jacquemin}}, \bibinfo {author} {\bibfnamefont {I.}~\bibnamefont {Duchemin}},
  \ and\ \bibinfo {author} {\bibfnamefont {X.}~\bibnamefont {Blase}},\ }\href
  {\doibase 10.1021/acs.jctc.5b00619} {\bibfield  {journal} {\bibinfo
  {journal} {J. Chem. Theory Comput.}\ }\textbf {\bibinfo {volume} {11}},\
  \bibinfo {pages} {5340} (\bibinfo {year} {2015}{\natexlab{b}})}\BibitemShut
  {NoStop}%
\bibitem [{\citenamefont {Hirose}, \citenamefont {Noguchi},\ and\ \citenamefont
  {Sugino}(2015)}]{Hirose_2015}%
  \BibitemOpen
  \bibfield  {author} {\bibinfo {author} {\bibfnamefont {D.}~\bibnamefont
  {Hirose}}, \bibinfo {author} {\bibfnamefont {Y.}~\bibnamefont {Noguchi}}, \
  and\ \bibinfo {author} {\bibfnamefont {O.}~\bibnamefont {Sugino}},\ }\href
  {\doibase 10.1103/PhysRevB.91.205111} {\bibfield  {journal} {\bibinfo
  {journal} {Phys. Rev. B}\ }\textbf {\bibinfo {volume} {91}},\ \bibinfo
  {pages} {205111} (\bibinfo {year} {2015})}\BibitemShut {NoStop}%
\bibitem [{\citenamefont {Jacquemin}, \citenamefont {Duchemin},\ and\
  \citenamefont {Blase}(2017)}]{Jacquemin_2017a}%
  \BibitemOpen
  \bibfield  {author} {\bibinfo {author} {\bibfnamefont {D.}~\bibnamefont
  {Jacquemin}}, \bibinfo {author} {\bibfnamefont {I.}~\bibnamefont {Duchemin}},
  \ and\ \bibinfo {author} {\bibfnamefont {X.}~\bibnamefont {Blase}},\ }\href
  {\doibase 10.1021/acs.jpclett.7b00381} {\bibfield  {journal} {\bibinfo
  {journal} {J. Phys. Chem. Lett.}\ }\textbf {\bibinfo {volume} {8}},\ \bibinfo
  {pages} {1524} (\bibinfo {year} {2017})}\BibitemShut {NoStop}%
\bibitem [{\citenamefont {Jacquemin}\ \emph {et~al.}(2017)\citenamefont
  {Jacquemin}, \citenamefont {Duchemin}, \citenamefont {Blondel},\ and\
  \citenamefont {Blase}}]{Jacquemin_2017b}%
  \BibitemOpen
  \bibfield  {author} {\bibinfo {author} {\bibfnamefont {D.}~\bibnamefont
  {Jacquemin}}, \bibinfo {author} {\bibfnamefont {I.}~\bibnamefont {Duchemin}},
  \bibinfo {author} {\bibfnamefont {A.}~\bibnamefont {Blondel}}, \ and\
  \bibinfo {author} {\bibfnamefont {X.}~\bibnamefont {Blase}},\ }\href
  {\doibase 10.1021/acs.jctc.6b01169} {\bibfield  {journal} {\bibinfo
  {journal} {J. Chem. Theory Comput.}\ }\textbf {\bibinfo {volume} {13}},\
  \bibinfo {pages} {767} (\bibinfo {year} {2017})}\BibitemShut {NoStop}%
\bibitem [{\citenamefont {Rangel}\ \emph {et~al.}(2017)\citenamefont {Rangel},
  \citenamefont {Hamed}, \citenamefont {Bruneval},\ and\ \citenamefont
  {Neaton}}]{Rangel_2017}%
  \BibitemOpen
  \bibfield  {author} {\bibinfo {author} {\bibfnamefont {T.}~\bibnamefont
  {Rangel}}, \bibinfo {author} {\bibfnamefont {S.~M.}\ \bibnamefont {Hamed}},
  \bibinfo {author} {\bibfnamefont {F.}~\bibnamefont {Bruneval}}, \ and\
  \bibinfo {author} {\bibfnamefont {J.~B.}\ \bibnamefont {Neaton}},\ }\href
  {\doibase 10.1063/1.4983126} {\bibfield  {journal} {\bibinfo  {journal} {J.
  Chem. Phys.}\ }\textbf {\bibinfo {volume} {146}},\ \bibinfo {pages} {194108}
  (\bibinfo {year} {2017})}\BibitemShut {NoStop}%
\bibitem [{\citenamefont {Krause}\ and\ \citenamefont
  {Klopper}(2017)}]{Krause_2017}%
  \BibitemOpen
  \bibfield  {author} {\bibinfo {author} {\bibfnamefont {K.}~\bibnamefont
  {Krause}}\ and\ \bibinfo {author} {\bibfnamefont {W.}~\bibnamefont
  {Klopper}},\ }\href {\doibase 10.1002/jcc.24688} {\bibfield  {journal}
  {\bibinfo  {journal} {J. Comput. Chem.}\ }\textbf {\bibinfo {volume} {38}},\
  \bibinfo {pages} {383} (\bibinfo {year} {2017})}\BibitemShut {NoStop}%
\bibitem [{\citenamefont {Gui}, \citenamefont {Holzer},\ and\ \citenamefont
  {Klopper}(2018)}]{Gui_2018}%
  \BibitemOpen
  \bibfield  {author} {\bibinfo {author} {\bibfnamefont {X.}~\bibnamefont
  {Gui}}, \bibinfo {author} {\bibfnamefont {C.}~\bibnamefont {Holzer}}, \ and\
  \bibinfo {author} {\bibfnamefont {W.}~\bibnamefont {Klopper}},\ }\href
  {\doibase 10.1021/acs.jctc.8b00014} {\bibfield  {journal} {\bibinfo
  {journal} {J. Chem. Theory Comput.}\ }\textbf {\bibinfo {volume} {14}},\
  \bibinfo {pages} {2127} (\bibinfo {year} {2018})}\BibitemShut {NoStop}%
\bibitem [{\citenamefont {Liu}\ \emph {et~al.}(2020)\citenamefont {Liu},
  \citenamefont {Kloppenburg}, \citenamefont {Yao}, \citenamefont {Ren},
  \citenamefont {Appel}, \citenamefont {Kanai},\ and\ \citenamefont
  {Blum}}]{Liu_2020}%
  \BibitemOpen
  \bibfield  {author} {\bibinfo {author} {\bibfnamefont {C.}~\bibnamefont
  {Liu}}, \bibinfo {author} {\bibfnamefont {J.}~\bibnamefont {Kloppenburg}},
  \bibinfo {author} {\bibfnamefont {Y.}~\bibnamefont {Yao}}, \bibinfo {author}
  {\bibfnamefont {X.}~\bibnamefont {Ren}}, \bibinfo {author} {\bibfnamefont
  {H.}~\bibnamefont {Appel}}, \bibinfo {author} {\bibfnamefont
  {Y.}~\bibnamefont {Kanai}}, \ and\ \bibinfo {author} {\bibfnamefont
  {V.}~\bibnamefont {Blum}},\ }\href {\doibase 10.1063/1.5123290} {\bibfield
  {journal} {\bibinfo  {journal} {J. Chem. Phys.}\ }\textbf {\bibinfo {volume}
  {152}},\ \bibinfo {pages} {044105} (\bibinfo {year} {2020})}\BibitemShut
  {NoStop}%
\bibitem [{\citenamefont {Holzer}\ and\ \citenamefont
  {Klopper}(2018)}]{Holzer_2018a}%
  \BibitemOpen
  \bibfield  {author} {\bibinfo {author} {\bibfnamefont {C.}~\bibnamefont
  {Holzer}}\ and\ \bibinfo {author} {\bibfnamefont {W.}~\bibnamefont
  {Klopper}},\ }\href {\doibase 10.1063/1.5051028} {\bibfield  {journal}
  {\bibinfo  {journal} {J. Chem. Phys.}\ }\textbf {\bibinfo {volume} {149}},\
  \bibinfo {pages} {101101} (\bibinfo {year} {2018})}\BibitemShut {NoStop}%
\bibitem [{\citenamefont {Holzer}\ \emph {et~al.}(2018)\citenamefont {Holzer},
  \citenamefont {Gui}, \citenamefont {Harding}, \citenamefont {Kresse},
  \citenamefont {Helgaker},\ and\ \citenamefont {Klopper}}]{Holzer_2018b}%
  \BibitemOpen
  \bibfield  {author} {\bibinfo {author} {\bibfnamefont {C.}~\bibnamefont
  {Holzer}}, \bibinfo {author} {\bibfnamefont {X.}~\bibnamefont {Gui}},
  \bibinfo {author} {\bibfnamefont {M.~E.}\ \bibnamefont {Harding}}, \bibinfo
  {author} {\bibfnamefont {G.}~\bibnamefont {Kresse}}, \bibinfo {author}
  {\bibfnamefont {T.}~\bibnamefont {Helgaker}}, \ and\ \bibinfo {author}
  {\bibfnamefont {W.}~\bibnamefont {Klopper}},\ }\href {\doibase
  10.1063/1.5047030} {\bibfield  {journal} {\bibinfo  {journal} {J. Chem.
  Phys.}\ }\textbf {\bibinfo {volume} {149}},\ \bibinfo {pages} {144106}
  (\bibinfo {year} {2018})}\BibitemShut {NoStop}%
\bibitem [{\citenamefont {Loos}\ \emph
  {et~al.}(2020{\natexlab{a}})\citenamefont {Loos}, \citenamefont {Scemama},
  \citenamefont {Duchemin}, \citenamefont {Jacquemin},\ and\ \citenamefont
  {Blase}}]{Loos_2020e}%
  \BibitemOpen
  \bibfield  {author} {\bibinfo {author} {\bibfnamefont {P.-F.}\ \bibnamefont
  {Loos}}, \bibinfo {author} {\bibfnamefont {A.}~\bibnamefont {Scemama}},
  \bibinfo {author} {\bibfnamefont {I.}~\bibnamefont {Duchemin}}, \bibinfo
  {author} {\bibfnamefont {D.}~\bibnamefont {Jacquemin}}, \ and\ \bibinfo
  {author} {\bibfnamefont {X.}~\bibnamefont {Blase}},\ }\href {\doibase
  10.1021/acs.jpclett.0c00460} {\bibfield  {journal} {\bibinfo  {journal} {J.
  Phys. Chem. Lett.}\ }\textbf {\bibinfo {volume} {11}},\ \bibinfo {pages}
  {3536} (\bibinfo {year} {2020}{\natexlab{a}})}\BibitemShut {NoStop}%
\bibitem [{\citenamefont {Bruneval}\ \emph {et~al.}(2016)\citenamefont
  {Bruneval}, \citenamefont {Rangel}, \citenamefont {Hamed}, \citenamefont
  {Shao}, \citenamefont {Yang},\ and\ \citenamefont {Neaton}}]{Bruneval_2016}%
  \BibitemOpen
  \bibfield  {author} {\bibinfo {author} {\bibfnamefont {F.}~\bibnamefont
  {Bruneval}}, \bibinfo {author} {\bibfnamefont {T.}~\bibnamefont {Rangel}},
  \bibinfo {author} {\bibfnamefont {S.~M.}\ \bibnamefont {Hamed}}, \bibinfo
  {author} {\bibfnamefont {M.}~\bibnamefont {Shao}}, \bibinfo {author}
  {\bibfnamefont {C.}~\bibnamefont {Yang}}, \ and\ \bibinfo {author}
  {\bibfnamefont {J.~B.}\ \bibnamefont {Neaton}},\ }\href {\doibase
  10.1016/j.cpc.2016.06.019} {\bibfield  {journal} {\bibinfo  {journal}
  {Comput. Phys. Commun.}\ }\textbf {\bibinfo {volume} {208}},\ \bibinfo
  {pages} {149} (\bibinfo {year} {2016})}\BibitemShut {NoStop}%
\bibitem [{\citenamefont {Runge}\ and\ \citenamefont
  {Gross}(1984)}]{Runge_1984}%
  \BibitemOpen
  \bibfield  {author} {\bibinfo {author} {\bibfnamefont {E.}~\bibnamefont
  {Runge}}\ and\ \bibinfo {author} {\bibfnamefont {E.~K.~U.}\ \bibnamefont
  {Gross}},\ }\href {\doibase 10.1103/PhysRevLett.52.997} {\bibfield  {journal}
  {\bibinfo  {journal} {Phys. Rev. Lett.}\ }\textbf {\bibinfo {volume} {52}},\
  \bibinfo {pages} {997} (\bibinfo {year} {1984})}\BibitemShut {NoStop}%
\bibitem [{\citenamefont {Casida}(1995)}]{Casida_1995}%
  \BibitemOpen
  \bibfield  {author} {\bibinfo {author} {\bibfnamefont {M.~E.}\ \bibnamefont
  {Casida}},\ }\enquote {\bibinfo {title} {Time-dependent density functional
  response theory for molecules},}\ \ (\bibinfo  {publisher} {World Scientific,
  Singapore},\ \bibinfo {year} {1995})\ pp.\ \bibinfo {pages}
  {155--192}\BibitemShut {NoStop}%
\bibitem [{\citenamefont {Petersilka}, \citenamefont {Gossmann},\ and\
  \citenamefont {Gross}(1996)}]{Petersilka_1996}%
  \BibitemOpen
  \bibfield  {author} {\bibinfo {author} {\bibfnamefont {M.}~\bibnamefont
  {Petersilka}}, \bibinfo {author} {\bibfnamefont {U.~J.}\ \bibnamefont
  {Gossmann}}, \ and\ \bibinfo {author} {\bibfnamefont {E.~K.~U.}\ \bibnamefont
  {Gross}},\ }\href {\doibase 10.1103/PhysRevLett.76.1212} {\bibfield
  {journal} {\bibinfo  {journal} {Phys. Rev. Lett.}\ }\textbf {\bibinfo
  {volume} {76}},\ \bibinfo {pages} {1212} (\bibinfo {year}
  {1996})}\BibitemShut {NoStop}%
\bibitem [{\citenamefont {Ullrich}(2012)}]{UlrichBook}%
  \BibitemOpen
  \bibfield  {author} {\bibinfo {author} {\bibfnamefont {C.}~\bibnamefont
  {Ullrich}},\ }\href@noop {} {\emph {\bibinfo {title} {Time-Dependent
  Density-Functional Theory: Concepts and Applications}}},\ Oxford Graduate
  Texts\ (\bibinfo  {publisher} {Oxford University Press},\ \bibinfo {address}
  {New York},\ \bibinfo {year} {2012})\BibitemShut {NoStop}%
\bibitem [{\citenamefont {Maitra}, \citenamefont {F.~Zhang},\ and\
  \citenamefont {Burke}(2004)}]{Maitra_2004}%
  \BibitemOpen
  \bibfield  {author} {\bibinfo {author} {\bibfnamefont {N.~T.}\ \bibnamefont
  {Maitra}}, \bibinfo {author} {\bibfnamefont {R.~J.~C.}\ \bibnamefont
  {F.~Zhang}}, \ and\ \bibinfo {author} {\bibfnamefont {K.}~\bibnamefont
  {Burke}},\ }\href {\doibase 10.1063/1.1651060} {\bibfield  {journal}
  {\bibinfo  {journal} {J. Chem. Phys.}\ }\textbf {\bibinfo {volume} {120}},\
  \bibinfo {pages} {5932} (\bibinfo {year} {2004})}\BibitemShut {NoStop}%
\bibitem [{\citenamefont {Cave}\ \emph {et~al.}(2004)\citenamefont {Cave},
  \citenamefont {Zhang}, \citenamefont {Maitra},\ and\ \citenamefont
  {Burke}}]{Cave_2004}%
  \BibitemOpen
  \bibfield  {author} {\bibinfo {author} {\bibfnamefont {R.~J.}\ \bibnamefont
  {Cave}}, \bibinfo {author} {\bibfnamefont {F.}~\bibnamefont {Zhang}},
  \bibinfo {author} {\bibfnamefont {N.~T.}\ \bibnamefont {Maitra}}, \ and\
  \bibinfo {author} {\bibfnamefont {K.}~\bibnamefont {Burke}},\ }\href
  {\doibase 10.1016/j.cplett.2004.03.051} {\bibfield  {journal} {\bibinfo
  {journal} {Chem. Phys. Lett.}\ }\textbf {\bibinfo {volume} {389}},\ \bibinfo
  {pages} {39} (\bibinfo {year} {2004})}\BibitemShut {NoStop}%
\bibitem [{\citenamefont {Saha}, \citenamefont {Ehara},\ and\ \citenamefont
  {Nakatsuji}(2006)}]{Saha_2006}%
  \BibitemOpen
  \bibfield  {author} {\bibinfo {author} {\bibfnamefont {B.}~\bibnamefont
  {Saha}}, \bibinfo {author} {\bibfnamefont {M.}~\bibnamefont {Ehara}}, \ and\
  \bibinfo {author} {\bibfnamefont {H.}~\bibnamefont {Nakatsuji}},\ }\href
  {\doibase 10.1063/1.2200344} {\bibfield  {journal} {\bibinfo  {journal} {J.
  Chem. Phys.}\ }\textbf {\bibinfo {volume} {125}},\ \bibinfo {pages} {014316}
  (\bibinfo {year} {2006})}\BibitemShut {NoStop}%
\bibitem [{\citenamefont {Watson}\ and\ \citenamefont
  {Chan}(2012)}]{Watson_2012}%
  \BibitemOpen
  \bibfield  {author} {\bibinfo {author} {\bibfnamefont {M.~A.}\ \bibnamefont
  {Watson}}\ and\ \bibinfo {author} {\bibfnamefont {G.~K.-L.}\ \bibnamefont
  {Chan}},\ }\href {\doibase 10.1021/ct300591z} {\bibfield  {journal} {\bibinfo
   {journal} {J. Chem. Theory Comput.}\ }\textbf {\bibinfo {volume} {8}},\
  \bibinfo {pages} {4013} (\bibinfo {year} {2012})}\BibitemShut {NoStop}%
\bibitem [{\citenamefont {Shu}\ and\ \citenamefont {Truhlar}(2017)}]{Shu_2017}%
  \BibitemOpen
  \bibfield  {author} {\bibinfo {author} {\bibfnamefont {Y.}~\bibnamefont
  {Shu}}\ and\ \bibinfo {author} {\bibfnamefont {D.~G.}\ \bibnamefont
  {Truhlar}},\ }\href {\doibase 10.1021/jacs.7b06283} {\bibfield  {journal}
  {\bibinfo  {journal} {J. Am. Chem. Soc.}\ }\textbf {\bibinfo {volume}
  {139}},\ \bibinfo {pages} {13770} (\bibinfo {year} {2017})}\BibitemShut
  {NoStop}%
\bibitem [{\citenamefont {Barca}, \citenamefont {Gilbert},\ and\ \citenamefont
  {Gill}(2018{\natexlab{a}})}]{Barca_2018a}%
  \BibitemOpen
  \bibfield  {author} {\bibinfo {author} {\bibfnamefont {G.~M.~J.}\
  \bibnamefont {Barca}}, \bibinfo {author} {\bibfnamefont {A.~T.~B.}\
  \bibnamefont {Gilbert}}, \ and\ \bibinfo {author} {\bibfnamefont {P.~M.~W.}\
  \bibnamefont {Gill}},\ }\href {\doibase 10.1021/acs.jctc.7b00994} {\bibfield
  {journal} {\bibinfo  {journal} {J. Chem. Theory. Comput.}\ }\textbf {\bibinfo
  {volume} {14}},\ \bibinfo {pages} {1501} (\bibinfo {year}
  {2018}{\natexlab{a}})}\BibitemShut {NoStop}%
\bibitem [{\citenamefont {Barca}, \citenamefont {Gilbert},\ and\ \citenamefont
  {Gill}(2018{\natexlab{b}})}]{Barca_2018b}%
  \BibitemOpen
  \bibfield  {author} {\bibinfo {author} {\bibfnamefont {G.~M.~J.}\
  \bibnamefont {Barca}}, \bibinfo {author} {\bibfnamefont {A.~T.~B.}\
  \bibnamefont {Gilbert}}, \ and\ \bibinfo {author} {\bibfnamefont {P.~M.~W.}\
  \bibnamefont {Gill}},\ }\href {\doibase 10.1021/acs.jctc.7b00963} {\bibfield
  {journal} {\bibinfo  {journal} {J. Chem. Theory. Comput.}\ }\textbf {\bibinfo
  {volume} {14}},\ \bibinfo {pages} {9} (\bibinfo {year}
  {2018}{\natexlab{b}})}\BibitemShut {NoStop}%
\bibitem [{\citenamefont {Loos}\ \emph {et~al.}(2019)\citenamefont {Loos},
  \citenamefont {Boggio-Pasqua}, \citenamefont {Scemama}, \citenamefont
  {Caffarel},\ and\ \citenamefont {Jacquemin}}]{Loos_2019}%
  \BibitemOpen
  \bibfield  {author} {\bibinfo {author} {\bibfnamefont {P.-F.}\ \bibnamefont
  {Loos}}, \bibinfo {author} {\bibfnamefont {M.}~\bibnamefont {Boggio-Pasqua}},
  \bibinfo {author} {\bibfnamefont {A.}~\bibnamefont {Scemama}}, \bibinfo
  {author} {\bibfnamefont {M.}~\bibnamefont {Caffarel}}, \ and\ \bibinfo
  {author} {\bibfnamefont {D.}~\bibnamefont {Jacquemin}},\ }\href {\doibase
  10.1021/acs.jctc.8b01205} {\bibfield  {journal} {\bibinfo  {journal} {J.
  Chem. Theory Comput.}\ }\textbf {\bibinfo {volume} {15}},\ \bibinfo {pages}
  {1939} (\bibinfo {year} {2019})}\BibitemShut {NoStop}%
\bibitem [{\citenamefont {Casida}(2005)}]{Casida_2005}%
  \BibitemOpen
  \bibfield  {author} {\bibinfo {author} {\bibfnamefont {M.~E.}\ \bibnamefont
  {Casida}},\ }\href {\doibase 10.1063/1.1836757} {\bibfield  {journal}
  {\bibinfo  {journal} {J. Chem. Phys.}\ }\textbf {\bibinfo {volume} {122}},\
  \bibinfo {pages} {054111} (\bibinfo {year} {2005})}\BibitemShut {NoStop}%
\bibitem [{\citenamefont {{Huix-Rotllant}}\ \emph {et~al.}(2011)\citenamefont
  {{Huix-Rotllant}}, \citenamefont {Ipatov}, \citenamefont {Rubio},\ and\
  \citenamefont {Casida}}]{Huix-Rotllant_2011}%
  \BibitemOpen
  \bibfield  {author} {\bibinfo {author} {\bibfnamefont {M.}~\bibnamefont
  {{Huix-Rotllant}}}, \bibinfo {author} {\bibfnamefont {A.}~\bibnamefont
  {Ipatov}}, \bibinfo {author} {\bibfnamefont {A.}~\bibnamefont {Rubio}}, \
  and\ \bibinfo {author} {\bibfnamefont {M.~E.}\ \bibnamefont {Casida}},\
  }\href {\doibase 10.1016/j.chemphys.2011.03.019} {\bibfield  {journal}
  {\bibinfo  {journal} {Chem. Phys.}\ }\textbf {\bibinfo {volume} {391}},\
  \bibinfo {pages} {120} (\bibinfo {year} {2011})}\BibitemShut {NoStop}%
\bibitem [{\citenamefont {Loos}\ \emph
  {et~al.}(2020{\natexlab{b}})\citenamefont {Loos}, \citenamefont {Scemama},
  \citenamefont {Boggio-Pasqua},\ and\ \citenamefont {Jacquemin}}]{Loos_2020f}%
  \BibitemOpen
  \bibfield  {author} {\bibinfo {author} {\bibfnamefont {P.-F.}\ \bibnamefont
  {Loos}}, \bibinfo {author} {\bibfnamefont {A.}~\bibnamefont {Scemama}},
  \bibinfo {author} {\bibfnamefont {M.}~\bibnamefont {Boggio-Pasqua}}, \ and\
  \bibinfo {author} {\bibfnamefont {D.}~\bibnamefont {Jacquemin}},\ }\href
  {\doibase 10.1021/acs.jctc.0c00227} {\bibfield  {journal} {\bibinfo
  {journal} {J. Chem. Theory Comput.}\ }\textbf {\bibinfo {volume} {16}},\
  \bibinfo {pages} {3720} (\bibinfo {year} {2020}{\natexlab{b}})}\BibitemShut
  {NoStop}%
\bibitem [{\citenamefont {Levine}\ \emph {et~al.}(2006)\citenamefont {Levine},
  \citenamefont {Ko}, \citenamefont {Quenneville},\ and\ \citenamefont
  {Martinez}}]{Levine_2006}%
  \BibitemOpen
  \bibfield  {author} {\bibinfo {author} {\bibfnamefont {B.~G.}\ \bibnamefont
  {Levine}}, \bibinfo {author} {\bibfnamefont {C.}~\bibnamefont {Ko}}, \bibinfo
  {author} {\bibfnamefont {J.}~\bibnamefont {Quenneville}}, \ and\ \bibinfo
  {author} {\bibfnamefont {T.~J.}\ \bibnamefont {Martinez}},\ }\href {\doibase
  10.1080/00268970500417762} {\bibfield  {journal} {\bibinfo  {journal} {Mol.
  Phys.}\ }\textbf {\bibinfo {volume} {104}},\ \bibinfo {pages} {1039}
  (\bibinfo {year} {2006})}\BibitemShut {NoStop}%
\bibitem [{\citenamefont {Tozer}\ and\ \citenamefont
  {Handy}(2000)}]{Tozer_2000}%
  \BibitemOpen
  \bibfield  {author} {\bibinfo {author} {\bibfnamefont {D.~J.}\ \bibnamefont
  {Tozer}}\ and\ \bibinfo {author} {\bibfnamefont {N.~C.}\ \bibnamefont
  {Handy}},\ }\href {\doibase 10.1039/a910321j} {\bibfield  {journal} {\bibinfo
   {journal} {Phys. Chem. Chem. Phys.}\ }\textbf {\bibinfo {volume} {2}},\
  \bibinfo {pages} {2117} (\bibinfo {year} {2000})}\BibitemShut {NoStop}%
\bibitem [{\citenamefont {Elliott}\ \emph {et~al.}(2011)\citenamefont
  {Elliott}, \citenamefont {Goldson}, \citenamefont {Canahui},\ and\
  \citenamefont {Maitra}}]{Elliott_2011}%
  \BibitemOpen
  \bibfield  {author} {\bibinfo {author} {\bibfnamefont {P.}~\bibnamefont
  {Elliott}}, \bibinfo {author} {\bibfnamefont {S.}~\bibnamefont {Goldson}},
  \bibinfo {author} {\bibfnamefont {C.}~\bibnamefont {Canahui}}, \ and\
  \bibinfo {author} {\bibfnamefont {N.~T.}\ \bibnamefont {Maitra}},\ }\href
  {\doibase 10.1016/j.chemphys.2011.03.020} {\bibfield  {journal} {\bibinfo
  {journal} {Chem. Phys.}\ }\textbf {\bibinfo {volume} {391}},\ \bibinfo
  {pages} {110} (\bibinfo {year} {2011})}\BibitemShut {NoStop}%
\bibitem [{\citenamefont {Maitra}(2012)}]{Maitra_2012}%
  \BibitemOpen
  \bibfield  {author} {\bibinfo {author} {\bibfnamefont {N.~T.}\ \bibnamefont
  {Maitra}},\ }\enquote {\bibinfo {title} {Memory: History , initial-state
  dependence , and double-excitations},}\ in\ \href {\doibase
  10.1007/978-3-642-23518-4_8} {\emph {\bibinfo {booktitle} {Fundamentals of
  Time-Dependent Density Functional Theory}}},\ Vol.\ \bibinfo {volume} {837},\
  \bibinfo {editor} {edited by\ \bibinfo {editor} {\bibfnamefont {M.~A.}\
  \bibnamefont {Marques}}, \bibinfo {editor} {\bibfnamefont {N.~T.}\
  \bibnamefont {Maitra}}, \bibinfo {editor} {\bibfnamefont {F.~M.}\
  \bibnamefont {Nogueira}}, \bibinfo {editor} {\bibfnamefont {E.}~\bibnamefont
  {Gross}}, \ and\ \bibinfo {editor} {\bibfnamefont {A.}~\bibnamefont
  {Rubio}}}\ (\bibinfo  {publisher} {Springer Berlin Heidelberg},\ \bibinfo
  {address} {Berlin, Heidelberg},\ \bibinfo {year} {2012})\ pp.\ \bibinfo
  {pages} {167--184}\BibitemShut {NoStop}%
\bibitem [{\citenamefont {Maitra}(2016)}]{Maitra_2016}%
  \BibitemOpen
  \bibfield  {author} {\bibinfo {author} {\bibfnamefont {N.~T.}\ \bibnamefont
  {Maitra}},\ }\href {\doibase 10.1063/1.4953039} {\bibfield  {journal}
  {\bibinfo  {journal} {J. Chem. Phys.}\ }\textbf {\bibinfo {volume} {144}},\
  \bibinfo {pages} {220901} (\bibinfo {year} {2016})}\BibitemShut {NoStop}%
\bibitem [{\citenamefont {Martin}, \citenamefont {Reining},\ and\ \citenamefont
  {Ceperley}(2016{\natexlab{b}})}]{ReiningBook}%
  \BibitemOpen
  \bibfield  {author} {\bibinfo {author} {\bibfnamefont {R.}~\bibnamefont
  {Martin}}, \bibinfo {author} {\bibfnamefont {L.}~\bibnamefont {Reining}}, \
  and\ \bibinfo {author} {\bibfnamefont {D.}~\bibnamefont {Ceperley}},\
  }\href@noop {} {\emph {\bibinfo {title} {Interacting Electrons: Theory and
  Computational Approaches}}}\ (\bibinfo  {publisher} {Cambridge University
  Press},\ \bibinfo {year} {2016})\BibitemShut {NoStop}%
\bibitem [{\citenamefont {Romaniello}\ \emph {et~al.}(2009)\citenamefont
  {Romaniello}, \citenamefont {Sangalli}, \citenamefont {Berger}, \citenamefont
  {Sottile}, \citenamefont {Molinari}, \citenamefont {Reining},\ and\
  \citenamefont {Onida}}]{Romaniello_2009b}%
  \BibitemOpen
  \bibfield  {author} {\bibinfo {author} {\bibfnamefont {P.}~\bibnamefont
  {Romaniello}}, \bibinfo {author} {\bibfnamefont {D.}~\bibnamefont
  {Sangalli}}, \bibinfo {author} {\bibfnamefont {J.~A.}\ \bibnamefont
  {Berger}}, \bibinfo {author} {\bibfnamefont {F.}~\bibnamefont {Sottile}},
  \bibinfo {author} {\bibfnamefont {L.~G.}\ \bibnamefont {Molinari}}, \bibinfo
  {author} {\bibfnamefont {L.}~\bibnamefont {Reining}}, \ and\ \bibinfo
  {author} {\bibfnamefont {G.}~\bibnamefont {Onida}},\ }\href {\doibase
  10.1063/1.3065669} {\bibfield  {journal} {\bibinfo  {journal} {J. Chem.
  Phys.}\ }\textbf {\bibinfo {volume} {130}},\ \bibinfo {pages} {044108}
  (\bibinfo {year} {2009})}\BibitemShut {NoStop}%
\bibitem [{\citenamefont {Sangalli}\ \emph {et~al.}(2011)\citenamefont
  {Sangalli}, \citenamefont {Romaniello}, \citenamefont {Onida},\ and\
  \citenamefont {Marini}}]{Sangalli_2011}%
  \BibitemOpen
  \bibfield  {author} {\bibinfo {author} {\bibfnamefont {D.}~\bibnamefont
  {Sangalli}}, \bibinfo {author} {\bibfnamefont {P.}~\bibnamefont
  {Romaniello}}, \bibinfo {author} {\bibfnamefont {G.}~\bibnamefont {Onida}}, \
  and\ \bibinfo {author} {\bibfnamefont {A.}~\bibnamefont {Marini}},\ }\href
  {\doibase 10.1063/1.3518705} {\bibfield  {journal} {\bibinfo  {journal} {J.
  Chem. Phys.}\ }\textbf {\bibinfo {volume} {134}},\ \bibinfo {pages} {034115}
  (\bibinfo {year} {2011})}\BibitemShut {NoStop}%
\bibitem [{\citenamefont {Loos}\ and\ \citenamefont
  {Blase}(2020)}]{Loos_2020h}%
  \BibitemOpen
  \bibfield  {author} {\bibinfo {author} {\bibfnamefont {P.-F.}\ \bibnamefont
  {Loos}}\ and\ \bibinfo {author} {\bibfnamefont {X.}~\bibnamefont {Blase}},\
  }\href {\doibase 10.1063/5.0023168} {\bibfield  {journal} {\bibinfo
  {journal} {J. Chem. Phys.}\ }\textbf {\bibinfo {volume} {153}},\ \bibinfo
  {pages} {114120} (\bibinfo {year} {2020})}\BibitemShut {NoStop}%
\bibitem [{\citenamefont {Authier}\ and\ \citenamefont
  {Loos}(2020)}]{Authier_2020}%
  \BibitemOpen
  \bibfield  {author} {\bibinfo {author} {\bibfnamefont {J.}~\bibnamefont
  {Authier}}\ and\ \bibinfo {author} {\bibfnamefont {P.-F.}\ \bibnamefont
  {Loos}},\ }\href {\doibase 10.1063/5.0028040} {\bibfield  {journal} {\bibinfo
   {journal} {J. Chem. Phys.}\ }\textbf {\bibinfo {volume} {153}},\ \bibinfo
  {pages} {184105} (\bibinfo {year} {2020})}\BibitemShut {NoStop}%
\bibitem [{\citenamefont {Loos}\ \emph {et~al.}(2018)\citenamefont {Loos},
  \citenamefont {Scemama}, \citenamefont {Blondel}, \citenamefont {Garniron},
  \citenamefont {Caffarel},\ and\ \citenamefont {Jacquemin}}]{Loos_2018a}%
  \BibitemOpen
  \bibfield  {author} {\bibinfo {author} {\bibfnamefont {P.~F.}\ \bibnamefont
  {Loos}}, \bibinfo {author} {\bibfnamefont {A.}~\bibnamefont {Scemama}},
  \bibinfo {author} {\bibfnamefont {A.}~\bibnamefont {Blondel}}, \bibinfo
  {author} {\bibfnamefont {Y.}~\bibnamefont {Garniron}}, \bibinfo {author}
  {\bibfnamefont {M.}~\bibnamefont {Caffarel}}, \ and\ \bibinfo {author}
  {\bibfnamefont {D.}~\bibnamefont {Jacquemin}},\ }\href {\doibase
  10.1021/acs.jctc.8b00406} {\bibfield  {journal} {\bibinfo  {journal} {J.
  Chem. Theory Comput.}\ }\textbf {\bibinfo {volume} {14}},\ \bibinfo {pages}
  {4360} (\bibinfo {year} {2018})}\BibitemShut {NoStop}%
\bibitem [{\citenamefont {Loos}\ \emph
  {et~al.}(2020{\natexlab{c}})\citenamefont {Loos}, \citenamefont {Lipparini},
  \citenamefont {Boggio-Pasqua}, \citenamefont {Scemama},\ and\ \citenamefont
  {Jacquemin}}]{Loos_2020c}%
  \BibitemOpen
  \bibfield  {author} {\bibinfo {author} {\bibfnamefont {P.~F.}\ \bibnamefont
  {Loos}}, \bibinfo {author} {\bibfnamefont {F.}~\bibnamefont {Lipparini}},
  \bibinfo {author} {\bibfnamefont {M.}~\bibnamefont {Boggio-Pasqua}}, \bibinfo
  {author} {\bibfnamefont {A.}~\bibnamefont {Scemama}}, \ and\ \bibinfo
  {author} {\bibfnamefont {D.}~\bibnamefont {Jacquemin}},\ }\href {\doibase
  10.1021/acs.jctc.9b01216} {\bibfield  {journal} {\bibinfo  {journal} {J.
  Chem. Theory Comput.}\ }\textbf {\bibinfo {volume} {16}},\ \bibinfo {pages}
  {1711} (\bibinfo {year} {2020}{\natexlab{c}})}\BibitemShut {NoStop}%
\bibitem [{\citenamefont {V{\'e}ril}\ \emph {et~al.}(2020)\citenamefont
  {V{\'e}ril}, \citenamefont {Scemama}, \citenamefont {Caffarel}, \citenamefont
  {Lipparini}, \citenamefont {Boggio-Pasqua}, \citenamefont {Jacquemin},\ and\
  \citenamefont {Loos}}]{Veril_2020}%
  \BibitemOpen
  \bibfield  {author} {\bibinfo {author} {\bibfnamefont {M.}~\bibnamefont
  {V{\'e}ril}}, \bibinfo {author} {\bibfnamefont {A.}~\bibnamefont {Scemama}},
  \bibinfo {author} {\bibfnamefont {M.}~\bibnamefont {Caffarel}}, \bibinfo
  {author} {\bibfnamefont {F.}~\bibnamefont {Lipparini}}, \bibinfo {author}
  {\bibfnamefont {M.}~\bibnamefont {Boggio-Pasqua}}, \bibinfo {author}
  {\bibfnamefont {D.}~\bibnamefont {Jacquemin}}, \ and\ \bibinfo {author}
  {\bibfnamefont {P.-F.}\ \bibnamefont {Loos}},\ }\href@noop {} {\enquote
  {\bibinfo {title} {Questdb: a database of highly-accurate excitation energies
  for the electronic structure community},}\ } (\bibinfo {year} {2020}),\
  \Eprint {http://arxiv.org/abs/2011.14675} {arXiv:2011.14675
  [physics.chem-ph]} \BibitemShut {NoStop}%
\bibitem [{\citenamefont {Christiansen}, \citenamefont {Koch},\ and\
  \citenamefont {J{\o}rgensen}(1995{\natexlab{a}})}]{Christiansen_1995b}%
  \BibitemOpen
  \bibfield  {author} {\bibinfo {author} {\bibfnamefont {O.}~\bibnamefont
  {Christiansen}}, \bibinfo {author} {\bibfnamefont {H.}~\bibnamefont {Koch}},
  \ and\ \bibinfo {author} {\bibfnamefont {P.}~\bibnamefont {J{\o}rgensen}},\
  }\href {\doibase http://dx.doi.org/10.1063/1.470315} {\bibfield  {journal}
  {\bibinfo  {journal} {J. Chem. Phys.}\ }\textbf {\bibinfo {volume} {103}},\
  \bibinfo {pages} {7429} (\bibinfo {year} {1995}{\natexlab{a}})}\BibitemShut
  {NoStop}%
\bibitem [{\citenamefont {Koch}\ \emph {et~al.}(1997)\citenamefont {Koch},
  \citenamefont {Christiansen}, \citenamefont {Jorgensen}, \citenamefont
  {Sanchez~de Mer{\'a}s},\ and\ \citenamefont {Helgaker}}]{Koch_1997}%
  \BibitemOpen
  \bibfield  {author} {\bibinfo {author} {\bibfnamefont {H.}~\bibnamefont
  {Koch}}, \bibinfo {author} {\bibfnamefont {O.}~\bibnamefont {Christiansen}},
  \bibinfo {author} {\bibfnamefont {P.}~\bibnamefont {Jorgensen}}, \bibinfo
  {author} {\bibfnamefont {A.~M.}\ \bibnamefont {Sanchez~de Mer{\'a}s}}, \ and\
  \bibinfo {author} {\bibfnamefont {T.}~\bibnamefont {Helgaker}},\ }\href
  {\doibase http://dx.doi.org/10.1063/1.473322} {\bibfield  {journal} {\bibinfo
   {journal} {J. Chem. Phys.}\ }\textbf {\bibinfo {volume} {106}},\ \bibinfo
  {pages} {1808} (\bibinfo {year} {1997})}\BibitemShut {NoStop}%
\bibitem [{\citenamefont {Kucharski}\ and\ \citenamefont
  {Bartlett}(1991)}]{Kucharski_1991}%
  \BibitemOpen
  \bibfield  {author} {\bibinfo {author} {\bibfnamefont {S.~A.}\ \bibnamefont
  {Kucharski}}\ and\ \bibinfo {author} {\bibfnamefont {R.~J.}\ \bibnamefont
  {Bartlett}},\ }\href {\doibase 10.1007/BF01117419} {\bibfield  {journal}
  {\bibinfo  {journal} {Theor. Chim. Acta}\ }\textbf {\bibinfo {volume} {80}},\
  \bibinfo {pages} {387} (\bibinfo {year} {1991})}\BibitemShut {NoStop}%
\bibitem [{\citenamefont {K{\'a}llay}\ and\ \citenamefont
  {Gauss}(2004)}]{Kallay_2004}%
  \BibitemOpen
  \bibfield  {author} {\bibinfo {author} {\bibfnamefont {M.}~\bibnamefont
  {K{\'a}llay}}\ and\ \bibinfo {author} {\bibfnamefont {J.}~\bibnamefont
  {Gauss}},\ }\href {\doibase http://dx.doi.org/10.1063/1.1805494} {\bibfield
  {journal} {\bibinfo  {journal} {J. Chem. Phys.}\ }\textbf {\bibinfo {volume}
  {121}},\ \bibinfo {pages} {9257} (\bibinfo {year} {2004})}\BibitemShut
  {NoStop}%
\bibitem [{\citenamefont {Hirata}, \citenamefont {Nooijen},\ and\ \citenamefont
  {Bartlett}(2000)}]{Hirata_2000}%
  \BibitemOpen
  \bibfield  {author} {\bibinfo {author} {\bibfnamefont {S.}~\bibnamefont
  {Hirata}}, \bibinfo {author} {\bibfnamefont {M.}~\bibnamefont {Nooijen}}, \
  and\ \bibinfo {author} {\bibfnamefont {R.~J.}\ \bibnamefont {Bartlett}},\
  }\href {\doibase 10.1016/S0009-2614(00)00772-7} {\bibfield  {journal}
  {\bibinfo  {journal} {Chem. Phys. Lett.}\ }\textbf {\bibinfo {volume}
  {326}},\ \bibinfo {pages} {255} (\bibinfo {year} {2000})}\BibitemShut
  {NoStop}%
\bibitem [{\citenamefont {Hirata}(2004)}]{Hirata_2004}%
  \BibitemOpen
  \bibfield  {author} {\bibinfo {author} {\bibfnamefont {S.}~\bibnamefont
  {Hirata}},\ }\href@noop {} {\bibfield  {journal} {\bibinfo  {journal} {J.
  Chem. Phys.}\ }\textbf {\bibinfo {volume} {121}},\ \bibinfo {pages} {51}
  (\bibinfo {year} {2004})}\BibitemShut {NoStop}%
\bibitem [{\citenamefont {Krylov}(2001{\natexlab{a}})}]{Krylov_2001a}%
  \BibitemOpen
  \bibfield  {author} {\bibinfo {author} {\bibfnamefont {A.~I.}\ \bibnamefont
  {Krylov}},\ }\href {\doibase https://doi.org/10.1016/S0009-2614(01)00287-1}
  {\bibfield  {journal} {\bibinfo  {journal} {Chem. Phys. Lett.}\ }\textbf
  {\bibinfo {volume} {338}},\ \bibinfo {pages} {375 } (\bibinfo {year}
  {2001}{\natexlab{a}})}\BibitemShut {NoStop}%
\bibitem [{\citenamefont {Krylov}(2001{\natexlab{b}})}]{Krylov_2001b}%
  \BibitemOpen
  \bibfield  {author} {\bibinfo {author} {\bibfnamefont {A.~I.}\ \bibnamefont
  {Krylov}},\ }\href {\doibase 10.1016/S0009-2614(01)01316-1} {\bibfield
  {journal} {\bibinfo  {journal} {Chem. Phys. Lett.}\ }\textbf {\bibinfo
  {volume} {350}},\ \bibinfo {pages} {522} (\bibinfo {year}
  {2001}{\natexlab{b}})}\BibitemShut {NoStop}%
\bibitem [{\citenamefont {Krylov}\ and\ \citenamefont
  {Sherrill}(2002)}]{Krylov_2002}%
  \BibitemOpen
  \bibfield  {author} {\bibinfo {author} {\bibfnamefont {A.~I.}\ \bibnamefont
  {Krylov}}\ and\ \bibinfo {author} {\bibfnamefont {C.~D.}\ \bibnamefont
  {Sherrill}},\ }\href {\doibase 10.1063/1.1445116} {\bibfield  {journal}
  {\bibinfo  {journal} {J. Chem. Phys.}\ }\textbf {\bibinfo {volume} {116}},\
  \bibinfo {pages} {3194} (\bibinfo {year} {2002})}\BibitemShut {NoStop}%
\bibitem [{\citenamefont {Bethe}(1931)}]{Bethe_1931}%
  \BibitemOpen
  \bibfield  {author} {\bibinfo {author} {\bibfnamefont {H.}~\bibnamefont
  {Bethe}},\ }\href {\doibase 10.1007/BF01341708} {\bibfield  {journal}
  {\bibinfo  {journal} {Z. Physik}\ }\textbf {\bibinfo {volume} {71}},\
  \bibinfo {pages} {205} (\bibinfo {year} {1931})}\BibitemShut {NoStop}%
\bibitem [{\citenamefont {Shibuya}\ and\ \citenamefont
  {McKoy}(1970)}]{Shibuya_1970}%
  \BibitemOpen
  \bibfield  {author} {\bibinfo {author} {\bibfnamefont {T.-I.}\ \bibnamefont
  {Shibuya}}\ and\ \bibinfo {author} {\bibfnamefont {V.}~\bibnamefont
  {McKoy}},\ }\href {\doibase 10.1103/PhysRevA.2.2208} {\bibfield  {journal}
  {\bibinfo  {journal} {Phys. Rev. A}\ }\textbf {\bibinfo {volume} {2}},\
  \bibinfo {pages} {2208} (\bibinfo {year} {1970})}\BibitemShut {NoStop}%
\bibitem [{\citenamefont {Krylov}(2008)}]{Krylov_2008}%
  \BibitemOpen
  \bibfield  {author} {\bibinfo {author} {\bibfnamefont {A.~I.}\ \bibnamefont
  {Krylov}},\ }\href {\doibase 10.1146/annurev.physchem.59.032607.093602}
  {\bibfield  {journal} {\bibinfo  {journal} {Annu. Rev. Phys. Chem.}\ }\textbf
  {\bibinfo {volume} {59}},\ \bibinfo {pages} {433} (\bibinfo {year}
  {2008})}\BibitemShut {NoStop}%
\bibitem [{\citenamefont {Peng}\ \emph {et~al.}(2013)\citenamefont {Peng},
  \citenamefont {Steinmann}, \citenamefont {{van Aggelen}},\ and\ \citenamefont
  {Yang}}]{Peng_2013}%
  \BibitemOpen
  \bibfield  {author} {\bibinfo {author} {\bibfnamefont {D.}~\bibnamefont
  {Peng}}, \bibinfo {author} {\bibfnamefont {S.~N.}\ \bibnamefont {Steinmann}},
  \bibinfo {author} {\bibfnamefont {H.}~\bibnamefont {{van Aggelen}}}, \ and\
  \bibinfo {author} {\bibfnamefont {W.}~\bibnamefont {Yang}},\ }\href {\doibase
  10.1063/1.4820556} {\bibfield  {journal} {\bibinfo  {journal} {J. Chem.
  Phys.}\ }\textbf {\bibinfo {volume} {139}},\ \bibinfo {pages} {104112}
  (\bibinfo {year} {2013})}\BibitemShut {NoStop}%
\bibitem [{\citenamefont {Yang}, \citenamefont {van Aggelen},\ and\
  \citenamefont {Yang}(2013)}]{Yang_2013b}%
  \BibitemOpen
  \bibfield  {author} {\bibinfo {author} {\bibfnamefont {Y.}~\bibnamefont
  {Yang}}, \bibinfo {author} {\bibfnamefont {H.}~\bibnamefont {van Aggelen}}, \
  and\ \bibinfo {author} {\bibfnamefont {W.}~\bibnamefont {Yang}},\ }\href
  {\doibase 10.1063/1.4834875} {\bibfield  {journal} {\bibinfo  {journal} {J.
  Chem. Phys.}\ }\textbf {\bibinfo {volume} {139}},\ \bibinfo {pages} {224105}
  (\bibinfo {year} {2013})}\BibitemShut {NoStop}%
\bibitem [{\citenamefont {Yang}\ \emph {et~al.}(2014)\citenamefont {Yang},
  \citenamefont {Peng}, \citenamefont {Lu},\ and\ \citenamefont
  {Yang}}]{Yang_2014a}%
  \BibitemOpen
  \bibfield  {author} {\bibinfo {author} {\bibfnamefont {Y.}~\bibnamefont
  {Yang}}, \bibinfo {author} {\bibfnamefont {D.}~\bibnamefont {Peng}}, \bibinfo
  {author} {\bibfnamefont {J.}~\bibnamefont {Lu}}, \ and\ \bibinfo {author}
  {\bibfnamefont {W.}~\bibnamefont {Yang}},\ }\href {\doibase
  10.1063/1.4895792} {\bibfield  {journal} {\bibinfo  {journal} {J. Chem.
  Phys.}\ }\textbf {\bibinfo {volume} {141}},\ \bibinfo {pages} {124104}
  (\bibinfo {year} {2014})}\BibitemShut {NoStop}%
\bibitem [{\citenamefont {Peng}\ \emph {et~al.}(2014)\citenamefont {Peng},
  \citenamefont {Yang}, \citenamefont {Zhang},\ and\ \citenamefont
  {Yang}}]{Peng_2014}%
  \BibitemOpen
  \bibfield  {author} {\bibinfo {author} {\bibfnamefont {D.}~\bibnamefont
  {Peng}}, \bibinfo {author} {\bibfnamefont {Y.}~\bibnamefont {Yang}}, \bibinfo
  {author} {\bibfnamefont {P.}~\bibnamefont {Zhang}}, \ and\ \bibinfo {author}
  {\bibfnamefont {W.}~\bibnamefont {Yang}},\ }\href {\doibase
  10.1063/1.4901716} {\bibfield  {journal} {\bibinfo  {journal} {J. Chem.
  Phys.}\ }\textbf {\bibinfo {volume} {141}},\ \bibinfo {pages} {214102}
  (\bibinfo {year} {2014})}\BibitemShut {NoStop}%
\bibitem [{\citenamefont {Zhang}\ and\ \citenamefont
  {Yang}(2016)}]{Zhang_2016}%
  \BibitemOpen
  \bibfield  {author} {\bibinfo {author} {\bibfnamefont {D.}~\bibnamefont
  {Zhang}}\ and\ \bibinfo {author} {\bibfnamefont {W.}~\bibnamefont {Yang}},\
  }\href {\doibase 10.1063/1.4964501} {\bibfield  {journal} {\bibinfo
  {journal} {J. Chem. Phys.}\ }\textbf {\bibinfo {volume} {145}},\ \bibinfo
  {pages} {144105} (\bibinfo {year} {2016})}\BibitemShut {NoStop}%
\bibitem [{\citenamefont {Sutton}\ \emph {et~al.}(2018)\citenamefont {Sutton},
  \citenamefont {Yang}, \citenamefont {Zhang},\ and\ \citenamefont
  {Yang}}]{Sutton_2018}%
  \BibitemOpen
  \bibfield  {author} {\bibinfo {author} {\bibfnamefont {C.}~\bibnamefont
  {Sutton}}, \bibinfo {author} {\bibfnamefont {Y.}~\bibnamefont {Yang}},
  \bibinfo {author} {\bibfnamefont {D.}~\bibnamefont {Zhang}}, \ and\ \bibinfo
  {author} {\bibfnamefont {W.}~\bibnamefont {Yang}},\ }\href {\doibase
  10.1021/acs.jpclett.8b01366} {\bibfield  {journal} {\bibinfo  {journal} {J.
  Phys. Chem. Lett.}\ }\textbf {\bibinfo {volume} {9}},\ \bibinfo {pages}
  {4029} (\bibinfo {year} {2018})}\BibitemShut {NoStop}%
\bibitem [{\citenamefont {Krylov}(2000)}]{Krylov_2000b}%
  \BibitemOpen
  \bibfield  {author} {\bibinfo {author} {\bibfnamefont {A.~I.}\ \bibnamefont
  {Krylov}},\ }\href {\doibase 10.1063/1.1308557} {\bibfield  {journal}
  {\bibinfo  {journal} {J. Chem. Phys.}\ }\textbf {\bibinfo {volume} {113}},\
  \bibinfo {pages} {6052} (\bibinfo {year} {2000})}\BibitemShut {NoStop}%
\bibitem [{\citenamefont {Sears}, \citenamefont {Sherrill},\ and\ \citenamefont
  {Krylov}(2003)}]{Sears_2003}%
  \BibitemOpen
  \bibfield  {author} {\bibinfo {author} {\bibfnamefont {J.~S.}\ \bibnamefont
  {Sears}}, \bibinfo {author} {\bibfnamefont {C.~D.}\ \bibnamefont {Sherrill}},
  \ and\ \bibinfo {author} {\bibfnamefont {A.~I.}\ \bibnamefont {Krylov}},\
  }\href {\doibase 10.1063/1.1568735} {\bibfield  {journal} {\bibinfo
  {journal} {J. Chem. Phys.}\ }\textbf {\bibinfo {volume} {118}},\ \bibinfo
  {pages} {9084} (\bibinfo {year} {2003})}\BibitemShut {NoStop}%
\bibitem [{\citenamefont {Casanova}\ and\ \citenamefont
  {Head-Gordon}(2008)}]{Casanova_2008}%
  \BibitemOpen
  \bibfield  {author} {\bibinfo {author} {\bibfnamefont {D.}~\bibnamefont
  {Casanova}}\ and\ \bibinfo {author} {\bibfnamefont {M.}~\bibnamefont
  {Head-Gordon}},\ }\href {\doibase 10.1063/1.2965131} {\bibfield  {journal}
  {\bibinfo  {journal} {J. Chem. Phys.}\ }\textbf {\bibinfo {volume} {129}},\
  \bibinfo {pages} {064104} (\bibinfo {year} {2008})}\BibitemShut {NoStop}%
\bibitem [{\citenamefont {{Huix-Rotllant}}\ \emph {et~al.}(2010)\citenamefont
  {{Huix-Rotllant}}, \citenamefont {Natarajan}, \citenamefont {Ipatov},
  \citenamefont {Muhavini~Wawire}, \citenamefont {Deutsch},\ and\ \citenamefont
  {Casida}}]{Huix-Rotllant_2010}%
  \BibitemOpen
  \bibfield  {author} {\bibinfo {author} {\bibfnamefont {M.}~\bibnamefont
  {{Huix-Rotllant}}}, \bibinfo {author} {\bibfnamefont {B.}~\bibnamefont
  {Natarajan}}, \bibinfo {author} {\bibfnamefont {A.}~\bibnamefont {Ipatov}},
  \bibinfo {author} {\bibfnamefont {C.}~\bibnamefont {Muhavini~Wawire}},
  \bibinfo {author} {\bibfnamefont {T.}~\bibnamefont {Deutsch}}, \ and\
  \bibinfo {author} {\bibfnamefont {M.~E.}\ \bibnamefont {Casida}},\ }\href
  {\doibase 10.1039/c0cp00273a} {\bibfield  {journal} {\bibinfo  {journal}
  {Phys. Chem. Chem. Phys.}\ }\textbf {\bibinfo {volume} {12}},\ \bibinfo
  {pages} {12811} (\bibinfo {year} {2010})}\BibitemShut {NoStop}%
\bibitem [{\citenamefont {Li}\ and\ \citenamefont {Liu}(2010)}]{Li_2010}%
  \BibitemOpen
  \bibfield  {author} {\bibinfo {author} {\bibfnamefont {Z.}~\bibnamefont
  {Li}}\ and\ \bibinfo {author} {\bibfnamefont {W.}~\bibnamefont {Liu}},\
  }\href {\doibase 10.1063/1.3463799} {\bibfield  {journal} {\bibinfo
  {journal} {J. Chem. Phys.}\ }\textbf {\bibinfo {volume} {133}},\ \bibinfo
  {pages} {064106} (\bibinfo {year} {2010})}\BibitemShut {NoStop}%
\bibitem [{\citenamefont {Li}\ \emph {et~al.}(2011)\citenamefont {Li},
  \citenamefont {Liu}, \citenamefont {Zhang},\ and\ \citenamefont
  {Suo}}]{Li_2011a}%
  \BibitemOpen
  \bibfield  {author} {\bibinfo {author} {\bibfnamefont {Z.}~\bibnamefont
  {Li}}, \bibinfo {author} {\bibfnamefont {W.}~\bibnamefont {Liu}}, \bibinfo
  {author} {\bibfnamefont {Y.}~\bibnamefont {Zhang}}, \ and\ \bibinfo {author}
  {\bibfnamefont {B.}~\bibnamefont {Suo}},\ }\href {\doibase 10.1063/1.3573374}
  {\bibfield  {journal} {\bibinfo  {journal} {J. Chem. Phys.}\ }\textbf
  {\bibinfo {volume} {134}},\ \bibinfo {pages} {134101} (\bibinfo {year}
  {2011})}\BibitemShut {NoStop}%
\bibitem [{\citenamefont {Li}\ and\ \citenamefont {Liu}(2011)}]{Li_2011b}%
  \BibitemOpen
  \bibfield  {author} {\bibinfo {author} {\bibfnamefont {Z.}~\bibnamefont
  {Li}}\ and\ \bibinfo {author} {\bibfnamefont {W.}~\bibnamefont {Liu}},\
  }\href {\doibase 10.1063/1.3660688} {\bibfield  {journal} {\bibinfo
  {journal} {J. Chem. Phys.}\ }\textbf {\bibinfo {volume} {135}},\ \bibinfo
  {pages} {194106} (\bibinfo {year} {2011})}\BibitemShut {NoStop}%
\bibitem [{\citenamefont {Zhang}\ and\ \citenamefont
  {Herbert}(2015)}]{Zhang_2015}%
  \BibitemOpen
  \bibfield  {author} {\bibinfo {author} {\bibfnamefont {X.}~\bibnamefont
  {Zhang}}\ and\ \bibinfo {author} {\bibfnamefont {J.~M.}\ \bibnamefont
  {Herbert}},\ }\href {\doibase 10.1063/1.4907376} {\bibfield  {journal}
  {\bibinfo  {journal} {J. Chem. Phys.}\ }\textbf {\bibinfo {volume} {142}},\
  \bibinfo {pages} {064109} (\bibinfo {year} {2015})}\BibitemShut {NoStop}%
\bibitem [{\citenamefont {Lee}\ \emph {et~al.}(2018)\citenamefont {Lee},
  \citenamefont {Filatov}, \citenamefont {Lee},\ and\ \citenamefont
  {Choi}}]{Lee_2018}%
  \BibitemOpen
  \bibfield  {author} {\bibinfo {author} {\bibfnamefont {S.}~\bibnamefont
  {Lee}}, \bibinfo {author} {\bibfnamefont {M.}~\bibnamefont {Filatov}},
  \bibinfo {author} {\bibfnamefont {S.}~\bibnamefont {Lee}}, \ and\ \bibinfo
  {author} {\bibfnamefont {C.~H.}\ \bibnamefont {Choi}},\ }\href {\doibase
  10.1063/1.5044202} {\bibfield  {journal} {\bibinfo  {journal} {J. Chem.
  Phys.}\ }\textbf {\bibinfo {volume} {149}},\ \bibinfo {pages} {104101}
  (\bibinfo {year} {2018})}\BibitemShut {NoStop}%
\bibitem [{\citenamefont {Levchenko}\ and\ \citenamefont
  {Krylov}(2004)}]{Levchenko_2004}%
  \BibitemOpen
  \bibfield  {author} {\bibinfo {author} {\bibfnamefont {S.~V.}\ \bibnamefont
  {Levchenko}}\ and\ \bibinfo {author} {\bibfnamefont {A.~I.}\ \bibnamefont
  {Krylov}},\ }\href {\doibase 10.1063/1.1630018} {\bibfield  {journal}
  {\bibinfo  {journal} {J. Chem. Phys.}\ }\textbf {\bibinfo {volume} {120}},\
  \bibinfo {pages} {175} (\bibinfo {year} {2004})}\BibitemShut {NoStop}%
\bibitem [{\citenamefont {Manohar}\ and\ \citenamefont
  {Krylov}(2008)}]{Manohar_2008}%
  \BibitemOpen
  \bibfield  {author} {\bibinfo {author} {\bibfnamefont {P.~U.}\ \bibnamefont
  {Manohar}}\ and\ \bibinfo {author} {\bibfnamefont {A.~I.}\ \bibnamefont
  {Krylov}},\ }\href {\doibase 10.1063/1.3013087} {\bibfield  {journal}
  {\bibinfo  {journal} {J. Chem. Phys.}\ }\textbf {\bibinfo {volume} {129}},\
  \bibinfo {pages} {194105} (\bibinfo {year} {2008})}\BibitemShut {NoStop}%
\bibitem [{\citenamefont {Casanova}\ \emph {et~al.}(2009)\citenamefont
  {Casanova}, \citenamefont {Slipchenko}, \citenamefont {Krylov},\ and\
  \citenamefont {Head-Gordon}}]{Casanova_2009a}%
  \BibitemOpen
  \bibfield  {author} {\bibinfo {author} {\bibfnamefont {D.}~\bibnamefont
  {Casanova}}, \bibinfo {author} {\bibfnamefont {L.~V.}\ \bibnamefont
  {Slipchenko}}, \bibinfo {author} {\bibfnamefont {A.~I.}\ \bibnamefont
  {Krylov}}, \ and\ \bibinfo {author} {\bibfnamefont {M.}~\bibnamefont
  {Head-Gordon}},\ }\href {\doibase 10.1063/1.3066652} {\bibfield  {journal}
  {\bibinfo  {journal} {J. Chem. Phys.}\ }\textbf {\bibinfo {volume} {130}},\
  \bibinfo {pages} {044103} (\bibinfo {year} {2009})}\BibitemShut {NoStop}%
\bibitem [{\citenamefont {Dutta}, \citenamefont {Pal},\ and\ \citenamefont
  {Ghosh}(2013)}]{Dutta_2013}%
  \BibitemOpen
  \bibfield  {author} {\bibinfo {author} {\bibfnamefont {A.~K.}\ \bibnamefont
  {Dutta}}, \bibinfo {author} {\bibfnamefont {S.}~\bibnamefont {Pal}}, \ and\
  \bibinfo {author} {\bibfnamefont {D.}~\bibnamefont {Ghosh}},\ }\href
  {\doibase 10.1063/1.4821936} {\bibfield  {journal} {\bibinfo  {journal} {J.
  Chem. Phys.}\ }\textbf {\bibinfo {volume} {139}},\ \bibinfo {pages} {124116}
  (\bibinfo {year} {2013})}\BibitemShut {NoStop}%
\bibitem [{\citenamefont {Mato}\ and\ \citenamefont
  {Gordon}(2018)}]{Mato_2018}%
  \BibitemOpen
  \bibfield  {author} {\bibinfo {author} {\bibfnamefont {J.}~\bibnamefont
  {Mato}}\ and\ \bibinfo {author} {\bibfnamefont {M.~S.}\ \bibnamefont
  {Gordon}},\ }\href {\doibase 10.1039/C7CP06837A} {\bibfield  {journal}
  {\bibinfo  {journal} {Phys. Chem. Chem. Phys.}\ }\textbf {\bibinfo {volume}
  {20}},\ \bibinfo {pages} {2615} (\bibinfo {year} {2018})}\BibitemShut
  {NoStop}%
\bibitem [{\citenamefont {Casanova}\ and\ \citenamefont
  {Head-Gordon}(2009)}]{Casanova_2009b}%
  \BibitemOpen
  \bibfield  {author} {\bibinfo {author} {\bibfnamefont {D.}~\bibnamefont
  {Casanova}}\ and\ \bibinfo {author} {\bibfnamefont {M.}~\bibnamefont
  {Head-Gordon}},\ }\href {\doibase 10.1039/B911513G} {\bibfield  {journal}
  {\bibinfo  {journal} {Phys. Chem. Chem. Phys.}\ }\textbf {\bibinfo {volume}
  {11}},\ \bibinfo {pages} {9779} (\bibinfo {year} {2009})}\BibitemShut
  {NoStop}%
\bibitem [{\citenamefont {Shao}, \citenamefont {Head-Gordon},\ and\
  \citenamefont {Krylov}(2003)}]{Shao_2003}%
  \BibitemOpen
  \bibfield  {author} {\bibinfo {author} {\bibfnamefont {Y.}~\bibnamefont
  {Shao}}, \bibinfo {author} {\bibfnamefont {M.}~\bibnamefont {Head-Gordon}}, \
  and\ \bibinfo {author} {\bibfnamefont {A.~I.}\ \bibnamefont {Krylov}},\
  }\href {\doibase 10.1063/1.1545679} {\bibfield  {journal} {\bibinfo
  {journal} {J. Chem. Phys.}\ }\textbf {\bibinfo {volume} {118}},\ \bibinfo
  {pages} {4807} (\bibinfo {year} {2003})}\BibitemShut {NoStop}%
\bibitem [{\citenamefont {Wang}\ and\ \citenamefont
  {Ziegler}(2004)}]{Wang_2004}%
  \BibitemOpen
  \bibfield  {author} {\bibinfo {author} {\bibfnamefont {F.}~\bibnamefont
  {Wang}}\ and\ \bibinfo {author} {\bibfnamefont {T.}~\bibnamefont {Ziegler}},\
  }\href {\doibase 10.1063/1.1821494} {\bibfield  {journal} {\bibinfo
  {journal} {J. Chem. Phys.}\ }\textbf {\bibinfo {volume} {121}},\ \bibinfo
  {pages} {12191} (\bibinfo {year} {2004})}\BibitemShut {NoStop}%
\bibitem [{\citenamefont {Bernard}, \citenamefont {Shao},\ and\ \citenamefont
  {Krylov}(2012)}]{Bernard_2012}%
  \BibitemOpen
  \bibfield  {author} {\bibinfo {author} {\bibfnamefont {Y.~A.}\ \bibnamefont
  {Bernard}}, \bibinfo {author} {\bibfnamefont {Y.}~\bibnamefont {Shao}}, \
  and\ \bibinfo {author} {\bibfnamefont {A.~I.}\ \bibnamefont {Krylov}},\
  }\href {\doibase 10.1063/1.4714499} {\bibfield  {journal} {\bibinfo
  {journal} {J. Chem. Phys.}\ }\textbf {\bibinfo {volume} {136}},\ \bibinfo
  {pages} {204103} (\bibinfo {year} {2012})}\BibitemShut {NoStop}%
\bibitem [{\citenamefont {Lefrancois}, \citenamefont {Wormit},\ and\
  \citenamefont {Dreuw}(2015)}]{Lefrancois_2015}%
  \BibitemOpen
  \bibfield  {author} {\bibinfo {author} {\bibfnamefont {D.}~\bibnamefont
  {Lefrancois}}, \bibinfo {author} {\bibfnamefont {M.}~\bibnamefont {Wormit}},
  \ and\ \bibinfo {author} {\bibfnamefont {A.}~\bibnamefont {Dreuw}},\ }\href
  {\doibase 10.1063/1.4931653} {\bibfield  {journal} {\bibinfo  {journal} {J.
  Chem. Phys.}\ }\textbf {\bibinfo {volume} {143}},\ \bibinfo {pages} {124107}
  (\bibinfo {year} {2015})}\BibitemShut {NoStop}%
\bibitem [{\citenamefont {Lefrancois}, \citenamefont {Rehn},\ and\
  \citenamefont {Dreuw}(2016)}]{Lefrancois_2016}%
  \BibitemOpen
  \bibfield  {author} {\bibinfo {author} {\bibfnamefont {D.}~\bibnamefont
  {Lefrancois}}, \bibinfo {author} {\bibfnamefont {D.~R.}\ \bibnamefont
  {Rehn}}, \ and\ \bibinfo {author} {\bibfnamefont {A.}~\bibnamefont {Dreuw}},\
  }\href {\doibase 10.1063/1.4961298} {\bibfield  {journal} {\bibinfo
  {journal} {J. Chem. Phys.}\ }\textbf {\bibinfo {volume} {145}},\ \bibinfo
  {pages} {084102} (\bibinfo {year} {2016})}\BibitemShut {NoStop}%
\bibitem [{\citenamefont {Mayhall}\ and\ \citenamefont
  {Head-Gordon}(2014)}]{Mayhall_2014a}%
  \BibitemOpen
  \bibfield  {author} {\bibinfo {author} {\bibfnamefont {N.~J.}\ \bibnamefont
  {Mayhall}}\ and\ \bibinfo {author} {\bibfnamefont {M.}~\bibnamefont
  {Head-Gordon}},\ }\href {\doibase 10.1063/1.4889918} {\bibfield  {journal}
  {\bibinfo  {journal} {J. Chem. Phys.}\ }\textbf {\bibinfo {volume} {141}},\
  \bibinfo {pages} {044112} (\bibinfo {year} {2014})}\BibitemShut {NoStop}%
\bibitem [{\citenamefont {Mayhall}, \citenamefont {Goldey},\ and\ \citenamefont
  {Head-Gordon}(2014)}]{Mayhall_2014b}%
  \BibitemOpen
  \bibfield  {author} {\bibinfo {author} {\bibfnamefont {N.~J.}\ \bibnamefont
  {Mayhall}}, \bibinfo {author} {\bibfnamefont {M.}~\bibnamefont {Goldey}}, \
  and\ \bibinfo {author} {\bibfnamefont {M.}~\bibnamefont {Head-Gordon}},\
  }\href {\doibase 10.1021/ct400898p} {\bibfield  {journal} {\bibinfo
  {journal} {J. Chem. Phys.}\ }\textbf {\bibinfo {volume} {10}},\ \bibinfo
  {pages} {589} (\bibinfo {year} {2014})}\BibitemShut {NoStop}%
\bibitem [{\citenamefont {Bell}\ \emph {et~al.}(2013)\citenamefont {Bell},
  \citenamefont {Zimmerman}, \citenamefont {Casanova}, \citenamefont {Goldey},\
  and\ \citenamefont {Head-Gordon}}]{Bell_2013}%
  \BibitemOpen
  \bibfield  {author} {\bibinfo {author} {\bibfnamefont {F.}~\bibnamefont
  {Bell}}, \bibinfo {author} {\bibfnamefont {P.~M.}\ \bibnamefont {Zimmerman}},
  \bibinfo {author} {\bibfnamefont {D.}~\bibnamefont {Casanova}}, \bibinfo
  {author} {\bibfnamefont {M.}~\bibnamefont {Goldey}}, \ and\ \bibinfo {author}
  {\bibfnamefont {M.}~\bibnamefont {Head-Gordon}},\ }\href {\doibase
  10.1039/C2CP43293E} {\bibfield  {journal} {\bibinfo  {journal} {Phys. Chem.
  Chem. Phys.}\ }\textbf {\bibinfo {volume} {15}},\ \bibinfo {pages} {358}
  (\bibinfo {year} {2013})}\BibitemShut {NoStop}%
\bibitem [{\citenamefont {Mayhall}\ \emph {et~al.}(2014)\citenamefont
  {Mayhall}, \citenamefont {Horn}, \citenamefont {Sundstrom},\ and\
  \citenamefont {Head-Gordon}}]{Mayhall_2014c}%
  \BibitemOpen
  \bibfield  {author} {\bibinfo {author} {\bibfnamefont {N.~J.}\ \bibnamefont
  {Mayhall}}, \bibinfo {author} {\bibfnamefont {P.~R.}\ \bibnamefont {Horn}},
  \bibinfo {author} {\bibfnamefont {E.~J.}\ \bibnamefont {Sundstrom}}, \ and\
  \bibinfo {author} {\bibfnamefont {M.}~\bibnamefont {Head-Gordon}},\ }\href
  {\doibase 10.1039/C4CP02818J} {\bibfield  {journal} {\bibinfo  {journal}
  {Phys. Chem. Chem. Phys.}\ }\textbf {\bibinfo {volume} {16}},\ \bibinfo
  {pages} {22694} (\bibinfo {year} {2014})}\BibitemShut {NoStop}%
\bibitem [{\citenamefont {Golubeva}\ \emph {et~al.}(2007)\citenamefont
  {Golubeva}, \citenamefont {Nemukhin}, \citenamefont {Klippenstein},
  \citenamefont {Harding},\ and\ \citenamefont {Krylov}}]{Golubeva_2007}%
  \BibitemOpen
  \bibfield  {author} {\bibinfo {author} {\bibfnamefont {A.~A.}\ \bibnamefont
  {Golubeva}}, \bibinfo {author} {\bibfnamefont {A.~V.}\ \bibnamefont
  {Nemukhin}}, \bibinfo {author} {\bibfnamefont {S.~J.}\ \bibnamefont
  {Klippenstein}}, \bibinfo {author} {\bibfnamefont {L.~B.}\ \bibnamefont
  {Harding}}, \ and\ \bibinfo {author} {\bibfnamefont {A.~I.}\ \bibnamefont
  {Krylov}},\ }\href {\doibase 10.1021/jp0764079} {\bibfield  {journal}
  {\bibinfo  {journal} {J. Phys. Chem. A}\ }\textbf {\bibinfo {volume} {111}},\
  \bibinfo {pages} {13264} (\bibinfo {year} {2007})}\BibitemShut {NoStop}%
\bibitem [{\citenamefont {Slipchenko}\ and\ \citenamefont
  {Krylov}(2002)}]{Slipchenko_2002}%
  \BibitemOpen
  \bibfield  {author} {\bibinfo {author} {\bibfnamefont {L.~V.}\ \bibnamefont
  {Slipchenko}}\ and\ \bibinfo {author} {\bibfnamefont {A.~I.}\ \bibnamefont
  {Krylov}},\ }\href {\doibase 10.1063/1.1498819} {\bibfield  {journal}
  {\bibinfo  {journal} {J. Chem. Phys.}\ }\textbf {\bibinfo {volume} {117}},\
  \bibinfo {pages} {4694} (\bibinfo {year} {2002})}\BibitemShut {NoStop}%
\bibitem [{\citenamefont {Wang}\ and\ \citenamefont
  {Krylov}(2005)}]{Wang_2005}%
  \BibitemOpen
  \bibfield  {author} {\bibinfo {author} {\bibfnamefont {T.}~\bibnamefont
  {Wang}}\ and\ \bibinfo {author} {\bibfnamefont {A.~I.}\ \bibnamefont
  {Krylov}},\ }\href {\doibase 10.1063/1.2018645} {\bibfield  {journal}
  {\bibinfo  {journal} {J. Chem. Phys.}\ }\textbf {\bibinfo {volume} {123}},\
  \bibinfo {pages} {104304} (\bibinfo {year} {2005})}\BibitemShut {NoStop}%
\bibitem [{\citenamefont {Slipchenko}\ and\ \citenamefont
  {Krylov}(2003)}]{Slipchenko_2003}%
  \BibitemOpen
  \bibfield  {author} {\bibinfo {author} {\bibfnamefont {L.~V.}\ \bibnamefont
  {Slipchenko}}\ and\ \bibinfo {author} {\bibfnamefont {A.~I.}\ \bibnamefont
  {Krylov}},\ }\href {\doibase 10.1063/1.1561052} {\bibfield  {journal}
  {\bibinfo  {journal} {J. Chem. Phys.}\ }\textbf {\bibinfo {volume} {118}},\
  \bibinfo {pages} {6874} (\bibinfo {year} {2003})}\BibitemShut {NoStop}%
\bibitem [{\citenamefont {Rinkevicius}\ and\ \citenamefont
  {{\AA}gren}(2010)}]{Rinkevicius_2010}%
  \BibitemOpen
  \bibfield  {author} {\bibinfo {author} {\bibfnamefont {Z.}~\bibnamefont
  {Rinkevicius}}\ and\ \bibinfo {author} {\bibfnamefont {H.}~\bibnamefont
  {{\AA}gren}},\ }\href {\doibase https://doi.org/10.1016/j.cplett.2010.03.074}
  {\bibfield  {journal} {\bibinfo  {journal} {Chem. Phys. Lett.}\ }\textbf
  {\bibinfo {volume} {491}},\ \bibinfo {pages} {132 } (\bibinfo {year}
  {2010})}\BibitemShut {NoStop}%
\bibitem [{\citenamefont {Ibeji}\ and\ \citenamefont
  {Ghosh}(2015)}]{Ibeji_2015}%
  \BibitemOpen
  \bibfield  {author} {\bibinfo {author} {\bibfnamefont {C.~U.}\ \bibnamefont
  {Ibeji}}\ and\ \bibinfo {author} {\bibfnamefont {D.}~\bibnamefont {Ghosh}},\
  }\href {\doibase 10.1039/C5CP00214A} {\bibfield  {journal} {\bibinfo
  {journal} {Phys. Chem. Chem. Phys.}\ }\textbf {\bibinfo {volume} {17}},\
  \bibinfo {pages} {9849} (\bibinfo {year} {2015})}\BibitemShut {NoStop}%
\bibitem [{\citenamefont {Hossain}\ \emph {et~al.}(2017)\citenamefont
  {Hossain}, \citenamefont {Deng}, \citenamefont {Gozem}, \citenamefont
  {Krylov}, \citenamefont {Wang},\ and\ \citenamefont
  {Wenthold}}]{Hossain_2017}%
  \BibitemOpen
  \bibfield  {author} {\bibinfo {author} {\bibfnamefont {E.}~\bibnamefont
  {Hossain}}, \bibinfo {author} {\bibfnamefont {S.~M.}\ \bibnamefont {Deng}},
  \bibinfo {author} {\bibfnamefont {S.}~\bibnamefont {Gozem}}, \bibinfo
  {author} {\bibfnamefont {A.~I.}\ \bibnamefont {Krylov}}, \bibinfo {author}
  {\bibfnamefont {X.-B.}\ \bibnamefont {Wang}}, \ and\ \bibinfo {author}
  {\bibfnamefont {P.~G.}\ \bibnamefont {Wenthold}},\ }\href {\doibase
  10.1021/jacs.7b05197} {\bibfield  {journal} {\bibinfo  {journal} {J. Am.
  Chem. Soc.}\ }\textbf {\bibinfo {volume} {139}},\ \bibinfo {pages} {11138}
  (\bibinfo {year} {2017})}\BibitemShut {NoStop}%
\bibitem [{\citenamefont {Orms}\ \emph {et~al.}(2018)\citenamefont {Orms},
  \citenamefont {Rehn}, \citenamefont {Dreuw},\ and\ \citenamefont
  {Krylov}}]{Orms_2018}%
  \BibitemOpen
  \bibfield  {author} {\bibinfo {author} {\bibfnamefont {N.}~\bibnamefont
  {Orms}}, \bibinfo {author} {\bibfnamefont {D.~R.}\ \bibnamefont {Rehn}},
  \bibinfo {author} {\bibfnamefont {A.}~\bibnamefont {Dreuw}}, \ and\ \bibinfo
  {author} {\bibfnamefont {A.~I.}\ \bibnamefont {Krylov}},\ }\href {\doibase
  10.1021/acs.jctc.7b01012} {\bibfield  {journal} {\bibinfo  {journal} {J.
  Chem. Theory Comput.}\ }\textbf {\bibinfo {volume} {14}},\ \bibinfo {pages}
  {638} (\bibinfo {year} {2018})}\BibitemShut {NoStop}%
\bibitem [{\citenamefont {Luxon}\ \emph {et~al.}(2018)\citenamefont {Luxon},
  \citenamefont {Orms}, \citenamefont {Kanters}, \citenamefont {Krylov},\ and\
  \citenamefont {Parish}}]{Luxon_2018}%
  \BibitemOpen
  \bibfield  {author} {\bibinfo {author} {\bibfnamefont {A.~R.}\ \bibnamefont
  {Luxon}}, \bibinfo {author} {\bibfnamefont {N.}~\bibnamefont {Orms}},
  \bibinfo {author} {\bibfnamefont {R.}~\bibnamefont {Kanters}}, \bibinfo
  {author} {\bibfnamefont {A.~I.}\ \bibnamefont {Krylov}}, \ and\ \bibinfo
  {author} {\bibfnamefont {C.~A.}\ \bibnamefont {Parish}},\ }\href {\doibase
  10.1021/acs.jpca.7b10576} {\bibfield  {journal} {\bibinfo  {journal} {J.
  Phys. Chem. A}\ }\textbf {\bibinfo {volume} {122}},\ \bibinfo {pages} {420}
  (\bibinfo {year} {2018})}\BibitemShut {NoStop}%
\bibitem [{\citenamefont {Casanova}(2012)}]{Casanova_2012}%
  \BibitemOpen
  \bibfield  {author} {\bibinfo {author} {\bibfnamefont {D.}~\bibnamefont
  {Casanova}},\ }\href {\doibase 10.1063/1.4747341} {\bibfield  {journal}
  {\bibinfo  {journal} {J. Chem. Phys.}\ }\textbf {\bibinfo {volume} {137}},\
  \bibinfo {pages} {084105} (\bibinfo {year} {2012})}\BibitemShut {NoStop}%
\bibitem [{\citenamefont {Gozem}, \citenamefont {Krylov},\ and\ \citenamefont
  {Olivucci}(2013)}]{Gozem_2013}%
  \BibitemOpen
  \bibfield  {author} {\bibinfo {author} {\bibfnamefont {S.}~\bibnamefont
  {Gozem}}, \bibinfo {author} {\bibfnamefont {A.~I.}\ \bibnamefont {Krylov}}, \
  and\ \bibinfo {author} {\bibfnamefont {M.}~\bibnamefont {Olivucci}},\ }\href
  {\doibase 10.1021/ct300759z} {\bibfield  {journal} {\bibinfo  {journal} {J.
  Chem. Theory Comput.}\ }\textbf {\bibinfo {volume} {9}},\ \bibinfo {pages}
  {284} (\bibinfo {year} {2013})}\BibitemShut {NoStop}%
\bibitem [{\citenamefont {Nikiforov}\ \emph {et~al.}(2014)\citenamefont
  {Nikiforov}, \citenamefont {Gamez}, \citenamefont {Thiel}, \citenamefont
  {Huix-Rotllant},\ and\ \citenamefont {Filatov}}]{Nikiforov_2014}%
  \BibitemOpen
  \bibfield  {author} {\bibinfo {author} {\bibfnamefont {A.}~\bibnamefont
  {Nikiforov}}, \bibinfo {author} {\bibfnamefont {J.~A.}\ \bibnamefont
  {Gamez}}, \bibinfo {author} {\bibfnamefont {W.}~\bibnamefont {Thiel}},
  \bibinfo {author} {\bibfnamefont {M.}~\bibnamefont {Huix-Rotllant}}, \ and\
  \bibinfo {author} {\bibfnamefont {M.}~\bibnamefont {Filatov}},\ }\href
  {\doibase 10.1063/1.4896372} {\bibfield  {journal} {\bibinfo  {journal} {J.
  Chem. Phys.}\ }\textbf {\bibinfo {volume} {141}},\ \bibinfo {pages} {124122}
  (\bibinfo {year} {2014})}\BibitemShut {NoStop}%
\bibitem [{\citenamefont {Strinati}(1982)}]{Strinati_1982}%
  \BibitemOpen
  \bibfield  {author} {\bibinfo {author} {\bibfnamefont {G.}~\bibnamefont
  {Strinati}},\ }\href {\doibase 10.1103/PhysRevLett.49.1519} {\bibfield
  {journal} {\bibinfo  {journal} {Phys. Rev. Lett.}\ }\textbf {\bibinfo
  {volume} {49}},\ \bibinfo {pages} {1519} (\bibinfo {year}
  {1982})}\BibitemShut {NoStop}%
\bibitem [{\citenamefont {Rohlfing}\ and\ \citenamefont
  {Louie}(2000)}]{Rohlfing_2000}%
  \BibitemOpen
  \bibfield  {author} {\bibinfo {author} {\bibfnamefont {M.}~\bibnamefont
  {Rohlfing}}\ and\ \bibinfo {author} {\bibfnamefont {S.~G.}\ \bibnamefont
  {Louie}},\ }\href {\doibase 10.1103/PhysRevB.62.4927} {\bibfield  {journal}
  {\bibinfo  {journal} {Phys. Rev. B}\ }\textbf {\bibinfo {volume} {62}},\
  \bibinfo {pages} {4927} (\bibinfo {year} {2000})}\BibitemShut {NoStop}%
\bibitem [{\citenamefont {Ma}, \citenamefont {Rohlfing},\ and\ \citenamefont
  {Molteni}(2009{\natexlab{a}})}]{Ma_2009a}%
  \BibitemOpen
  \bibfield  {author} {\bibinfo {author} {\bibfnamefont {Y.}~\bibnamefont
  {Ma}}, \bibinfo {author} {\bibfnamefont {M.}~\bibnamefont {Rohlfing}}, \ and\
  \bibinfo {author} {\bibfnamefont {C.}~\bibnamefont {Molteni}},\ }\href
  {\doibase 10.1103/PhysRevB.80.241405} {\bibfield  {journal} {\bibinfo
  {journal} {Phys. Rev. B}\ }\textbf {\bibinfo {volume} {80}},\ \bibinfo
  {pages} {241405} (\bibinfo {year} {2009}{\natexlab{a}})}\BibitemShut
  {NoStop}%
\bibitem [{\citenamefont {Ma}, \citenamefont {Rohlfing},\ and\ \citenamefont
  {Molteni}(2009{\natexlab{b}})}]{Ma_2009b}%
  \BibitemOpen
  \bibfield  {author} {\bibinfo {author} {\bibfnamefont {Y.}~\bibnamefont
  {Ma}}, \bibinfo {author} {\bibfnamefont {M.}~\bibnamefont {Rohlfing}}, \ and\
  \bibinfo {author} {\bibfnamefont {C.}~\bibnamefont {Molteni}},\ }\href
  {\doibase 10.1021/ct900528h} {\bibfield  {journal} {\bibinfo  {journal} {J.
  Chem. Theory. Comput.}\ }\textbf {\bibinfo {volume} {6}},\ \bibinfo {pages}
  {257} (\bibinfo {year} {2009}{\natexlab{b}})}\BibitemShut {NoStop}%
\bibitem [{\citenamefont {Baumeier}\ \emph {et~al.}(2012)\citenamefont
  {Baumeier}, \citenamefont {Andrienko}, \citenamefont {Ma},\ and\
  \citenamefont {Rohlfing}}]{Baumeier_2012b}%
  \BibitemOpen
  \bibfield  {author} {\bibinfo {author} {\bibfnamefont {B.}~\bibnamefont
  {Baumeier}}, \bibinfo {author} {\bibfnamefont {D.}~\bibnamefont {Andrienko}},
  \bibinfo {author} {\bibfnamefont {Y.}~\bibnamefont {Ma}}, \ and\ \bibinfo
  {author} {\bibfnamefont {M.}~\bibnamefont {Rohlfing}},\ }\href {\doibase
  10.1021/ct2008999} {\bibfield  {journal} {\bibinfo  {journal} {J. Chem.
  Theory Comput.}\ }\textbf {\bibinfo {volume} {8}},\ \bibinfo {pages} {997}
  (\bibinfo {year} {2012})}\BibitemShut {NoStop}%
\bibitem [{\citenamefont {Lettmann}\ and\ \citenamefont
  {Rohlfing}(2019)}]{Lettmann_2019}%
  \BibitemOpen
  \bibfield  {author} {\bibinfo {author} {\bibfnamefont {T.}~\bibnamefont
  {Lettmann}}\ and\ \bibinfo {author} {\bibfnamefont {M.}~\bibnamefont
  {Rohlfing}},\ }\href {\doibase 10.1021/acs.jctc.9b00223} {\bibfield
  {journal} {\bibinfo  {journal} {J. Chem. Theory Comput.}\ }\textbf {\bibinfo
  {volume} {15}},\ \bibinfo {pages} {4547} (\bibinfo {year}
  {2019})}\BibitemShut {NoStop}%
\bibitem [{\citenamefont {Hohenberg}\ and\ \citenamefont
  {Kohn}(1964)}]{Hohenberg_1964}%
  \BibitemOpen
  \bibfield  {author} {\bibinfo {author} {\bibfnamefont {P.}~\bibnamefont
  {Hohenberg}}\ and\ \bibinfo {author} {\bibfnamefont {W.}~\bibnamefont
  {Kohn}},\ }\href {\doibase 10.1103/PhysRev.136.B864} {\bibfield  {journal}
  {\bibinfo  {journal} {Phys. Rev.}\ }\textbf {\bibinfo {volume} {136}},\
  \bibinfo {pages} {B864} (\bibinfo {year} {1964})}\BibitemShut {NoStop}%
\bibitem [{\citenamefont {Kohn}\ and\ \citenamefont {Sham}(1965)}]{Kohn_1965}%
  \BibitemOpen
  \bibfield  {author} {\bibinfo {author} {\bibfnamefont {W.}~\bibnamefont
  {Kohn}}\ and\ \bibinfo {author} {\bibfnamefont {L.~J.}\ \bibnamefont
  {Sham}},\ }\href {\doibase 10.1103/PhysRev.140.A1133} {\bibfield  {journal}
  {\bibinfo  {journal} {Phys. Rev.}\ }\textbf {\bibinfo {volume} {140}},\
  \bibinfo {pages} {A1133} (\bibinfo {year} {1965})}\BibitemShut {NoStop}%
\bibitem [{\citenamefont {Parr}\ and\ \citenamefont {Yang}(1989)}]{ParrBook}%
  \BibitemOpen
  \bibfield  {author} {\bibinfo {author} {\bibfnamefont {R.~G.}\ \bibnamefont
  {Parr}}\ and\ \bibinfo {author} {\bibfnamefont {W.}~\bibnamefont {Yang}},\
  }\href@noop {} {\emph {\bibinfo {title} {Density-Functional Theory of Atoms
  and Molecules}}}\ (\bibinfo  {publisher} {Oxford},\ \bibinfo {address}
  {Clarendon Press},\ \bibinfo {year} {1989})\BibitemShut {NoStop}%
\bibitem [{\citenamefont {Gill}(1994)}]{Gill_1994}%
  \BibitemOpen
  \bibfield  {author} {\bibinfo {author} {\bibfnamefont {P.~M.~W.}\
  \bibnamefont {Gill}},\ }\href {\doibase 10.1016/S0065-3276(08)60019-2}
  {\bibfield  {journal} {\bibinfo  {journal} {Adv. Quantum Chem.}\ }\textbf
  {\bibinfo {volume} {25}},\ \bibinfo {pages} {141} (\bibinfo {year}
  {1994})}\BibitemShut {NoStop}%
\bibitem [{\citenamefont {Strinati}, \citenamefont {Mattausch},\ and\
  \citenamefont {Hanke}(1980)}]{Strinati_1980}%
  \BibitemOpen
  \bibfield  {author} {\bibinfo {author} {\bibfnamefont {G.}~\bibnamefont
  {Strinati}}, \bibinfo {author} {\bibfnamefont {H.~J.}\ \bibnamefont
  {Mattausch}}, \ and\ \bibinfo {author} {\bibfnamefont {W.}~\bibnamefont
  {Hanke}},\ }\href {\doibase 10.1103/PhysRevLett.45.290} {\bibfield  {journal}
  {\bibinfo  {journal} {Phys. Rev. Lett.}\ }\textbf {\bibinfo {volume} {45}},\
  \bibinfo {pages} {290} (\bibinfo {year} {1980})}\BibitemShut {NoStop}%
\bibitem [{\citenamefont {Hybertsen}\ and\ \citenamefont
  {Louie}(1985)}]{Hybertsen_1985a}%
  \BibitemOpen
  \bibfield  {author} {\bibinfo {author} {\bibfnamefont {M.~S.}\ \bibnamefont
  {Hybertsen}}\ and\ \bibinfo {author} {\bibfnamefont {S.~G.}\ \bibnamefont
  {Louie}},\ }\href {\doibase 10.1103/PhysRevLett.55.1418} {\bibfield
  {journal} {\bibinfo  {journal} {Phys. Rev. Lett.}\ }\textbf {\bibinfo
  {volume} {55}},\ \bibinfo {pages} {1418} (\bibinfo {year}
  {1985})}\BibitemShut {NoStop}%
\bibitem [{\citenamefont {Hybertsen}\ and\ \citenamefont
  {Louie}(1986)}]{Hybertsen_1986}%
  \BibitemOpen
  \bibfield  {author} {\bibinfo {author} {\bibfnamefont {M.~S.}\ \bibnamefont
  {Hybertsen}}\ and\ \bibinfo {author} {\bibfnamefont {S.~G.}\ \bibnamefont
  {Louie}},\ }\href {\doibase 10.1103/PhysRevB.34.5390} {\bibfield  {journal}
  {\bibinfo  {journal} {Phys. Rev. B}\ }\textbf {\bibinfo {volume} {34}},\
  \bibinfo {pages} {5390} (\bibinfo {year} {1986})}\BibitemShut {NoStop}%
\bibitem [{\citenamefont {Godby}, \citenamefont {Schl\"uter},\ and\
  \citenamefont {Sham}(1988)}]{Godby_1988}%
  \BibitemOpen
  \bibfield  {author} {\bibinfo {author} {\bibfnamefont {R.~W.}\ \bibnamefont
  {Godby}}, \bibinfo {author} {\bibfnamefont {M.}~\bibnamefont {Schl\"uter}}, \
  and\ \bibinfo {author} {\bibfnamefont {L.~J.}\ \bibnamefont {Sham}},\ }\href
  {\doibase 10.1103/PhysRevB.37.10159} {\bibfield  {journal} {\bibinfo
  {journal} {Phys. Rev. B}\ }\textbf {\bibinfo {volume} {37}},\ \bibinfo
  {pages} {10159} (\bibinfo {year} {1988})}\BibitemShut {NoStop}%
\bibitem [{\citenamefont {von~der Linden}\ and\ \citenamefont
  {Horsch}(1988)}]{Linden_1988}%
  \BibitemOpen
  \bibfield  {author} {\bibinfo {author} {\bibfnamefont {W.}~\bibnamefont
  {von~der Linden}}\ and\ \bibinfo {author} {\bibfnamefont {P.}~\bibnamefont
  {Horsch}},\ }\href {\doibase 10.1103/PhysRevB.37.8351} {\bibfield  {journal}
  {\bibinfo  {journal} {Phys. Rev. B}\ }\textbf {\bibinfo {volume} {37}},\
  \bibinfo {pages} {8351} (\bibinfo {year} {1988})}\BibitemShut {NoStop}%
\bibitem [{\citenamefont {Northrup}, \citenamefont {Hybertsen},\ and\
  \citenamefont {Louie}(1991)}]{Northrup_1991}%
  \BibitemOpen
  \bibfield  {author} {\bibinfo {author} {\bibfnamefont {J.~E.}\ \bibnamefont
  {Northrup}}, \bibinfo {author} {\bibfnamefont {M.~S.}\ \bibnamefont
  {Hybertsen}}, \ and\ \bibinfo {author} {\bibfnamefont {S.~G.}\ \bibnamefont
  {Louie}},\ }\href {\doibase 10.1103/PhysRevLett.66.500} {\bibfield  {journal}
  {\bibinfo  {journal} {Phys. Rev. Lett.}\ }\textbf {\bibinfo {volume} {66}},\
  \bibinfo {pages} {500} (\bibinfo {year} {1991})}\BibitemShut {NoStop}%
\bibitem [{\citenamefont {Blase}, \citenamefont {Zhu},\ and\ \citenamefont
  {Louie}(1994)}]{Blase_1994}%
  \BibitemOpen
  \bibfield  {author} {\bibinfo {author} {\bibfnamefont {X.}~\bibnamefont
  {Blase}}, \bibinfo {author} {\bibfnamefont {X.}~\bibnamefont {Zhu}}, \ and\
  \bibinfo {author} {\bibfnamefont {S.~G.}\ \bibnamefont {Louie}},\ }\href
  {\doibase 10.1103/PhysRevB.49.4973} {\bibfield  {journal} {\bibinfo
  {journal} {Phys. Rev. B}\ }\textbf {\bibinfo {volume} {49}},\ \bibinfo
  {pages} {4973} (\bibinfo {year} {1994})}\BibitemShut {NoStop}%
\bibitem [{\citenamefont {Rohlfing}, \citenamefont {Kr{\"u}ger},\ and\
  \citenamefont {Pollmann}(1995)}]{Rohlfing_1995}%
  \BibitemOpen
  \bibfield  {author} {\bibinfo {author} {\bibfnamefont {M.}~\bibnamefont
  {Rohlfing}}, \bibinfo {author} {\bibfnamefont {P.}~\bibnamefont
  {Kr{\"u}ger}}, \ and\ \bibinfo {author} {\bibfnamefont {J.}~\bibnamefont
  {Pollmann}},\ }\href {\doibase 10.1103/PhysRevB.52.1905} {\bibfield
  {journal} {\bibinfo  {journal} {Phys. Rev. B}\ }\textbf {\bibinfo {volume}
  {52}},\ \bibinfo {pages} {1905} (\bibinfo {year} {1995})}\BibitemShut
  {NoStop}%
\bibitem [{\citenamefont {Shishkin}\ and\ \citenamefont
  {Kresse}(2007)}]{Shishkin_2007}%
  \BibitemOpen
  \bibfield  {author} {\bibinfo {author} {\bibfnamefont {M.}~\bibnamefont
  {Shishkin}}\ and\ \bibinfo {author} {\bibfnamefont {G.}~\bibnamefont
  {Kresse}},\ }\href {\doibase 10.1103/PhysRevB.75.235102} {\bibfield
  {journal} {\bibinfo  {journal} {Phys. Rev. B}\ }\textbf {\bibinfo {volume}
  {75}},\ \bibinfo {pages} {235102} (\bibinfo {year} {2007})}\BibitemShut
  {NoStop}%
\bibitem [{\citenamefont {Loos}, \citenamefont {Romaniello},\ and\
  \citenamefont {Berger}(2018)}]{Loos_2018b}%
  \BibitemOpen
  \bibfield  {author} {\bibinfo {author} {\bibfnamefont {P.~F.}\ \bibnamefont
  {Loos}}, \bibinfo {author} {\bibfnamefont {P.}~\bibnamefont {Romaniello}}, \
  and\ \bibinfo {author} {\bibfnamefont {J.~A.}\ \bibnamefont {Berger}},\
  }\href {\doibase 10.1021/acs.jctc.8b00260} {\bibfield  {journal} {\bibinfo
  {journal} {J. Chem. Theory Comput.}\ }\textbf {\bibinfo {volume} {14}},\
  \bibinfo {pages} {3071} (\bibinfo {year} {2018})}\BibitemShut {NoStop}%
\bibitem [{\citenamefont {Blase}\ and\ \citenamefont
  {Attaccalite}(2011)}]{Blase_2011}%
  \BibitemOpen
  \bibfield  {author} {\bibinfo {author} {\bibfnamefont {X.}~\bibnamefont
  {Blase}}\ and\ \bibinfo {author} {\bibfnamefont {C.}~\bibnamefont
  {Attaccalite}},\ }\href {\doibase 10.1063/1.3655352} {\bibfield  {journal}
  {\bibinfo  {journal} {Appl. Phys. Lett.}\ }\textbf {\bibinfo {volume} {99}},\
  \bibinfo {pages} {171909} (\bibinfo {year} {2011})}\BibitemShut {NoStop}%
\bibitem [{\citenamefont {Faber}\ \emph {et~al.}(2011)\citenamefont {Faber},
  \citenamefont {Attaccalite}, \citenamefont {Olevano}, \citenamefont {Runge},\
  and\ \citenamefont {Blase}}]{Faber_2011}%
  \BibitemOpen
  \bibfield  {author} {\bibinfo {author} {\bibfnamefont {C.}~\bibnamefont
  {Faber}}, \bibinfo {author} {\bibfnamefont {C.}~\bibnamefont {Attaccalite}},
  \bibinfo {author} {\bibfnamefont {V.}~\bibnamefont {Olevano}}, \bibinfo
  {author} {\bibfnamefont {E.}~\bibnamefont {Runge}}, \ and\ \bibinfo {author}
  {\bibfnamefont {X.}~\bibnamefont {Blase}},\ }\href {\doibase
  10.1103/PhysRevB.83.115123} {\bibfield  {journal} {\bibinfo  {journal} {Phys.
  Rev. B}\ }\textbf {\bibinfo {volume} {83}},\ \bibinfo {pages} {115123}
  (\bibinfo {year} {2011})}\BibitemShut {NoStop}%
\bibitem [{\citenamefont {Rangel}\ \emph {et~al.}(2016)\citenamefont {Rangel},
  \citenamefont {Hamed}, \citenamefont {Bruneval},\ and\ \citenamefont
  {Neaton}}]{Rangel_2016}%
  \BibitemOpen
  \bibfield  {author} {\bibinfo {author} {\bibfnamefont {T.}~\bibnamefont
  {Rangel}}, \bibinfo {author} {\bibfnamefont {S.~M.}\ \bibnamefont {Hamed}},
  \bibinfo {author} {\bibfnamefont {F.}~\bibnamefont {Bruneval}}, \ and\
  \bibinfo {author} {\bibfnamefont {J.~B.}\ \bibnamefont {Neaton}},\ }\href
  {\doibase 10.1021/acs.jctc.6b00163} {\bibfield  {journal} {\bibinfo
  {journal} {J. Chem. Theory Comput.}\ }\textbf {\bibinfo {volume} {12}},\
  \bibinfo {pages} {2834} (\bibinfo {year} {2016})}\BibitemShut {NoStop}%
\bibitem [{\citenamefont {Faleev}, \citenamefont {{van Schilfgaarde}},\ and\
  \citenamefont {Kotani}(2004)}]{Faleev_2004}%
  \BibitemOpen
  \bibfield  {author} {\bibinfo {author} {\bibfnamefont {S.~V.}\ \bibnamefont
  {Faleev}}, \bibinfo {author} {\bibfnamefont {M.}~\bibnamefont {{van
  Schilfgaarde}}}, \ and\ \bibinfo {author} {\bibfnamefont {T.}~\bibnamefont
  {Kotani}},\ }\href {\doibase 10.1103/PhysRevLett.93.126406} {\bibfield
  {journal} {\bibinfo  {journal} {Phys. Rev. Lett.}\ }\textbf {\bibinfo
  {volume} {93}},\ \bibinfo {pages} {126406} (\bibinfo {year}
  {2004})}\BibitemShut {NoStop}%
\bibitem [{\citenamefont {{van Schilfgaarde}}, \citenamefont {Kotani},\ and\
  \citenamefont {Faleev}(2006)}]{vanSchilfgaarde_2006}%
  \BibitemOpen
  \bibfield  {author} {\bibinfo {author} {\bibfnamefont {M.}~\bibnamefont {{van
  Schilfgaarde}}}, \bibinfo {author} {\bibfnamefont {T.}~\bibnamefont
  {Kotani}}, \ and\ \bibinfo {author} {\bibfnamefont {S.}~\bibnamefont
  {Faleev}},\ }\href {\doibase 10.1103/PhysRevLett.96.226402} {\bibfield
  {journal} {\bibinfo  {journal} {Phys. Rev. Lett.}\ }\textbf {\bibinfo
  {volume} {96}},\ \bibinfo {pages} {226402} (\bibinfo {year}
  {2006})}\BibitemShut {NoStop}%
\bibitem [{\citenamefont {Kotani}, \citenamefont {{van Schilfgaarde}},\ and\
  \citenamefont {Faleev}(2007)}]{Kotani_2007}%
  \BibitemOpen
  \bibfield  {author} {\bibinfo {author} {\bibfnamefont {T.}~\bibnamefont
  {Kotani}}, \bibinfo {author} {\bibfnamefont {M.}~\bibnamefont {{van
  Schilfgaarde}}}, \ and\ \bibinfo {author} {\bibfnamefont {S.~V.}\
  \bibnamefont {Faleev}},\ }\href {\doibase 10.1103/PhysRevB.76.165106}
  {\bibfield  {journal} {\bibinfo  {journal} {Phys. Rev. B}\ }\textbf {\bibinfo
  {volume} {76}},\ \bibinfo {pages} {165106} (\bibinfo {year}
  {2007})}\BibitemShut {NoStop}%
\bibitem [{\citenamefont {Ke}(2011)}]{Ke_2011}%
  \BibitemOpen
  \bibfield  {author} {\bibinfo {author} {\bibfnamefont {S.-H.}\ \bibnamefont
  {Ke}},\ }\href {\doibase 10.1103/PhysRevB.84.205415} {\bibfield  {journal}
  {\bibinfo  {journal} {Phys. Rev. B}\ }\textbf {\bibinfo {volume} {84}},\
  \bibinfo {pages} {205415} (\bibinfo {year} {2011})}\BibitemShut {NoStop}%
\bibitem [{\citenamefont {Kaplan}\ \emph {et~al.}(2016)\citenamefont {Kaplan},
  \citenamefont {Harding}, \citenamefont {Seiler}, \citenamefont {Weigend},
  \citenamefont {Evers},\ and\ \citenamefont {{van Setten}}}]{Kaplan_2016}%
  \BibitemOpen
  \bibfield  {author} {\bibinfo {author} {\bibfnamefont {F.}~\bibnamefont
  {Kaplan}}, \bibinfo {author} {\bibfnamefont {M.~E.}\ \bibnamefont {Harding}},
  \bibinfo {author} {\bibfnamefont {C.}~\bibnamefont {Seiler}}, \bibinfo
  {author} {\bibfnamefont {F.}~\bibnamefont {Weigend}}, \bibinfo {author}
  {\bibfnamefont {F.}~\bibnamefont {Evers}}, \ and\ \bibinfo {author}
  {\bibfnamefont {M.~J.}\ \bibnamefont {{van Setten}}},\ }\href {\doibase
  10.1021/acs.jctc.5b01238} {\bibfield  {journal} {\bibinfo  {journal} {J.
  Chem. Theory Comput.}\ }\textbf {\bibinfo {volume} {12}},\ \bibinfo {pages}
  {2528} (\bibinfo {year} {2016})}\BibitemShut {NoStop}%
\bibitem [{\citenamefont {Hanke}\ and\ \citenamefont
  {Sham}(1980)}]{Hanke_1980}%
  \BibitemOpen
  \bibfield  {author} {\bibinfo {author} {\bibfnamefont {W.}~\bibnamefont
  {Hanke}}\ and\ \bibinfo {author} {\bibfnamefont {L.~J.}\ \bibnamefont
  {Sham}},\ }\href {\doibase 10.1103/PhysRevB.21.4656} {\bibfield  {journal}
  {\bibinfo  {journal} {Phys. Rev. B}\ }\textbf {\bibinfo {volume} {21}},\
  \bibinfo {pages} {4656} (\bibinfo {year} {1980})}\BibitemShut {NoStop}%
\bibitem [{\citenamefont {Sottile}, \citenamefont {Olevano},\ and\
  \citenamefont {Reining}(2003)}]{Sottile_2003}%
  \BibitemOpen
  \bibfield  {author} {\bibinfo {author} {\bibfnamefont {F.}~\bibnamefont
  {Sottile}}, \bibinfo {author} {\bibfnamefont {V.}~\bibnamefont {Olevano}}, \
  and\ \bibinfo {author} {\bibfnamefont {L.}~\bibnamefont {Reining}},\ }\href
  {\doibase 10.1103/PhysRevLett.91.056402} {\bibfield  {journal} {\bibinfo
  {journal} {Phys. Rev. Lett.}\ }\textbf {\bibinfo {volume} {91}},\ \bibinfo
  {pages} {056402} (\bibinfo {year} {2003})}\BibitemShut {NoStop}%
\bibitem [{\citenamefont {My{\"o}h{\"a}nen}\ \emph {et~al.}(2008)\citenamefont
  {My{\"o}h{\"a}nen}, \citenamefont {Stan}, \citenamefont {Stefanucci},\ and\
  \citenamefont {van Leeuwen}}]{Myohanen_2008}%
  \BibitemOpen
  \bibfield  {author} {\bibinfo {author} {\bibfnamefont {P.}~\bibnamefont
  {My{\"o}h{\"a}nen}}, \bibinfo {author} {\bibfnamefont {A.}~\bibnamefont
  {Stan}}, \bibinfo {author} {\bibfnamefont {G.}~\bibnamefont {Stefanucci}}, \
  and\ \bibinfo {author} {\bibfnamefont {R.}~\bibnamefont {van Leeuwen}},\
  }\href {\doibase 10.1209/0295-5075/84/67001} {\bibfield  {journal} {\bibinfo
  {journal} {Europhys. Lett.}\ }\textbf {\bibinfo {volume} {84}},\ \bibinfo
  {pages} {67001} (\bibinfo {year} {2008})}\BibitemShut {NoStop}%
\bibitem [{\citenamefont {Sakkinen}, \citenamefont {Manninen},\ and\
  \citenamefont {van Leeuwen}(2012)}]{Sakkinen_2012}%
  \BibitemOpen
  \bibfield  {author} {\bibinfo {author} {\bibfnamefont {N.}~\bibnamefont
  {Sakkinen}}, \bibinfo {author} {\bibfnamefont {M.}~\bibnamefont {Manninen}},
  \ and\ \bibinfo {author} {\bibfnamefont {R.}~\bibnamefont {van Leeuwen}},\
  }\href {\doibase 10.1088/1367-2630/14/1/013032} {\bibfield  {journal}
  {\bibinfo  {journal} {New J. Phys.}\ }\textbf {\bibinfo {volume} {14}},\
  \bibinfo {pages} {013032} (\bibinfo {year} {2012})}\BibitemShut {NoStop}%
\bibitem [{\citenamefont {Zhang}, \citenamefont {Steinmann},\ and\
  \citenamefont {Yang}(2013)}]{Zhang_2013}%
  \BibitemOpen
  \bibfield  {author} {\bibinfo {author} {\bibfnamefont {D.}~\bibnamefont
  {Zhang}}, \bibinfo {author} {\bibfnamefont {S.~N.}\ \bibnamefont
  {Steinmann}}, \ and\ \bibinfo {author} {\bibfnamefont {W.}~\bibnamefont
  {Yang}},\ }\href {\doibase 10.1063/1.4824907} {\bibfield  {journal} {\bibinfo
   {journal} {J. Chem. Phys.}\ }\textbf {\bibinfo {volume} {139}},\ \bibinfo
  {pages} {154109} (\bibinfo {year} {2013})}\BibitemShut {NoStop}%
\bibitem [{\citenamefont {Rebolini}\ and\ \citenamefont
  {Toulouse}(2016)}]{Rebolini_2016}%
  \BibitemOpen
  \bibfield  {author} {\bibinfo {author} {\bibfnamefont {E.}~\bibnamefont
  {Rebolini}}\ and\ \bibinfo {author} {\bibfnamefont {J.}~\bibnamefont
  {Toulouse}},\ }\href {\doibase 10.1063/1.4943003} {\bibfield  {journal}
  {\bibinfo  {journal} {J. Chem. Phys.}\ }\textbf {\bibinfo {volume} {144}},\
  \bibinfo {pages} {094107} (\bibinfo {year} {2016})}\BibitemShut {NoStop}%
\bibitem [{\citenamefont {Olevano}, \citenamefont {Toulouse},\ and\
  \citenamefont {Schuck}(2019)}]{Olevano_2019}%
  \BibitemOpen
  \bibfield  {author} {\bibinfo {author} {\bibfnamefont {V.}~\bibnamefont
  {Olevano}}, \bibinfo {author} {\bibfnamefont {J.}~\bibnamefont {Toulouse}}, \
  and\ \bibinfo {author} {\bibfnamefont {P.}~\bibnamefont {Schuck}},\ }\href
  {\doibase 10.1063/1.5080330} {\bibfield  {journal} {\bibinfo  {journal} {J.
  Chem. Phys.}\ }\textbf {\bibinfo {volume} {150}},\ \bibinfo {pages} {084112}
  (\bibinfo {year} {2019})}\BibitemShut {NoStop}%
\bibitem [{\citenamefont {Casida}\ and\ \citenamefont
  {{Huix-Rotllant}}(2016)}]{Casida_2016}%
  \BibitemOpen
  \bibfield  {author} {\bibinfo {author} {\bibfnamefont {M.~E.}\ \bibnamefont
  {Casida}}\ and\ \bibinfo {author} {\bibfnamefont {M.}~\bibnamefont
  {{Huix-Rotllant}}},\ }\href {\doibase 10.1007/128_2015_632} {\bibfield
  {journal} {\bibinfo  {journal} {Top. Curr. Chem.}\ }\textbf {\bibinfo
  {volume} {368}},\ \bibinfo {pages} {1} (\bibinfo {year} {2016})}\BibitemShut
  {NoStop}%
\bibitem [{\citenamefont {Harbach}, \citenamefont {Wormit},\ and\ \citenamefont
  {Dreuw}(2014)}]{Harbach_2014}%
  \BibitemOpen
  \bibfield  {author} {\bibinfo {author} {\bibfnamefont {P.~H.~P.}\
  \bibnamefont {Harbach}}, \bibinfo {author} {\bibfnamefont {M.}~\bibnamefont
  {Wormit}}, \ and\ \bibinfo {author} {\bibfnamefont {A.}~\bibnamefont
  {Dreuw}},\ }\href {\doibase 10.1063/1.4892418} {\bibfield  {journal}
  {\bibinfo  {journal} {J. Chem. Phys.}\ }\textbf {\bibinfo {volume} {141}},\
  \bibinfo {pages} {064113} (\bibinfo {year} {2014})}\BibitemShut {NoStop}%
\bibitem [{\citenamefont {K{\'a}nn{\'a}r}\ and\ \citenamefont
  {Szalay}(2014)}]{Kannar_2014}%
  \BibitemOpen
  \bibfield  {author} {\bibinfo {author} {\bibfnamefont {D.}~\bibnamefont
  {K{\'a}nn{\'a}r}}\ and\ \bibinfo {author} {\bibfnamefont {P.~G.}\
  \bibnamefont {Szalay}},\ }\href {\doibase 10.1021/ct500495n} {\bibfield
  {journal} {\bibinfo  {journal} {J. Chem. Theory Comput.}\ }\textbf {\bibinfo
  {volume} {10}},\ \bibinfo {pages} {3757} (\bibinfo {year}
  {2014})}\BibitemShut {NoStop}%
\bibitem [{\citenamefont {Chrayteh}\ \emph {et~al.}(2021)\citenamefont
  {Chrayteh}, \citenamefont {Blondel}, \citenamefont {Loos},\ and\
  \citenamefont {Jacquemin}}]{Chrayteh_2021}%
  \BibitemOpen
  \bibfield  {author} {\bibinfo {author} {\bibfnamefont {A.}~\bibnamefont
  {Chrayteh}}, \bibinfo {author} {\bibfnamefont {A.}~\bibnamefont {Blondel}},
  \bibinfo {author} {\bibfnamefont {P.-F.}\ \bibnamefont {Loos}}, \ and\
  \bibinfo {author} {\bibfnamefont {D.}~\bibnamefont {Jacquemin}},\ }\href
  {\doibase 10.1021/acs.jctc.0c01111} {\bibfield  {journal} {\bibinfo
  {journal} {J. Chem. Theory Comput.}\ }\textbf {\bibinfo {volume} {17}},\
  \bibinfo {pages} {416} (\bibinfo {year} {2021})}\BibitemShut {NoStop}%
\bibitem [{\citenamefont {Sarkar}\ \emph {et~al.}(ress)\citenamefont {Sarkar},
  \citenamefont {Boggio-Pasqua}, \citenamefont {Loos},\ and\ \citenamefont
  {Jacquemin}}]{Sarkar_2021}%
  \BibitemOpen
  \bibfield  {author} {\bibinfo {author} {\bibfnamefont {R.}~\bibnamefont
  {Sarkar}}, \bibinfo {author} {\bibfnamefont {M.}~\bibnamefont
  {Boggio-Pasqua}}, \bibinfo {author} {\bibfnamefont {P.~F.}\ \bibnamefont
  {Loos}}, \ and\ \bibinfo {author} {\bibfnamefont {D.}~\bibnamefont
  {Jacquemin}},\ }\href@noop {} {\bibfield  {journal} {\bibinfo  {journal} {J.
  Chem. Theory Comput.}\ } (\bibinfo {year} {in press})}\BibitemShut {NoStop}%
\bibitem [{\citenamefont {V{\'e}ril}\ \emph {et~al.}(2018)\citenamefont
  {V{\'e}ril}, \citenamefont {Romaniello}, \citenamefont {Berger},\ and\
  \citenamefont {Loos}}]{Veril_2018}%
  \BibitemOpen
  \bibfield  {author} {\bibinfo {author} {\bibfnamefont {M.}~\bibnamefont
  {V{\'e}ril}}, \bibinfo {author} {\bibfnamefont {P.}~\bibnamefont
  {Romaniello}}, \bibinfo {author} {\bibfnamefont {J.~A.}\ \bibnamefont
  {Berger}}, \ and\ \bibinfo {author} {\bibfnamefont {P.~F.}\ \bibnamefont
  {Loos}},\ }\href {\doibase 10.1021/acs.jctc.8b00745} {\bibfield  {journal}
  {\bibinfo  {journal} {J. Chem. Theory Comput.}\ }\textbf {\bibinfo {volume}
  {14}},\ \bibinfo {pages} {5220} (\bibinfo {year} {2018})}\BibitemShut
  {NoStop}%
\bibitem [{\citenamefont {Berger}, \citenamefont {Loos},\ and\ \citenamefont
  {Romaniello}(2020)}]{Berger_2021}%
  \BibitemOpen
  \bibfield  {author} {\bibinfo {author} {\bibfnamefont {J.~A.}\ \bibnamefont
  {Berger}}, \bibinfo {author} {\bibfnamefont {P.-F.}\ \bibnamefont {Loos}}, \
  and\ \bibinfo {author} {\bibfnamefont {P.}~\bibnamefont {Romaniello}},\
  }\href {\doibase 10.1021/acs.jctc.0c00896} {\bibfield  {journal} {\bibinfo
  {journal} {J. Chem. Theory Comput.}\ }\textbf {\bibinfo {volume} {17}},\
  \bibinfo {pages} {191} (\bibinfo {year} {2020})}\BibitemShut {NoStop}%
\bibitem [{\citenamefont {Loos}\ \emph
  {et~al.}(2020{\natexlab{d}})\citenamefont {Loos}, \citenamefont {Pradines},
  \citenamefont {Scemama}, \citenamefont {Giner},\ and\ \citenamefont
  {Toulouse}}]{Loos_2020a}%
  \BibitemOpen
  \bibfield  {author} {\bibinfo {author} {\bibfnamefont {P.~F.}\ \bibnamefont
  {Loos}}, \bibinfo {author} {\bibfnamefont {B.}~\bibnamefont {Pradines}},
  \bibinfo {author} {\bibfnamefont {A.}~\bibnamefont {Scemama}}, \bibinfo
  {author} {\bibfnamefont {E.}~\bibnamefont {Giner}}, \ and\ \bibinfo {author}
  {\bibfnamefont {J.}~\bibnamefont {Toulouse}},\ }\href {\doibase
  10.1021/acs.jctc.9b01067} {\bibfield  {journal} {\bibinfo  {journal} {J.
  Chem. Theory Comput.}\ }\textbf {\bibinfo {volume} {16}},\ \bibinfo {pages}
  {1018} (\bibinfo {year} {2020}{\natexlab{d}})}\BibitemShut {NoStop}%
\bibitem [{\citenamefont {Stein}, \citenamefont {Kronik},\ and\ \citenamefont
  {Baer}(2009)}]{Stein_2009}%
  \BibitemOpen
  \bibfield  {author} {\bibinfo {author} {\bibfnamefont {T.}~\bibnamefont
  {Stein}}, \bibinfo {author} {\bibfnamefont {L.}~\bibnamefont {Kronik}}, \
  and\ \bibinfo {author} {\bibfnamefont {R.}~\bibnamefont {Baer}},\ }\href
  {\doibase 10.1021/ja8087482} {\bibfield  {journal} {\bibinfo  {journal} {J.
  Am. Chem. Soc.}\ }\textbf {\bibinfo {volume} {131}},\ \bibinfo {pages} {2818}
  (\bibinfo {year} {2009})}\BibitemShut {NoStop}%
\bibitem [{\citenamefont {Stein}\ \emph {et~al.}(2010)\citenamefont {Stein},
  \citenamefont {Eisenberg}, \citenamefont {Kronik},\ and\ \citenamefont
  {Baer}}]{Stein_2010}%
  \BibitemOpen
  \bibfield  {author} {\bibinfo {author} {\bibfnamefont {T.}~\bibnamefont
  {Stein}}, \bibinfo {author} {\bibfnamefont {H.}~\bibnamefont {Eisenberg}},
  \bibinfo {author} {\bibfnamefont {L.}~\bibnamefont {Kronik}}, \ and\ \bibinfo
  {author} {\bibfnamefont {R.}~\bibnamefont {Baer}},\ }\href {\doibase
  10.1103/PhysRevLett.105.266802} {\bibfield  {journal} {\bibinfo  {journal}
  {Phys. Rev. Lett.}\ }\textbf {\bibinfo {volume} {105}},\ \bibinfo {pages}
  {266802} (\bibinfo {year} {2010})}\BibitemShut {NoStop}%
\bibitem [{\citenamefont {Refaely-Abramson}\ \emph {et~al.}(2012)\citenamefont
  {Refaely-Abramson}, \citenamefont {Sharifzadeh}, \citenamefont {Govind},
  \citenamefont {Autschbach}, \citenamefont {Neaton}, \citenamefont {Baer},\
  and\ \citenamefont {Kronik}}]{Refaely-Abramson_2012}%
  \BibitemOpen
  \bibfield  {author} {\bibinfo {author} {\bibfnamefont {S.}~\bibnamefont
  {Refaely-Abramson}}, \bibinfo {author} {\bibfnamefont {S.}~\bibnamefont
  {Sharifzadeh}}, \bibinfo {author} {\bibfnamefont {N.}~\bibnamefont {Govind}},
  \bibinfo {author} {\bibfnamefont {J.}~\bibnamefont {Autschbach}}, \bibinfo
  {author} {\bibfnamefont {J.~B.}\ \bibnamefont {Neaton}}, \bibinfo {author}
  {\bibfnamefont {R.}~\bibnamefont {Baer}}, \ and\ \bibinfo {author}
  {\bibfnamefont {L.}~\bibnamefont {Kronik}},\ }\href {\doibase
  10.1103/PhysRevLett.109.226405} {\bibfield  {journal} {\bibinfo  {journal}
  {Phys. Rev. Lett.}\ }\textbf {\bibinfo {volume} {109}},\ \bibinfo {pages}
  {226405} (\bibinfo {year} {2012})}\BibitemShut {NoStop}%
\bibitem [{\citenamefont {Kronik}\ \emph {et~al.}(2012)\citenamefont {Kronik},
  \citenamefont {Stein}, \citenamefont {{Refaely-Abramson}},\ and\
  \citenamefont {Baer}}]{Kronik_2012}%
  \BibitemOpen
  \bibfield  {author} {\bibinfo {author} {\bibfnamefont {L.}~\bibnamefont
  {Kronik}}, \bibinfo {author} {\bibfnamefont {T.}~\bibnamefont {Stein}},
  \bibinfo {author} {\bibfnamefont {S.}~\bibnamefont {{Refaely-Abramson}}}, \
  and\ \bibinfo {author} {\bibfnamefont {R.}~\bibnamefont {Baer}},\ }\href
  {\doibase 10.1021/ct2009363} {\bibfield  {journal} {\bibinfo  {journal} {J.
  Chem. Theory Comput.}\ }\textbf {\bibinfo {volume} {8}},\ \bibinfo {pages}
  {1515} (\bibinfo {year} {2012})}\BibitemShut {NoStop}%
\bibitem [{\citenamefont {Loos}(2019)}]{QuAcK}%
  \BibitemOpen
  \bibfield  {author} {\bibinfo {author} {\bibfnamefont {P.~F.}\ \bibnamefont
  {Loos}},\ }\href {\doibase 10.5281/zenodo.3745928} {\enquote {\bibinfo
  {title} {{{QuAcK: a software for emerging quantum electronic structure
  methods}}},}\ } (\bibinfo {year} {2019}),\ \bibinfo {note}
  {\url{https://github.com/pfloos/QuAcK}}\BibitemShut {NoStop}%
\bibitem [{\citenamefont {Shao}\ \emph {et~al.}(2015)\citenamefont {Shao},
  \citenamefont {Gan}, \citenamefont {Epifanovsky}, \citenamefont {Gilbert},
  \citenamefont {Wormit}, \citenamefont {Kussmann}, \citenamefont {Lange},
  \citenamefont {Behn}, \citenamefont {Deng}, \citenamefont {Feng},
  \citenamefont {Ghosh}, \citenamefont {Goldey}, \citenamefont {Horn},
  \citenamefont {Jacobson}, \citenamefont {Kaliman}, \citenamefont
  {Khaliullin}, \citenamefont {Ku{\'s}}, \citenamefont {Landau}, \citenamefont
  {Liu}, \citenamefont {Proynov}, \citenamefont {Rhee}, \citenamefont
  {Richard}, \citenamefont {Rohrdanz}, \citenamefont {Steele}, \citenamefont
  {Sundstrom}, \citenamefont {Woodcock}, \citenamefont {Zimmerman},
  \citenamefont {Zuev}, \citenamefont {Albrecht}, \citenamefont {Alguire},
  \citenamefont {Austin}, \citenamefont {Beran}, \citenamefont {Bernard},
  \citenamefont {Berquist}, \citenamefont {Brandhorst}, \citenamefont
  {Bravaya}, \citenamefont {Brown}, \citenamefont {Casanova}, \citenamefont
  {Chang}, \citenamefont {Chen}, \citenamefont {Chien}, \citenamefont
  {Closser}, \citenamefont {Crittenden}, \citenamefont {Diedenhofen},
  \citenamefont {DiStasio}, \citenamefont {Do}, \citenamefont {Dutoi},
  \citenamefont {Edgar}, \citenamefont {Fatehi}, \citenamefont {Fusti-Molnar},
  \citenamefont {Ghysels}, \citenamefont {Golubeva-Zadorozhnaya}, \citenamefont
  {Gomes}, \citenamefont {Hanson-Heine}, \citenamefont {Harbach}, \citenamefont
  {Hauser}, \citenamefont {Hohenstein}, \citenamefont {Holden}, \citenamefont
  {Jagau}, \citenamefont {Ji}, \citenamefont {Kaduk}, \citenamefont
  {Khistyaev}, \citenamefont {Kim}, \citenamefont {Kim}, \citenamefont {King},
  \citenamefont {Klunzinger}, \citenamefont {Kosenkov}, \citenamefont
  {Kowalczyk}, \citenamefont {Krauter}, \citenamefont {Lao}, \citenamefont
  {Laurent}, \citenamefont {Lawler}, \citenamefont {Levchenko}, \citenamefont
  {Lin}, \citenamefont {Liu}, \citenamefont {Livshits}, \citenamefont {Lochan},
  \citenamefont {Luenser}, \citenamefont {Manohar}, \citenamefont {Manzer},
  \citenamefont {Mao}, \citenamefont {Mardirossian}, \citenamefont {Marenich},
  \citenamefont {Maurer}, \citenamefont {Mayhall}, \citenamefont {Neuscamman},
  \citenamefont {Oana}, \citenamefont {Olivares-Amaya}, \citenamefont
  {O'Neill}, \citenamefont {Parkhill}, \citenamefont {Perrine}, \citenamefont
  {Peverati}, \citenamefont {Prociuk}, \citenamefont {Rehn}, \citenamefont
  {Rosta}, \citenamefont {Russ}, \citenamefont {Sharada}, \citenamefont
  {Sharma}, \citenamefont {Small}, \citenamefont {Sodt}, \citenamefont {Stein},
  \citenamefont {St{\"u}ck}, \citenamefont {Su}, \citenamefont {Thom},
  \citenamefont {Tsuchimochi}, \citenamefont {Vanovschi}, \citenamefont {Vogt},
  \citenamefont {Vydrov}, \citenamefont {Wang}, \citenamefont {Watson},
  \citenamefont {Wenzel}, \citenamefont {White}, \citenamefont {Williams},
  \citenamefont {Yang}, \citenamefont {Yeganeh}, \citenamefont {Yost},
  \citenamefont {You}, \citenamefont {Zhang}, \citenamefont {Zhang},
  \citenamefont {Zhao}, \citenamefont {Brooks}, \citenamefont {Chan},
  \citenamefont {Chipman}, \citenamefont {Cramer}, \citenamefont {Goddard},
  \citenamefont {Gordon}, \citenamefont {Hehre}, \citenamefont {Klamt},
  \citenamefont {Schaefer}, \citenamefont {Schmidt}, \citenamefont {Sherrill},
  \citenamefont {Truhlar}, \citenamefont {Warshel}, \citenamefont {Xu},
  \citenamefont {Aspuru-Guzik}, \citenamefont {Baer}, \citenamefont {Bell},
  \citenamefont {Besley}, \citenamefont {Chai}, \citenamefont {Dreuw},
  \citenamefont {Dunietz}, \citenamefont {Furlani}, \citenamefont {Gwaltney},
  \citenamefont {Hsu}, \citenamefont {Jung}, \citenamefont {Kong},
  \citenamefont {Lambrecht}, \citenamefont {Liang}, \citenamefont {Ochsenfeld},
  \citenamefont {Rassolov}, \citenamefont {Slipchenko}, \citenamefont
  {Subotnik}, \citenamefont {Van~Voorhis}, \citenamefont {Herbert},
  \citenamefont {Krylov}, \citenamefont {Gill},\ and\ \citenamefont
  {Head-Gordon}}]{qchem4}%
  \BibitemOpen
  \bibfield  {author} {\bibinfo {author} {\bibfnamefont {Y.}~\bibnamefont
  {Shao}}, \bibinfo {author} {\bibfnamefont {Z.}~\bibnamefont {Gan}}, \bibinfo
  {author} {\bibfnamefont {E.}~\bibnamefont {Epifanovsky}}, \bibinfo {author}
  {\bibfnamefont {A.~T.}\ \bibnamefont {Gilbert}}, \bibinfo {author}
  {\bibfnamefont {M.}~\bibnamefont {Wormit}}, \bibinfo {author} {\bibfnamefont
  {J.}~\bibnamefont {Kussmann}}, \bibinfo {author} {\bibfnamefont {A.~W.}\
  \bibnamefont {Lange}}, \bibinfo {author} {\bibfnamefont {A.}~\bibnamefont
  {Behn}}, \bibinfo {author} {\bibfnamefont {J.}~\bibnamefont {Deng}}, \bibinfo
  {author} {\bibfnamefont {X.}~\bibnamefont {Feng}}, \bibinfo {author}
  {\bibfnamefont {D.}~\bibnamefont {Ghosh}}, \bibinfo {author} {\bibfnamefont
  {M.}~\bibnamefont {Goldey}}, \bibinfo {author} {\bibfnamefont {P.~R.}\
  \bibnamefont {Horn}}, \bibinfo {author} {\bibfnamefont {L.~D.}\ \bibnamefont
  {Jacobson}}, \bibinfo {author} {\bibfnamefont {I.}~\bibnamefont {Kaliman}},
  \bibinfo {author} {\bibfnamefont {R.~Z.}\ \bibnamefont {Khaliullin}},
  \bibinfo {author} {\bibfnamefont {T.}~\bibnamefont {Ku{\'s}}}, \bibinfo
  {author} {\bibfnamefont {A.}~\bibnamefont {Landau}}, \bibinfo {author}
  {\bibfnamefont {J.}~\bibnamefont {Liu}}, \bibinfo {author} {\bibfnamefont
  {E.~I.}\ \bibnamefont {Proynov}}, \bibinfo {author} {\bibfnamefont {Y.~M.}\
  \bibnamefont {Rhee}}, \bibinfo {author} {\bibfnamefont {R.~M.}\ \bibnamefont
  {Richard}}, \bibinfo {author} {\bibfnamefont {M.~A.}\ \bibnamefont
  {Rohrdanz}}, \bibinfo {author} {\bibfnamefont {R.~P.}\ \bibnamefont
  {Steele}}, \bibinfo {author} {\bibfnamefont {E.~J.}\ \bibnamefont
  {Sundstrom}}, \bibinfo {author} {\bibfnamefont {H.~L.}\ \bibnamefont
  {Woodcock}}, \bibinfo {author} {\bibfnamefont {P.~M.}\ \bibnamefont
  {Zimmerman}}, \bibinfo {author} {\bibfnamefont {D.}~\bibnamefont {Zuev}},
  \bibinfo {author} {\bibfnamefont {B.}~\bibnamefont {Albrecht}}, \bibinfo
  {author} {\bibfnamefont {E.}~\bibnamefont {Alguire}}, \bibinfo {author}
  {\bibfnamefont {B.}~\bibnamefont {Austin}}, \bibinfo {author} {\bibfnamefont
  {G.~J.~O.}\ \bibnamefont {Beran}}, \bibinfo {author} {\bibfnamefont {Y.~A.}\
  \bibnamefont {Bernard}}, \bibinfo {author} {\bibfnamefont {E.}~\bibnamefont
  {Berquist}}, \bibinfo {author} {\bibfnamefont {K.}~\bibnamefont
  {Brandhorst}}, \bibinfo {author} {\bibfnamefont {K.~B.}\ \bibnamefont
  {Bravaya}}, \bibinfo {author} {\bibfnamefont {S.~T.}\ \bibnamefont {Brown}},
  \bibinfo {author} {\bibfnamefont {D.}~\bibnamefont {Casanova}}, \bibinfo
  {author} {\bibfnamefont {C.-M.}\ \bibnamefont {Chang}}, \bibinfo {author}
  {\bibfnamefont {Y.}~\bibnamefont {Chen}}, \bibinfo {author} {\bibfnamefont
  {S.~H.}\ \bibnamefont {Chien}}, \bibinfo {author} {\bibfnamefont {K.~D.}\
  \bibnamefont {Closser}}, \bibinfo {author} {\bibfnamefont {D.~L.}\
  \bibnamefont {Crittenden}}, \bibinfo {author} {\bibfnamefont
  {M.}~\bibnamefont {Diedenhofen}}, \bibinfo {author} {\bibfnamefont {R.~A.}\
  \bibnamefont {DiStasio}}, \bibinfo {author} {\bibfnamefont {H.}~\bibnamefont
  {Do}}, \bibinfo {author} {\bibfnamefont {A.~D.}\ \bibnamefont {Dutoi}},
  \bibinfo {author} {\bibfnamefont {R.~G.}\ \bibnamefont {Edgar}}, \bibinfo
  {author} {\bibfnamefont {S.}~\bibnamefont {Fatehi}}, \bibinfo {author}
  {\bibfnamefont {L.}~\bibnamefont {Fusti-Molnar}}, \bibinfo {author}
  {\bibfnamefont {A.}~\bibnamefont {Ghysels}}, \bibinfo {author} {\bibfnamefont
  {A.}~\bibnamefont {Golubeva-Zadorozhnaya}}, \bibinfo {author} {\bibfnamefont
  {J.}~\bibnamefont {Gomes}}, \bibinfo {author} {\bibfnamefont {M.~W.}\
  \bibnamefont {Hanson-Heine}}, \bibinfo {author} {\bibfnamefont {P.~H.}\
  \bibnamefont {Harbach}}, \bibinfo {author} {\bibfnamefont {A.~W.}\
  \bibnamefont {Hauser}}, \bibinfo {author} {\bibfnamefont {E.~G.}\
  \bibnamefont {Hohenstein}}, \bibinfo {author} {\bibfnamefont {Z.~C.}\
  \bibnamefont {Holden}}, \bibinfo {author} {\bibfnamefont {T.-C.}\
  \bibnamefont {Jagau}}, \bibinfo {author} {\bibfnamefont {H.}~\bibnamefont
  {Ji}}, \bibinfo {author} {\bibfnamefont {B.}~\bibnamefont {Kaduk}}, \bibinfo
  {author} {\bibfnamefont {K.}~\bibnamefont {Khistyaev}}, \bibinfo {author}
  {\bibfnamefont {J.}~\bibnamefont {Kim}}, \bibinfo {author} {\bibfnamefont
  {J.}~\bibnamefont {Kim}}, \bibinfo {author} {\bibfnamefont {R.~A.}\
  \bibnamefont {King}}, \bibinfo {author} {\bibfnamefont {P.}~\bibnamefont
  {Klunzinger}}, \bibinfo {author} {\bibfnamefont {D.}~\bibnamefont
  {Kosenkov}}, \bibinfo {author} {\bibfnamefont {T.}~\bibnamefont {Kowalczyk}},
  \bibinfo {author} {\bibfnamefont {C.~M.}\ \bibnamefont {Krauter}}, \bibinfo
  {author} {\bibfnamefont {K.~U.}\ \bibnamefont {Lao}}, \bibinfo {author}
  {\bibfnamefont {A.~D.}\ \bibnamefont {Laurent}}, \bibinfo {author}
  {\bibfnamefont {K.~V.}\ \bibnamefont {Lawler}}, \bibinfo {author}
  {\bibfnamefont {S.~V.}\ \bibnamefont {Levchenko}}, \bibinfo {author}
  {\bibfnamefont {C.~Y.}\ \bibnamefont {Lin}}, \bibinfo {author} {\bibfnamefont
  {F.}~\bibnamefont {Liu}}, \bibinfo {author} {\bibfnamefont {E.}~\bibnamefont
  {Livshits}}, \bibinfo {author} {\bibfnamefont {R.~C.}\ \bibnamefont
  {Lochan}}, \bibinfo {author} {\bibfnamefont {A.}~\bibnamefont {Luenser}},
  \bibinfo {author} {\bibfnamefont {P.}~\bibnamefont {Manohar}}, \bibinfo
  {author} {\bibfnamefont {S.~F.}\ \bibnamefont {Manzer}}, \bibinfo {author}
  {\bibfnamefont {S.-P.}\ \bibnamefont {Mao}}, \bibinfo {author} {\bibfnamefont
  {N.}~\bibnamefont {Mardirossian}}, \bibinfo {author} {\bibfnamefont {A.~V.}\
  \bibnamefont {Marenich}}, \bibinfo {author} {\bibfnamefont {S.~A.}\
  \bibnamefont {Maurer}}, \bibinfo {author} {\bibfnamefont {N.~J.}\
  \bibnamefont {Mayhall}}, \bibinfo {author} {\bibfnamefont {E.}~\bibnamefont
  {Neuscamman}}, \bibinfo {author} {\bibfnamefont {C.~M.}\ \bibnamefont
  {Oana}}, \bibinfo {author} {\bibfnamefont {R.}~\bibnamefont
  {Olivares-Amaya}}, \bibinfo {author} {\bibfnamefont {D.~P.}\ \bibnamefont
  {O'Neill}}, \bibinfo {author} {\bibfnamefont {J.~A.}\ \bibnamefont
  {Parkhill}}, \bibinfo {author} {\bibfnamefont {T.~M.}\ \bibnamefont
  {Perrine}}, \bibinfo {author} {\bibfnamefont {R.}~\bibnamefont {Peverati}},
  \bibinfo {author} {\bibfnamefont {A.}~\bibnamefont {Prociuk}}, \bibinfo
  {author} {\bibfnamefont {D.~R.}\ \bibnamefont {Rehn}}, \bibinfo {author}
  {\bibfnamefont {E.}~\bibnamefont {Rosta}}, \bibinfo {author} {\bibfnamefont
  {N.~J.}\ \bibnamefont {Russ}}, \bibinfo {author} {\bibfnamefont {S.~M.}\
  \bibnamefont {Sharada}}, \bibinfo {author} {\bibfnamefont {S.}~\bibnamefont
  {Sharma}}, \bibinfo {author} {\bibfnamefont {D.~W.}\ \bibnamefont {Small}},
  \bibinfo {author} {\bibfnamefont {A.}~\bibnamefont {Sodt}}, \bibinfo {author}
  {\bibfnamefont {T.}~\bibnamefont {Stein}}, \bibinfo {author} {\bibfnamefont
  {D.}~\bibnamefont {St{\"u}ck}}, \bibinfo {author} {\bibfnamefont {Y.-C.}\
  \bibnamefont {Su}}, \bibinfo {author} {\bibfnamefont {A.~J.}\ \bibnamefont
  {Thom}}, \bibinfo {author} {\bibfnamefont {T.}~\bibnamefont {Tsuchimochi}},
  \bibinfo {author} {\bibfnamefont {V.}~\bibnamefont {Vanovschi}}, \bibinfo
  {author} {\bibfnamefont {L.}~\bibnamefont {Vogt}}, \bibinfo {author}
  {\bibfnamefont {O.}~\bibnamefont {Vydrov}}, \bibinfo {author} {\bibfnamefont
  {T.}~\bibnamefont {Wang}}, \bibinfo {author} {\bibfnamefont {M.~A.}\
  \bibnamefont {Watson}}, \bibinfo {author} {\bibfnamefont {J.}~\bibnamefont
  {Wenzel}}, \bibinfo {author} {\bibfnamefont {A.}~\bibnamefont {White}},
  \bibinfo {author} {\bibfnamefont {C.~F.}\ \bibnamefont {Williams}}, \bibinfo
  {author} {\bibfnamefont {J.}~\bibnamefont {Yang}}, \bibinfo {author}
  {\bibfnamefont {S.}~\bibnamefont {Yeganeh}}, \bibinfo {author} {\bibfnamefont
  {S.~R.}\ \bibnamefont {Yost}}, \bibinfo {author} {\bibfnamefont {Z.-Q.}\
  \bibnamefont {You}}, \bibinfo {author} {\bibfnamefont {I.~Y.}\ \bibnamefont
  {Zhang}}, \bibinfo {author} {\bibfnamefont {X.}~\bibnamefont {Zhang}},
  \bibinfo {author} {\bibfnamefont {Y.}~\bibnamefont {Zhao}}, \bibinfo {author}
  {\bibfnamefont {B.~R.}\ \bibnamefont {Brooks}}, \bibinfo {author}
  {\bibfnamefont {G.~K.}\ \bibnamefont {Chan}}, \bibinfo {author}
  {\bibfnamefont {D.~M.}\ \bibnamefont {Chipman}}, \bibinfo {author}
  {\bibfnamefont {C.~J.}\ \bibnamefont {Cramer}}, \bibinfo {author}
  {\bibfnamefont {W.~A.}\ \bibnamefont {Goddard}}, \bibinfo {author}
  {\bibfnamefont {M.~S.}\ \bibnamefont {Gordon}}, \bibinfo {author}
  {\bibfnamefont {W.~J.}\ \bibnamefont {Hehre}}, \bibinfo {author}
  {\bibfnamefont {A.}~\bibnamefont {Klamt}}, \bibinfo {author} {\bibfnamefont
  {H.~F.}\ \bibnamefont {Schaefer}}, \bibinfo {author} {\bibfnamefont {M.~W.}\
  \bibnamefont {Schmidt}}, \bibinfo {author} {\bibfnamefont {C.~D.}\
  \bibnamefont {Sherrill}}, \bibinfo {author} {\bibfnamefont {D.~G.}\
  \bibnamefont {Truhlar}}, \bibinfo {author} {\bibfnamefont {A.}~\bibnamefont
  {Warshel}}, \bibinfo {author} {\bibfnamefont {X.}~\bibnamefont {Xu}},
  \bibinfo {author} {\bibfnamefont {A.}~\bibnamefont {Aspuru-Guzik}}, \bibinfo
  {author} {\bibfnamefont {R.}~\bibnamefont {Baer}}, \bibinfo {author}
  {\bibfnamefont {A.~T.}\ \bibnamefont {Bell}}, \bibinfo {author}
  {\bibfnamefont {N.~A.}\ \bibnamefont {Besley}}, \bibinfo {author}
  {\bibfnamefont {J.-D.}\ \bibnamefont {Chai}}, \bibinfo {author}
  {\bibfnamefont {A.}~\bibnamefont {Dreuw}}, \bibinfo {author} {\bibfnamefont
  {B.~D.}\ \bibnamefont {Dunietz}}, \bibinfo {author} {\bibfnamefont {T.~R.}\
  \bibnamefont {Furlani}}, \bibinfo {author} {\bibfnamefont {S.~R.}\
  \bibnamefont {Gwaltney}}, \bibinfo {author} {\bibfnamefont {C.-P.}\
  \bibnamefont {Hsu}}, \bibinfo {author} {\bibfnamefont {Y.}~\bibnamefont
  {Jung}}, \bibinfo {author} {\bibfnamefont {J.}~\bibnamefont {Kong}}, \bibinfo
  {author} {\bibfnamefont {D.~S.}\ \bibnamefont {Lambrecht}}, \bibinfo {author}
  {\bibfnamefont {W.}~\bibnamefont {Liang}}, \bibinfo {author} {\bibfnamefont
  {C.}~\bibnamefont {Ochsenfeld}}, \bibinfo {author} {\bibfnamefont {V.~A.}\
  \bibnamefont {Rassolov}}, \bibinfo {author} {\bibfnamefont {L.~V.}\
  \bibnamefont {Slipchenko}}, \bibinfo {author} {\bibfnamefont {J.~E.}\
  \bibnamefont {Subotnik}}, \bibinfo {author} {\bibfnamefont {T.}~\bibnamefont
  {Van~Voorhis}}, \bibinfo {author} {\bibfnamefont {J.~M.}\ \bibnamefont
  {Herbert}}, \bibinfo {author} {\bibfnamefont {A.~I.}\ \bibnamefont {Krylov}},
  \bibinfo {author} {\bibfnamefont {P.~M.}\ \bibnamefont {Gill}}, \ and\
  \bibinfo {author} {\bibfnamefont {M.}~\bibnamefont {Head-Gordon}},\ }\href
  {\doibase 10.1080/00268976.2014.952696} {\bibfield  {journal} {\bibinfo
  {journal} {Mol. Phys.}\ }\textbf {\bibinfo {volume} {113}},\ \bibinfo {pages}
  {184} (\bibinfo {year} {2015})}\BibitemShut {NoStop}%
\bibitem [{\citenamefont {Becke}(1988)}]{Becke_1988}%
  \BibitemOpen
  \bibfield  {author} {\bibinfo {author} {\bibfnamefont {A.~D.}\ \bibnamefont
  {Becke}},\ }\href {\doibase 10.1103/PhysRevA.38.3098} {\bibfield  {journal}
  {\bibinfo  {journal} {Phys. Rev. A}\ }\textbf {\bibinfo {volume} {38}},\
  \bibinfo {pages} {3098} (\bibinfo {year} {1988})}\BibitemShut {NoStop}%
\bibitem [{\citenamefont {Lee}, \citenamefont {Yang},\ and\ \citenamefont
  {Parr}(1988)}]{Lee_1988}%
  \BibitemOpen
  \bibfield  {author} {\bibinfo {author} {\bibfnamefont {C.}~\bibnamefont
  {Lee}}, \bibinfo {author} {\bibfnamefont {W.}~\bibnamefont {Yang}}, \ and\
  \bibinfo {author} {\bibfnamefont {R.~G.}\ \bibnamefont {Parr}},\ }\href
  {\doibase 10.1103/PhysRevB.37.785} {\bibfield  {journal} {\bibinfo  {journal}
  {Phys. Rev. B}\ }\textbf {\bibinfo {volume} {37}},\ \bibinfo {pages} {785}
  (\bibinfo {year} {1988})}\BibitemShut {NoStop}%
\bibitem [{\citenamefont {Becke}(1993{\natexlab{a}})}]{Becke_1993a}%
  \BibitemOpen
  \bibfield  {author} {\bibinfo {author} {\bibfnamefont {A.~D.}\ \bibnamefont
  {Becke}},\ }\href {\doibase 10.1063/1.464913} {\bibfield  {journal} {\bibinfo
   {journal} {J. Chem. Phys.}\ }\textbf {\bibinfo {volume} {98}},\ \bibinfo
  {pages} {5648} (\bibinfo {year} {1993}{\natexlab{a}})}\BibitemShut {NoStop}%
\bibitem [{\citenamefont {Becke}(1993{\natexlab{b}})}]{Becke_1993b}%
  \BibitemOpen
  \bibfield  {author} {\bibinfo {author} {\bibfnamefont {A.~D.}\ \bibnamefont
  {Becke}},\ }\href {\doibase 10.1063/1.464304} {\bibfield  {journal} {\bibinfo
   {journal} {J. Chem. Phys.}\ }\textbf {\bibinfo {volume} {98}},\ \bibinfo
  {pages} {1372} (\bibinfo {year} {1993}{\natexlab{b}})}\BibitemShut {NoStop}%
\bibitem [{\citenamefont {Yanai}, \citenamefont {Tew},\ and\ \citenamefont
  {Handy}(2004)}]{Yanai_2004}%
  \BibitemOpen
  \bibfield  {author} {\bibinfo {author} {\bibfnamefont {T.}~\bibnamefont
  {Yanai}}, \bibinfo {author} {\bibfnamefont {D.~P.}\ \bibnamefont {Tew}}, \
  and\ \bibinfo {author} {\bibfnamefont {N.~C.}\ \bibnamefont {Handy}},\ }\href
  {\doibase 10.1016/j.cplett.2004.06.011} {\bibfield  {journal} {\bibinfo
  {journal} {Chem. Phys. Lett.}\ }\textbf {\bibinfo {volume} {393}},\ \bibinfo
  {pages} {51} (\bibinfo {year} {2004})}\BibitemShut {NoStop}%
\bibitem [{\citenamefont {Weintraub}, \citenamefont {Henderson},\ and\
  \citenamefont {Scuseria}(2009)}]{Weintraub_2009}%
  \BibitemOpen
  \bibfield  {author} {\bibinfo {author} {\bibfnamefont {E.}~\bibnamefont
  {Weintraub}}, \bibinfo {author} {\bibfnamefont {T.~M.}\ \bibnamefont
  {Henderson}}, \ and\ \bibinfo {author} {\bibfnamefont {G.~E.}\ \bibnamefont
  {Scuseria}},\ }\href {\doibase 10.1021/ct800530u} {\bibfield  {journal}
  {\bibinfo  {journal} {J. Chem. Theory Comput.}\ }\textbf {\bibinfo {volume}
  {5}},\ \bibinfo {pages} {754} (\bibinfo {year} {2009})},\ \bibinfo {note}
  {pMID: 26609580},\ \Eprint
  {http://arxiv.org/abs/https://doi.org/10.1021/ct800530u}
  {https://doi.org/10.1021/ct800530u} \BibitemShut {NoStop}%
\bibitem [{\citenamefont {Chai}\ and\ \citenamefont
  {Head-Gordon}(2008{\natexlab{a}})}]{Chai_2008a}%
  \BibitemOpen
  \bibfield  {author} {\bibinfo {author} {\bibfnamefont {J.~D.}\ \bibnamefont
  {Chai}}\ and\ \bibinfo {author} {\bibfnamefont {M.}~\bibnamefont
  {Head-Gordon}},\ }\href@noop {} {\bibfield  {journal} {\bibinfo  {journal}
  {J. Chem. Phys.}\ }\textbf {\bibinfo {volume} {128}},\ \bibinfo {pages}
  {084106} (\bibinfo {year} {2008}{\natexlab{a}})}\BibitemShut {NoStop}%
\bibitem [{\citenamefont {Chai}\ and\ \citenamefont
  {Head-Gordon}(2008{\natexlab{b}})}]{Chai_2008b}%
  \BibitemOpen
  \bibfield  {author} {\bibinfo {author} {\bibfnamefont {J.~D.}\ \bibnamefont
  {Chai}}\ and\ \bibinfo {author} {\bibfnamefont {M.}~\bibnamefont
  {Head-Gordon}},\ }\href@noop {} {\bibfield  {journal} {\bibinfo  {journal}
  {Phys. Chem. Chem. Phys.}\ }\textbf {\bibinfo {volume} {10}},\ \bibinfo
  {pages} {6615} (\bibinfo {year} {2008}{\natexlab{b}})}\BibitemShut {NoStop}%
\bibitem [{\citenamefont {Koch}\ \emph {et~al.}(1990)\citenamefont {Koch},
  \citenamefont {Jensen}, \citenamefont {Jorgensen},\ and\ \citenamefont
  {Helgaker}}]{Koch_1990}%
  \BibitemOpen
  \bibfield  {author} {\bibinfo {author} {\bibfnamefont {H.}~\bibnamefont
  {Koch}}, \bibinfo {author} {\bibfnamefont {H.~J.~A.}\ \bibnamefont {Jensen}},
  \bibinfo {author} {\bibfnamefont {P.}~\bibnamefont {Jorgensen}}, \ and\
  \bibinfo {author} {\bibfnamefont {T.}~\bibnamefont {Helgaker}},\ }\href
  {\doibase 10.1063/1.458815} {\bibfield  {journal} {\bibinfo  {journal} {J.
  Chem. Phys.}\ }\textbf {\bibinfo {volume} {93}},\ \bibinfo {pages} {3345}
  (\bibinfo {year} {1990})}\BibitemShut {NoStop}%
\bibitem [{\citenamefont {Stanton}\ and\ \citenamefont
  {Bartlett}(1993)}]{Stanton_1993}%
  \BibitemOpen
  \bibfield  {author} {\bibinfo {author} {\bibfnamefont {J.~F.}\ \bibnamefont
  {Stanton}}\ and\ \bibinfo {author} {\bibfnamefont {R.~J.}\ \bibnamefont
  {Bartlett}},\ }\href {\doibase 10.1063/1.464746} {\bibfield  {journal}
  {\bibinfo  {journal} {J. Chem. Phys.}\ }\textbf {\bibinfo {volume} {98}},\
  \bibinfo {pages} {7029} (\bibinfo {year} {1993})}\BibitemShut {NoStop}%
\bibitem [{\citenamefont {Koch}\ \emph {et~al.}(1994)\citenamefont {Koch},
  \citenamefont {Kobayashi}, \citenamefont {Sanchez~de Mer{\'a}s},\ and\
  \citenamefont {Jorgensen}}]{Koch_1994}%
  \BibitemOpen
  \bibfield  {author} {\bibinfo {author} {\bibfnamefont {H.}~\bibnamefont
  {Koch}}, \bibinfo {author} {\bibfnamefont {R.}~\bibnamefont {Kobayashi}},
  \bibinfo {author} {\bibfnamefont {A.}~\bibnamefont {Sanchez~de Mer{\'a}s}}, \
  and\ \bibinfo {author} {\bibfnamefont {P.}~\bibnamefont {Jorgensen}},\ }\href
  {\doibase 10.1063/1.466321} {\bibfield  {journal} {\bibinfo  {journal} {J.
  Chem. Phys.}\ }\textbf {\bibinfo {volume} {100}},\ \bibinfo {pages} {4393}
  (\bibinfo {year} {1994})}\BibitemShut {NoStop}%
\bibitem [{\citenamefont {Frisch}\ \emph {et~al.}()\citenamefont {Frisch},
  \citenamefont {Trucks}, \citenamefont {Schlegel}, \citenamefont {Scuseria},
  \citenamefont {Robb}, \citenamefont {Cheeseman}, \citenamefont {Scalmani},
  \citenamefont {Barone}, \citenamefont {Mennucci}, \citenamefont {Petersson},
  \citenamefont {Nakatsuji}, \citenamefont {Caricato}, \citenamefont {Li},
  \citenamefont {Hratchian}, \citenamefont {Izmaylov}, \citenamefont {Bloino},
  \citenamefont {Zheng}, \citenamefont {Sonnenberg}, \citenamefont {Hada},
  \citenamefont {Ehara}, \citenamefont {Toyota}, \citenamefont {Fukuda},
  \citenamefont {Hasegawa}, \citenamefont {Ishida}, \citenamefont {Nakajima},
  \citenamefont {Honda}, \citenamefont {Kitao}, \citenamefont {Nakai},
  \citenamefont {Vreven}, \citenamefont {Montgomery}, \citenamefont {Peralta},
  \citenamefont {Ogliaro}, \citenamefont {Bearpark}, \citenamefont {Heyd},
  \citenamefont {Brothers}, \citenamefont {Kudin}, \citenamefont {Staroverov},
  \citenamefont {Kobayashi}, \citenamefont {Normand}, \citenamefont
  {Raghavachari}, \citenamefont {Rendell}, \citenamefont {Burant},
  \citenamefont {Iyengar}, \citenamefont {Tomasi}, \citenamefont {Cossi},
  \citenamefont {Rega}, \citenamefont {Millam}, \citenamefont {Klene},
  \citenamefont {Knox}, \citenamefont {Cross}, \citenamefont {Bakken},
  \citenamefont {Adamo}, \citenamefont {Jaramillo}, \citenamefont {Gomperts},
  \citenamefont {Stratmann}, \citenamefont {Yazyev}, \citenamefont {Austin},
  \citenamefont {Cammi}, \citenamefont {Pomelli}, \citenamefont {Ochterski},
  \citenamefont {Martin}, \citenamefont {Morokuma}, \citenamefont {Zakrzewski},
  \citenamefont {Voth}, \citenamefont {Salvador}, \citenamefont {Dannenberg},
  \citenamefont {Dapprich}, \citenamefont {Daniels}, \citenamefont {Farkas},
  \citenamefont {Foresman}, \citenamefont {Ortiz}, \citenamefont {Cioslowski},\
  and\ \citenamefont {Fox}}]{g09}%
  \BibitemOpen
  \bibfield  {author} {\bibinfo {author} {\bibfnamefont {M.~J.}\ \bibnamefont
  {Frisch}}, \bibinfo {author} {\bibfnamefont {G.~W.}\ \bibnamefont {Trucks}},
  \bibinfo {author} {\bibfnamefont {H.~B.}\ \bibnamefont {Schlegel}}, \bibinfo
  {author} {\bibfnamefont {G.~E.}\ \bibnamefont {Scuseria}}, \bibinfo {author}
  {\bibfnamefont {M.~A.}\ \bibnamefont {Robb}}, \bibinfo {author}
  {\bibfnamefont {J.~R.}\ \bibnamefont {Cheeseman}}, \bibinfo {author}
  {\bibfnamefont {G.}~\bibnamefont {Scalmani}}, \bibinfo {author}
  {\bibfnamefont {V.}~\bibnamefont {Barone}}, \bibinfo {author} {\bibfnamefont
  {B.}~\bibnamefont {Mennucci}}, \bibinfo {author} {\bibfnamefont {G.~A.}\
  \bibnamefont {Petersson}}, \bibinfo {author} {\bibfnamefont {H.}~\bibnamefont
  {Nakatsuji}}, \bibinfo {author} {\bibfnamefont {M.}~\bibnamefont {Caricato}},
  \bibinfo {author} {\bibfnamefont {X.}~\bibnamefont {Li}}, \bibinfo {author}
  {\bibfnamefont {H.~P.}\ \bibnamefont {Hratchian}}, \bibinfo {author}
  {\bibfnamefont {A.~F.}\ \bibnamefont {Izmaylov}}, \bibinfo {author}
  {\bibfnamefont {J.}~\bibnamefont {Bloino}}, \bibinfo {author} {\bibfnamefont
  {G.}~\bibnamefont {Zheng}}, \bibinfo {author} {\bibfnamefont {J.~L.}\
  \bibnamefont {Sonnenberg}}, \bibinfo {author} {\bibfnamefont
  {M.}~\bibnamefont {Hada}}, \bibinfo {author} {\bibfnamefont {M.}~\bibnamefont
  {Ehara}}, \bibinfo {author} {\bibfnamefont {K.}~\bibnamefont {Toyota}},
  \bibinfo {author} {\bibfnamefont {R.}~\bibnamefont {Fukuda}}, \bibinfo
  {author} {\bibfnamefont {J.}~\bibnamefont {Hasegawa}}, \bibinfo {author}
  {\bibfnamefont {M.}~\bibnamefont {Ishida}}, \bibinfo {author} {\bibfnamefont
  {T.}~\bibnamefont {Nakajima}}, \bibinfo {author} {\bibfnamefont
  {Y.}~\bibnamefont {Honda}}, \bibinfo {author} {\bibfnamefont
  {O.}~\bibnamefont {Kitao}}, \bibinfo {author} {\bibfnamefont
  {H.}~\bibnamefont {Nakai}}, \bibinfo {author} {\bibfnamefont
  {T.}~\bibnamefont {Vreven}}, \bibinfo {author} {\bibfnamefont {J.~A.}\
  \bibnamefont {Montgomery}, \bibfnamefont {{Jr.}}}, \bibinfo {author}
  {\bibfnamefont {J.~E.}\ \bibnamefont {Peralta}}, \bibinfo {author}
  {\bibfnamefont {F.}~\bibnamefont {Ogliaro}}, \bibinfo {author} {\bibfnamefont
  {M.}~\bibnamefont {Bearpark}}, \bibinfo {author} {\bibfnamefont {J.~J.}\
  \bibnamefont {Heyd}}, \bibinfo {author} {\bibfnamefont {E.}~\bibnamefont
  {Brothers}}, \bibinfo {author} {\bibfnamefont {K.~N.}\ \bibnamefont {Kudin}},
  \bibinfo {author} {\bibfnamefont {V.~N.}\ \bibnamefont {Staroverov}},
  \bibinfo {author} {\bibfnamefont {R.}~\bibnamefont {Kobayashi}}, \bibinfo
  {author} {\bibfnamefont {J.}~\bibnamefont {Normand}}, \bibinfo {author}
  {\bibfnamefont {K.}~\bibnamefont {Raghavachari}}, \bibinfo {author}
  {\bibfnamefont {A.}~\bibnamefont {Rendell}}, \bibinfo {author} {\bibfnamefont
  {J.~C.}\ \bibnamefont {Burant}}, \bibinfo {author} {\bibfnamefont {S.~S.}\
  \bibnamefont {Iyengar}}, \bibinfo {author} {\bibfnamefont {J.}~\bibnamefont
  {Tomasi}}, \bibinfo {author} {\bibfnamefont {M.}~\bibnamefont {Cossi}},
  \bibinfo {author} {\bibfnamefont {N.}~\bibnamefont {Rega}}, \bibinfo {author}
  {\bibfnamefont {J.~M.}\ \bibnamefont {Millam}}, \bibinfo {author}
  {\bibfnamefont {M.}~\bibnamefont {Klene}}, \bibinfo {author} {\bibfnamefont
  {J.~E.}\ \bibnamefont {Knox}}, \bibinfo {author} {\bibfnamefont {J.~B.}\
  \bibnamefont {Cross}}, \bibinfo {author} {\bibfnamefont {V.}~\bibnamefont
  {Bakken}}, \bibinfo {author} {\bibfnamefont {C.}~\bibnamefont {Adamo}},
  \bibinfo {author} {\bibfnamefont {J.}~\bibnamefont {Jaramillo}}, \bibinfo
  {author} {\bibfnamefont {R.}~\bibnamefont {Gomperts}}, \bibinfo {author}
  {\bibfnamefont {R.~E.}\ \bibnamefont {Stratmann}}, \bibinfo {author}
  {\bibfnamefont {O.}~\bibnamefont {Yazyev}}, \bibinfo {author} {\bibfnamefont
  {A.~J.}\ \bibnamefont {Austin}}, \bibinfo {author} {\bibfnamefont
  {R.}~\bibnamefont {Cammi}}, \bibinfo {author} {\bibfnamefont
  {C.}~\bibnamefont {Pomelli}}, \bibinfo {author} {\bibfnamefont {J.~W.}\
  \bibnamefont {Ochterski}}, \bibinfo {author} {\bibfnamefont {R.~L.}\
  \bibnamefont {Martin}}, \bibinfo {author} {\bibfnamefont {K.}~\bibnamefont
  {Morokuma}}, \bibinfo {author} {\bibfnamefont {V.~G.}\ \bibnamefont
  {Zakrzewski}}, \bibinfo {author} {\bibfnamefont {G.~A.}\ \bibnamefont
  {Voth}}, \bibinfo {author} {\bibfnamefont {P.}~\bibnamefont {Salvador}},
  \bibinfo {author} {\bibfnamefont {J.~J.}\ \bibnamefont {Dannenberg}},
  \bibinfo {author} {\bibfnamefont {S.}~\bibnamefont {Dapprich}}, \bibinfo
  {author} {\bibfnamefont {A.~D.}\ \bibnamefont {Daniels}}, \bibinfo {author}
  {\bibfnamefont {{\"O}.}~\bibnamefont {Farkas}}, \bibinfo {author}
  {\bibfnamefont {J.~B.}\ \bibnamefont {Foresman}}, \bibinfo {author}
  {\bibfnamefont {J.~V.}\ \bibnamefont {Ortiz}}, \bibinfo {author}
  {\bibfnamefont {J.}~\bibnamefont {Cioslowski}}, \ and\ \bibinfo {author}
  {\bibfnamefont {D.~J.}\ \bibnamefont {Fox}},\ }\href@noop {} {\enquote
  {\bibinfo {title} {Gaussin~09 {R}evision {E}.01},}\ }\bibinfo {note}
  {Gaussian Inc. Wallingford CT 2009}\BibitemShut {NoStop}%
\bibitem [{\citenamefont {Caruso}\ \emph {et~al.}(2013)\citenamefont {Caruso},
  \citenamefont {Rohr}, \citenamefont {Hellgren}, \citenamefont {Ren},
  \citenamefont {Rinke}, \citenamefont {Rubio},\ and\ \citenamefont
  {Scheffler}}]{Caruso_2013}%
  \BibitemOpen
  \bibfield  {author} {\bibinfo {author} {\bibfnamefont {F.}~\bibnamefont
  {Caruso}}, \bibinfo {author} {\bibfnamefont {D.~R.}\ \bibnamefont {Rohr}},
  \bibinfo {author} {\bibfnamefont {M.}~\bibnamefont {Hellgren}}, \bibinfo
  {author} {\bibfnamefont {X.}~\bibnamefont {Ren}}, \bibinfo {author}
  {\bibfnamefont {P.}~\bibnamefont {Rinke}}, \bibinfo {author} {\bibfnamefont
  {A.}~\bibnamefont {Rubio}}, \ and\ \bibinfo {author} {\bibfnamefont
  {M.}~\bibnamefont {Scheffler}},\ }\href {\doibase
  10.1103/PhysRevLett.110.146403} {\bibfield  {journal} {\bibinfo  {journal}
  {Phys. Rev. Lett.}\ }\textbf {\bibinfo {volume} {110}},\ \bibinfo {pages}
  {146403} (\bibinfo {year} {2013})}\BibitemShut {NoStop}%
\bibitem [{\citenamefont {Barca}, \citenamefont {Gilbert},\ and\ \citenamefont
  {Gill}(2014)}]{Barca_2014}%
  \BibitemOpen
  \bibfield  {author} {\bibinfo {author} {\bibfnamefont {G.~M.~J.}\
  \bibnamefont {Barca}}, \bibinfo {author} {\bibfnamefont {A.~T.~B.}\
  \bibnamefont {Gilbert}}, \ and\ \bibinfo {author} {\bibfnamefont {P.~M.~W.}\
  \bibnamefont {Gill}},\ }\href {\doibase 10.1063/1.4896182} {\bibfield
  {journal} {\bibinfo  {journal} {J. Chem. Phys.}\ }\textbf {\bibinfo {volume}
  {141}},\ \bibinfo {pages} {111104} (\bibinfo {year} {2014})}\BibitemShut
  {NoStop}%
\bibitem [{\citenamefont {Vuckovic}\ and\ \citenamefont
  {Gori-Giorgi}(2017)}]{Vuckovic_2017}%
  \BibitemOpen
  \bibfield  {author} {\bibinfo {author} {\bibfnamefont {S.}~\bibnamefont
  {Vuckovic}}\ and\ \bibinfo {author} {\bibfnamefont {P.}~\bibnamefont
  {Gori-Giorgi}},\ }\href {\doibase 10.1021/acs.jpclett.7b01113} {\bibfield
  {journal} {\bibinfo  {journal} {J. Phys. Chem. Lett.}\ }\textbf {\bibinfo
  {volume} {8}},\ \bibinfo {pages} {2799} (\bibinfo {year} {2017})}\BibitemShut
  {NoStop}%
\bibitem [{\citenamefont {Li}\ and\ \citenamefont {Olevano}(2021)}]{Li_2021}%
  \BibitemOpen
  \bibfield  {author} {\bibinfo {author} {\bibfnamefont {J.}~\bibnamefont
  {Li}}\ and\ \bibinfo {author} {\bibfnamefont {V.}~\bibnamefont {Olevano}},\
  }\href {\doibase 10.1103/PhysRevA.103.012809} {\bibfield  {journal} {\bibinfo
   {journal} {Phys. Rev. A}\ }\textbf {\bibinfo {volume} {103}},\ \bibinfo
  {pages} {012809} (\bibinfo {year} {2021})}\BibitemShut {NoStop}%
\bibitem [{\citenamefont {Cohen}, \citenamefont {{Mori-S\'anchez}},\ and\
  \citenamefont {Yang}(2008)}]{Cohen_2008a}%
  \BibitemOpen
  \bibfield  {author} {\bibinfo {author} {\bibfnamefont {A.~J.}\ \bibnamefont
  {Cohen}}, \bibinfo {author} {\bibfnamefont {P.}~\bibnamefont
  {{Mori-S\'anchez}}}, \ and\ \bibinfo {author} {\bibfnamefont
  {W.}~\bibnamefont {Yang}},\ }\href {\doibase 10.1063/1.2987202} {\bibfield
  {journal} {\bibinfo  {journal} {J. Chem. Phys.}\ }\textbf {\bibinfo {volume}
  {129}},\ \bibinfo {pages} {121104} (\bibinfo {year} {2008})}\BibitemShut
  {NoStop}%
\bibitem [{\citenamefont {Cohen}, \citenamefont {Mori-S{\'a}nchez},\ and\
  \citenamefont {Yang}(2008)}]{Cohen_2008c}%
  \BibitemOpen
  \bibfield  {author} {\bibinfo {author} {\bibfnamefont {A.~J.}\ \bibnamefont
  {Cohen}}, \bibinfo {author} {\bibfnamefont {P.}~\bibnamefont
  {Mori-S{\'a}nchez}}, \ and\ \bibinfo {author} {\bibfnamefont
  {W.}~\bibnamefont {Yang}},\ }\href {\doibase 10.1126/science.1158722}
  {\bibfield  {journal} {\bibinfo  {journal} {Science}\ }\textbf {\bibinfo
  {volume} {321}},\ \bibinfo {pages} {792} (\bibinfo {year}
  {2008})}\BibitemShut {NoStop}%
\bibitem [{\citenamefont {Cohen}, \citenamefont {{Mori-S\'anchez}},\ and\
  \citenamefont {Yang}(2012)}]{Cohen_2012}%
  \BibitemOpen
  \bibfield  {author} {\bibinfo {author} {\bibfnamefont {A.~J.}\ \bibnamefont
  {Cohen}}, \bibinfo {author} {\bibfnamefont {P.}~\bibnamefont
  {{Mori-S\'anchez}}}, \ and\ \bibinfo {author} {\bibfnamefont
  {W.}~\bibnamefont {Yang}},\ }\href {\doibase 10.1021/cr200107z} {\bibfield
  {journal} {\bibinfo  {journal} {Chem. Rev.}\ }\textbf {\bibinfo {volume}
  {112}},\ \bibinfo {pages} {289} (\bibinfo {year} {2012})}\BibitemShut
  {NoStop}%
\bibitem [{\citenamefont {Coulson}\ and\ \citenamefont
  {Fischer}(1949)}]{Coulson_1949}%
  \BibitemOpen
  \bibfield  {author} {\bibinfo {author} {\bibfnamefont {C.}~\bibnamefont
  {Coulson}}\ and\ \bibinfo {author} {\bibfnamefont {I.}~\bibnamefont
  {Fischer}},\ }\href {\doibase 10.1080/14786444908521726} {\bibfield
  {journal} {\bibinfo  {journal} {Philos. Mag.}\ }\textbf {\bibinfo {volume}
  {40}},\ \bibinfo {pages} {386} (\bibinfo {year} {1949})}\BibitemShut
  {NoStop}%
\bibitem [{\citenamefont {Balkov{\'a}}\ and\ \citenamefont
  {Bartlett}(1994)}]{Balkova_1994}%
  \BibitemOpen
  \bibfield  {author} {\bibinfo {author} {\bibfnamefont {A.}~\bibnamefont
  {Balkov{\'a}}}\ and\ \bibinfo {author} {\bibfnamefont {R.~J.}\ \bibnamefont
  {Bartlett}},\ }\href {\doibase 10.1063/1.468025} {\bibfield  {journal}
  {\bibinfo  {journal} {J. Chem. Phys.}\ }\textbf {\bibinfo {volume} {101}},\
  \bibinfo {pages} {8972} (\bibinfo {year} {1994})}\BibitemShut {NoStop}%
\bibitem [{\citenamefont {Karadakov}(2008)}]{Karadakov_2008}%
  \BibitemOpen
  \bibfield  {author} {\bibinfo {author} {\bibfnamefont {P.~B.}\ \bibnamefont
  {Karadakov}},\ }\href {\doibase 10.1021/jp8037335} {\bibfield  {journal}
  {\bibinfo  {journal} {J. Phys. Chem. A}\ }\textbf {\bibinfo {volume} {112}},\
  \bibinfo {pages} {7303} (\bibinfo {year} {2008})}\BibitemShut {NoStop}%
\bibitem [{\citenamefont {Li}\ and\ \citenamefont {Paldus}(2009)}]{Li_2009}%
  \BibitemOpen
  \bibfield  {author} {\bibinfo {author} {\bibfnamefont {X.}~\bibnamefont
  {Li}}\ and\ \bibinfo {author} {\bibfnamefont {J.}~\bibnamefont {Paldus}},\
  }\href {\doibase 10.1063/1.3225203} {\bibfield  {journal} {\bibinfo
  {journal} {J. Chem. Phys.}\ }\textbf {\bibinfo {volume} {131}},\ \bibinfo
  {pages} {114103} (\bibinfo {year} {2009})}\BibitemShut {NoStop}%
\bibitem [{\citenamefont {Shen}\ and\ \citenamefont
  {Piecuch}(2012)}]{Shen_2012}%
  \BibitemOpen
  \bibfield  {author} {\bibinfo {author} {\bibfnamefont {J.}~\bibnamefont
  {Shen}}\ and\ \bibinfo {author} {\bibfnamefont {P.}~\bibnamefont {Piecuch}},\
  }\href {\doibase 10.1063/1.3700802} {\bibfield  {journal} {\bibinfo
  {journal} {J. Chem. Phys.}\ }\textbf {\bibinfo {volume} {136}},\ \bibinfo
  {pages} {144104} (\bibinfo {year} {2012})}\BibitemShut {NoStop}%
\bibitem [{\citenamefont {Vitale}, \citenamefont {Alavi},\ and\ \citenamefont
  {Kats}(2020)}]{Vitale_2020}%
  \BibitemOpen
  \bibfield  {author} {\bibinfo {author} {\bibfnamefont {E.}~\bibnamefont
  {Vitale}}, \bibinfo {author} {\bibfnamefont {A.}~\bibnamefont {Alavi}}, \
  and\ \bibinfo {author} {\bibfnamefont {D.}~\bibnamefont {Kats}},\ }\href
  {\doibase 10.1021/acs.jctc.0c00470} {\bibfield  {journal} {\bibinfo
  {journal} {J. Chem. Theory Comput.}\ }\textbf {\bibinfo {volume} {16}},\
  \bibinfo {pages} {5621} (\bibinfo {year} {2020})}\BibitemShut {NoStop}%
\bibitem [{\citenamefont {Head-Gordon}\ \emph {et~al.}(1994)\citenamefont
  {Head-Gordon}, \citenamefont {Rico}, \citenamefont {Oumi},\ and\
  \citenamefont {Lee}}]{Head-Gordon_1994}%
  \BibitemOpen
  \bibfield  {author} {\bibinfo {author} {\bibfnamefont {M.}~\bibnamefont
  {Head-Gordon}}, \bibinfo {author} {\bibfnamefont {R.~J.}\ \bibnamefont
  {Rico}}, \bibinfo {author} {\bibfnamefont {M.}~\bibnamefont {Oumi}}, \ and\
  \bibinfo {author} {\bibfnamefont {T.~J.}\ \bibnamefont {Lee}},\ }\href
  {\doibase 10.1016/0009-2614(94)00070-0} {\bibfield  {journal} {\bibinfo
  {journal} {Chem. Phys. Lett.}\ }\textbf {\bibinfo {volume} {219}},\ \bibinfo
  {pages} {21} (\bibinfo {year} {1994})}\BibitemShut {NoStop}%
\bibitem [{\citenamefont {Head-Gordon}, \citenamefont {Maurice},\ and\
  \citenamefont {Oumi}(1995)}]{Head-Gordon_1995}%
  \BibitemOpen
  \bibfield  {author} {\bibinfo {author} {\bibfnamefont {M.}~\bibnamefont
  {Head-Gordon}}, \bibinfo {author} {\bibfnamefont {D.}~\bibnamefont
  {Maurice}}, \ and\ \bibinfo {author} {\bibfnamefont {M.}~\bibnamefont
  {Oumi}},\ }\href {\doibase 10.1016/0009-2614(95)01111-L} {\bibfield
  {journal} {\bibinfo  {journal} {Chem. Phys. Lett.}\ }\textbf {\bibinfo
  {volume} {246}},\ \bibinfo {pages} {114} (\bibinfo {year}
  {1995})}\BibitemShut {NoStop}%
\bibitem [{\citenamefont {Trofimov}\ and\ \citenamefont
  {Schirmer}(1997)}]{Trofimov_1997}%
  \BibitemOpen
  \bibfield  {author} {\bibinfo {author} {\bibfnamefont {A.}~\bibnamefont
  {Trofimov}}\ and\ \bibinfo {author} {\bibfnamefont {J.}~\bibnamefont
  {Schirmer}},\ }\href {\doibase https://doi.org/10.1016/S0301-0104(96)00303-5}
  {\bibfield  {journal} {\bibinfo  {journal} {Chem. Phys.}\ }\textbf {\bibinfo
  {volume} {214}},\ \bibinfo {pages} {153} (\bibinfo {year}
  {1997})}\BibitemShut {NoStop}%
\bibitem [{\citenamefont {Dreuw}\ and\ \citenamefont
  {Wormit}(2015)}]{Dreuw_2015}%
  \BibitemOpen
  \bibfield  {author} {\bibinfo {author} {\bibfnamefont {A.}~\bibnamefont
  {Dreuw}}\ and\ \bibinfo {author} {\bibfnamefont {M.}~\bibnamefont {Wormit}},\
  }\href {\doibase 10.1002/wcms.1206} {\bibfield  {journal} {\bibinfo
  {journal} {Wiley Interdiscip. Rev. Comput. Mol. Sci.}\ }\textbf {\bibinfo
  {volume} {5}},\ \bibinfo {pages} {82} (\bibinfo {year} {2015})}\BibitemShut
  {NoStop}%
\bibitem [{\citenamefont {Christiansen}, \citenamefont {Koch},\ and\
  \citenamefont {J{\o}rgensen}(1995{\natexlab{b}})}]{Christiansen_1995a}%
  \BibitemOpen
  \bibfield  {author} {\bibinfo {author} {\bibfnamefont {O.}~\bibnamefont
  {Christiansen}}, \bibinfo {author} {\bibfnamefont {H.}~\bibnamefont {Koch}},
  \ and\ \bibinfo {author} {\bibfnamefont {P.}~\bibnamefont {J{\o}rgensen}},\
  }\href {\doibase http://dx.doi.org/10.1016/0009-2614(95)00841-Q} {\bibfield
  {journal} {\bibinfo  {journal} {Chem. Phys. Lett.}\ }\textbf {\bibinfo
  {volume} {243}},\ \bibinfo {pages} {409} (\bibinfo {year}
  {1995}{\natexlab{b}})}\BibitemShut {NoStop}%
\bibitem [{\citenamefont {Trofimov}, \citenamefont {Stelter},\ and\
  \citenamefont {Schirmer}(2002)}]{Trofimov_2002}%
  \BibitemOpen
  \bibfield  {author} {\bibinfo {author} {\bibfnamefont {A.~B.}\ \bibnamefont
  {Trofimov}}, \bibinfo {author} {\bibfnamefont {G.}~\bibnamefont {Stelter}}, \
  and\ \bibinfo {author} {\bibfnamefont {J.}~\bibnamefont {Schirmer}},\ }\href
  {\doibase 10.1063/1.1504708} {\bibfield  {journal} {\bibinfo  {journal} {J.
  Chem. Phys.}\ }\textbf {\bibinfo {volume} {117}},\ \bibinfo {pages} {6402}
  (\bibinfo {year} {2002})}\BibitemShut {NoStop}%
\end{thebibliography}%

\end{document}